\documentclass[3p]{elsarticle}
\makeatletter
\def\ps@pprintTitle{%
 \let\@oddhead\@empty
 \let\@evenhead\@empty
 \def\@oddfoot{\centerline{\thepage}}%
 \let\@evenfoot\@oddfoot}
\makeatother

\usepackage{lineno}

\usepackage{graphbox}
\usepackage{hyperref,doi}

\usepackage[utf8]{inputenc}
\usepackage[english]{babel}

\usepackage{amsmath} 
\usepackage{amssymb}
\usepackage{amsfonts}

\usepackage{ifthen} 
\newboolean{uprightparticles}
\setboolean{uprightparticles}{false} 
 \usepackage{xspace} 
 \usepackage{upgreek}
 \def\lhcb {\mbox{LHCb}\xspace}

\def\MagUp {\mbox{\em Mag\kern -0.05em Up}\xspace}

\ifthenelse{\boolean{uprightparticles}}
{

 \def\Peta        {\ensuremath{\upeta}\xspace}

 \def\Pmu         {\ensuremath{\upmu}\xspace}

 \def\Ppi         {\ensuremath{\uppi}\xspace}

 \def\Pphi        {\ensuremath{\upphi}\xspace}                 
                  
 \def\Pchi        {\ensuremath{\upchi}\xspace}                 
 \def\Ppsi        {\ensuremath{\uppsi}\xspace}

 \def\PDelta      {\ensuremath{\Delta}\xspace}                 
 \def\PXi      {\ensuremath{\Xi}\xspace}                 
 \def\PLambda      {\ensuremath{\Lambda}\xspace}                 
 \def\PSigma      {\ensuremath{\Sigma}\xspace}                 
 \def\POmega      {\ensuremath{\Omega}\xspace}                 
 \def\PUpsilon      {\ensuremath{\Upsilon}\xspace}

 \def\PB      {\ensuremath{\mathrm{B}}\xspace}                 
                  
 \def\PD      {\ensuremath{\mathrm{D}}\xspace}

 \def\PJ      {\ensuremath{\mathrm{J}}\xspace}                 
 \def\PK      {\ensuremath{\mathrm{K}}\xspace}

 \def\PZ      {\ensuremath{\mathrm{Z}}\xspace}                 
                  
 \def\Pb      {\ensuremath{\mathrm{b}}\xspace}                 
 \def\Pc      {\ensuremath{\mathrm{c}}\xspace}                 
 \def\Pd      {\ensuremath{\mathrm{d}}\xspace}                 
 \def\Pe      {\ensuremath{\mathrm{e}}\xspace}

 \def\Pi      {\ensuremath{\mathrm{i}}\xspace}

 \def\Pn      {\ensuremath{\mathrm{n}}\xspace}                 
                  
 \def\Pp      {\ensuremath{\mathrm{p}}\xspace}                 
 \def\Pq      {\ensuremath{\mathrm{q}}\xspace}                 
                  
 \def\Ps      {\ensuremath{\mathrm{s}}\xspace}                 
 \def\Pt      {\ensuremath{\mathrm{t}}\xspace}                 
 \def\Pu      {\ensuremath{\mathrm{u}}\xspace}

}
{

 \def\Peta        {\ensuremath{\eta}\xspace}

 \def\Pmu         {\ensuremath{\mu}\xspace}

 \def\Ppi         {\ensuremath{\pi}\xspace}

 \def\Pphi        {\ensuremath{\phi}\xspace}                 
                  
 \def\Pchi        {\ensuremath{\chi}\xspace}                 
 \def\Ppsi        {\ensuremath{\psi}\xspace}                 
                  
 \mathchardef\PDelta="7101
 \mathchardef\PXi="7104
 \mathchardef\PLambda="7103
 \mathchardef\PSigma="7106
 \mathchardef\POmega="710A
 \mathchardef\PUpsilon="7107
                  
 \def\PB      {\ensuremath{B}\xspace}                 
                  
 \def\PD      {\ensuremath{D}\xspace}

 \def\PJ      {\ensuremath{J}\xspace}                 
 \def\PK      {\ensuremath{K}\xspace}

 \def\PZ      {\ensuremath{Z}\xspace}                 
                  
 \def\Pb      {\ensuremath{b}\xspace}                 
 \def\Pc      {\ensuremath{c}\xspace}                 
 \def\Pd      {\ensuremath{d}\xspace}                 
 \def\Pe      {\ensuremath{e}\xspace}

 \def\Pi      {\ensuremath{i}\xspace}

 \def\Pn      {\ensuremath{n}\xspace}                 
                  
 \def\Pp      {\ensuremath{p}\xspace}                 
 \def\Pq      {\ensuremath{q}\xspace}                 
                  
 \def\Ps      {\ensuremath{s}\xspace}                 
 \def\Pt      {\ensuremath{t}\xspace}                 
 \def\Pu      {\ensuremath{u}\xspace}

}

\makeatletter
\ifcase \@ptsize \relax
  \newcommand{\miniscule}{\@setfontsize\miniscule{4}{5}}
\or
  \newcommand{\miniscule}{\@setfontsize\miniscule{5}{6}}
\or
  \newcommand{\miniscule}{\@setfontsize\miniscule{5}{6}}
\fi
\makeatother

\DeclareRobustCommand{\optbar}[1]{\shortstack{{\miniscule (\rule[.5ex]{1.25em}{.18mm})}
  \\ [-.7ex] $#1$}}

\def\en         {{\ensuremath{\Pe^-}}\xspace}   
\def\ep         {{\ensuremath{\Pe^+}}\xspace}

\def\mun        {{\ensuremath{\Pmu^-}}\xspace}

\def\Z      {{\ensuremath{\PZ}}\xspace}

\def\quark     {{\ensuremath{\Pq}}\xspace}
\def\quarkbar  {{\ensuremath{\overline \quark}}\xspace}

\def\uquark    {{\ensuremath{\Pu}}\xspace}
\def\uquarkbar {{\ensuremath{\overline \uquark}}\xspace}

\def\dquark    {{\ensuremath{\Pd}}\xspace}
\def\dquarkbar {{\ensuremath{\overline \dquark}}\xspace}

\def\squark    {{\ensuremath{\Ps}}\xspace}
\def\squarkbar {{\ensuremath{\overline \squark}}\xspace}

\def\cquark    {{\ensuremath{\Pc}}\xspace}
\def\cquarkbar {{\ensuremath{\overline \cquark}}\xspace}

\def\bquark    {{\ensuremath{\Pb}}\xspace}
\def\bquarkbar {{\ensuremath{\overline \bquark}}\xspace}

\def\tquark    {{\ensuremath{\Pt}}\xspace}

\def\pion   {{\ensuremath{\Ppi}}\xspace}
\def\piz    {{\ensuremath{\pion^0}}\xspace}

\def\pip    {{\ensuremath{\pion^+}}\xspace}
\def\pim    {{\ensuremath{\pion^-}}\xspace}
\def\pipm   {{\ensuremath{\pion^\pm}}\xspace}

\def\kaon    {{\ensuremath{\PK}}\xspace}
  \def\Kbar    {{\kern 0.2em\overline{\kern -0.2em \PK}{}}\xspace}

\def\KorKbar    {\kern 0.18em\optbar{\kern -0.18em K}{}\xspace}
\def\Kz      {{\ensuremath{\kaon^0}}\xspace}
\def\Kzb     {{\ensuremath{\Kbar{}^0}}\xspace}
\def\Kp      {{\ensuremath{\kaon^+}}\xspace}
\def\Km      {{\ensuremath{\kaon^-}}\xspace}
\def\Kpm     {{\ensuremath{\kaon^\pm}}\xspace}

\def\KS      {{\ensuremath{\kaon^0_{\rm\scriptscriptstyle S}}}\xspace}
\def\KL      {{\ensuremath{\kaon^0_{\rm\scriptscriptstyle L}}}\xspace}
\def\Kstarz  {{\ensuremath{\kaon^{*0}}}\xspace}

\def\Kstar   {{\ensuremath{\kaon^*}}\xspace}

\def\Kstarp  {{\ensuremath{\kaon^{*+}}}\xspace}

\newcommand{\etapr}{\ensuremath{\Peta^{\prime}}\xspace}

  \def\Dbar    {{\kern 0.2em\overline{\kern -0.2em \PD}{}}\xspace}
\def\D       {{\ensuremath{\PD}}\xspace}
\def\Db      {{\ensuremath{\Dbar}}\xspace}
\def\DorDbar    {\kern 0.18em\optbar{\kern -0.18em D}{}\xspace}
\def\Dz      {{\ensuremath{\D^0}}\xspace}
\def\Dzb     {{\ensuremath{\Dbar{}^0}}\xspace}
\def\Dp      {{\ensuremath{\D^+}}\xspace}

\def\Dpm     {{\ensuremath{\D^\pm}}\xspace}

\def\Dstar   {{\ensuremath{\D^*}}\xspace}

\def\Dstarp  {{\ensuremath{\D^{*+}}}\xspace}

\def\Ds      {{\ensuremath{\D^+_\squark}}\xspace}
\def\Dsp     {{\ensuremath{\D^+_\squark}}\xspace}
\def\Dsm     {{\ensuremath{\D^-_\squark}}\xspace}
\def\Dspm    {{\ensuremath{\D^{\pm}_\squark}}\xspace}

\def\B       {{\ensuremath{\PB}}\xspace}
\def\Bbar    {{\ensuremath{\kern 0.18em\overline{\kern -0.18em \PB}{}}}\xspace}
\def\Bb      {{\ensuremath{\Bbar}}\xspace}
\def\BorBbar    {\kern 0.18em\optbar{\kern -0.18em B}{}\xspace}
\def\Bz      {{\ensuremath{\B^0}}\xspace}
\def\Bzb     {{\ensuremath{\Bbar{}^0}}\xspace}
\def\Bu      {{\ensuremath{\B^+}}\xspace}
\def\Bub     {{\ensuremath{\B^-}}\xspace}
\def\Bp      {{\ensuremath{\Bu}}\xspace}
\def\Bm      {{\ensuremath{\Bub}}\xspace}
\def\Bpm     {{\ensuremath{\B^\pm}}\xspace}

\def\Bs      {{\ensuremath{\B^0_\squark}}\xspace}
\def\Bsb     {{\ensuremath{\Bbar{}^0_\squark}}\xspace}

\def\Bc      {{\ensuremath{\B_\cquark^+}}\xspace}
\def\Bcp     {{\ensuremath{\B_\cquark^+}}\xspace}

\def\BdorBs  {{\ensuremath{\B^0_{(\squark)}}}\xspace}

\def\jpsi     {{\ensuremath{{\PJ\mskip -3mu/\mskip -2mu\Ppsi\mskip 2mu}}}\xspace}

\def\chiczero {{\ensuremath{\Pchi_{\cquark 0}}}\xspace}

  \def\Y#1S{\ensuremath{\PUpsilon{(#1S)}}\xspace}

\def\TwoS  {{\Y2S}}
\def\ThreeS{{\Y3S}}
\def\FourS {{\Y4S}}
\def\FiveS {{\Y5S}}

\def\proton      {{\ensuremath{\Pp}}\xspace}
\def\antiproton  {{\ensuremath{\overline \proton}}\xspace}
\def\neutron     {{\ensuremath{\Pn}}\xspace}

\def\Lz          {{\ensuremath{\PLambda}}\xspace}
\def\Lbar        {{\ensuremath{\kern 0.1em\overline{\kern -0.1em\PLambda}}}\xspace}
\def\LorLbar     {\kern \thebaroffset\optbar{\kern -\thebaroffset \PLambda}\xspace}

\def\Lambdaresbar{{\ensuremath{\Lbar}}\xspace}

\def\Xires       {{\ensuremath{\PXi}}\xspace}

\def\Omegares    {{\ensuremath{\POmega}}\xspace}

\def\Lc          {{\ensuremath{\Lz^+_\cquark}}\xspace}

\def\Xic         {{\ensuremath{\Xires_\cquark}}\xspace}
\def\Xicz        {{\ensuremath{\Xires^0_\cquark}}\xspace}
\def\Xicp        {{\ensuremath{\Xires^+_\cquark}}\xspace}

\def\Omegac      {{\ensuremath{\Omegares^0_\cquark}}\xspace}

\def\Xiccpp      {{\ensuremath{\Xires^{++}_{\cquark\cquark}}}\xspace}

\def\Lb           {{\ensuremath{\Lz^0_\bquark}}\xspace}

\def\Xib          {{\ensuremath{\Xires_\bquark}}\xspace}
\def\Xibz         {{\ensuremath{\Xires^0_\bquark}}\xspace}

\newcommand{\decay}[2]{\ensuremath{#1\!\to #2}\xspace}         

\def\to                 {\ensuremath{\rightarrow}\xspace}

\def\CP                {{\ensuremath{C\!P}}\xspace}
\def\CPT               {{\ensuremath{C\!PT}}\xspace}

\def\AT#1     {\ensuremath{A_{\mathrm{T}}^{#1}}\xspace}

\def\C#1      {\ensuremath{\mathcal{C}_{#1}}\xspace}                       
\def\Cp#1     {\ensuremath{\mathcal{C}_{#1}^{'}}\xspace}                    
\def\Ceff#1   {\ensuremath{\mathcal{C}_{#1}^{\mathrm{(eff)}}}\xspace}        
\def\Cpeff#1  {\ensuremath{\mathcal{C}_{#1}^{'\mathrm{(eff)}}}\xspace}       
\def\Ope#1    {\ensuremath{\mathcal{O}_{#1}}\xspace}                       
\def\Opep#1   {\ensuremath{\mathcal{O}_{#1}^{'}}\xspace}

\newcommand{\tev}{\ifthenelse{\boolean{inbibliography}}{\ensuremath{~T\kern -0.05em eV}\xspace}{\ensuremath{\mathrm{\,Te\kern -0.1em V}}}\xspace}
\newcommand{\gev}{\ensuremath{\mathrm{\,Ge\kern -0.1em V}}\xspace}
\newcommand{\Tev}{\ensuremath{\mathrm{\,Te\kern -0.1em V}}\xspace}
\newcommand{\gevgev}{\ensuremath{\mathrm{\,Ge\kern -0.1em V^2}}\xspace}
\newcommand{\mev}{\ensuremath{\mathrm{\,Me\kern -0.1em V}}\xspace}
\newcommand{\kev}{\ensuremath{\mathrm{\,ke\kern -0.1em V}}\xspace}
\newcommand{\ev}{\ensuremath{\mathrm{\,e\kern -0.1em V}}\xspace}
\newcommand{\gevc}{\ensuremath{{\mathrm{\,Ge\kern -0.1em V\!/}c}}\xspace}
\newcommand{\mevc}{\ensuremath{{\mathrm{\,Me\kern -0.1em V\!/}c}}\xspace}
\newcommand{\gevcc}{\ensuremath{{\mathrm{\,Ge\kern -0.1em V\!/}c^2}}\xspace}
\newcommand{\gevgevcccc}{\ensuremath{{\mathrm{\,Ge\kern -0.1em V^2\!/}c^4}}\xspace}
\newcommand{\mevcc}{\ensuremath{{\mathrm{\,Me\kern -0.1em V\!/}c^2}}\xspace}

\def\invfb   {\ensuremath{\mbox{\,fb}^{-1}}\xspace}

\def\gsim{{~\raise.15em\hbox{$>$}\kern-.85em
          \lower.35em\hbox{$\sim$}~}\xspace}
\def\lsim{{~\raise.15em\hbox{$<$}\kern-.85em
          \lower.35em\hbox{$\sim$}~}\xspace}

\def\tell1  {TELL1\xspace}
\def\ukl1   {UKL1\xspace}

\def\beq{\begin{equation}}
\def\eeq{\end{equation}}
\def\bea{\begin{eqnarray}}
\def\eea{\end{eqnarray}}

\def\Db {\bar{D}}

\def\Bb {\bar{B}}

\def\mr2 {m_\rho^2 }

  \newcommand{\Ap}{{\cal A}}
\newcommand{\Apsq}{|{\cal A}|^2}
\newcommand{\Am}{\overline{\cal A}}
\newcommand{\Amsq}{|\overline{\cal A}|^2}
\newcommand{\Ad}{\Delta |{\cal A}|^2}

\newcommand{\arp}{\ensuremath{a_\rho }}
\newcommand{\arm}{\ensuremath{{\bar a}_\rho }}
\newcommand{\drp}{\ensuremath{\delta_\rho }}
\newcommand{\drm}{\ensuremath{\bar\delta_\rho }}

\newcommand{\mrr}{\ensuremath{m_{\rho}}}

\newcommand{\BWr}{\ensuremath{BW_{\rho}}}
\newcommand{\BWrsq}{\ensuremath{|BW_{\rho}|^2}}

\newcommand{\gr}{\ensuremath{\Gamma_{\rho}}}

\newcommand{\afp}{\ensuremath{a_{f_0} }}
\newcommand{\afm}{\ensuremath{{\bar a}_{f_0} }}
\newcommand{\dfp}{\ensuremath{\delta_{f_0} }}
\newcommand{\dfm}{\ensuremath{\bar\delta_{f_0} }}

\newcommand{\BWf}{\ensuremath{BW_{f_0}}}
\newcommand{\BWfsq}{\ensuremath{|BW_{f_0}|^2}}

\newcommand{\mf}{\ensuremath{m_{f_0}}}
\newcommand{\gf}{\ensuremath{\Gamma_{f_0}}}

 \usepackage{xcolor}

\biboptions{sort&compress}

\makeindex
\graphicspath{{figs/}} 
\begin{document}

\begin{frontmatter}

\title{Direct \CP violation in beauty and charm hadron decays}

\author{Ignacio Bediaga}
\address{Centro Brasileiro de Pesquisas F\'\i sicas \\ Rua Dr. Xavier Sigaud 150, Urca, CEP 22290-180 Rio de Janeiro , Brazil}
\author{Carla G\"obel\corref{mycorrespondingauthor}}
\cortext[mycorrespondingauthor]{Corresponding author}
\ead{carla.gobel@puc-rio.br}

\address{Pontif\'\i cia Universidade Cat\'olica do Rio de Janeiro\\ Rua Marqu\^es de São Vicente 225, Gávea, CEP 22451-900 Rio de Janeiro, Brazil}

\begin{abstract}
Since the discovery of \CP violation more than 5 decades ago, this phenomenon is still attracting a lot of interest. Among the many fascinating aspects of this subject, this review is dedicated to  direct \CP violation in non-leptonic decays. The advances within the last decade have been enormous, driven by the increasingly large samples of $b$- and $c$-hadron decays, and have led to very interesting results such as large \CP asymmetries in charmless \B decays and the observation of direct \CP violation in the charm sector. We address  the quest for understanding the origin of strong phases, the importance of final state interactions and the relation with \CPT symmetry, and different approaches to measure direct \CP violation in these decays. The main experimental results and their implications are then discussed.  \end{abstract}


\end{frontmatter}
\tableofcontents

\let\oldequation\equation
\let\oldendequation\endequation

\renewenvironment{equation}
  {\linenomathNonumbers\oldequation}
  {\oldendequation\endlinenomath}

\section{Introduction}
\label{introduction}

The experimental observation of charge-parity (\CP) violation in the neutral-kaon system in 1964 by Cronin, Fitch {\it et al.}~\cite{Cronin1964} caused a real shock in the physics community. Just a few years before, Landau presented a convincing theory showing \CP conservation to play  a key role in understanding particle-antiparticle symmetry~\cite{Landau}.  At that time, no one had disagreed with this fundamental idea and, as a consequence, there was no theory predicting the  \CP symmetry breaking.  Among several arguments for \CP to be a good symmetry,  a compelling one was that, assuming \CPT symmetry to be exact (a consequence of Lorentz invariance),  the reversibility of time would naturally lead to \CP conservation.  The observation of \CP violation meant, for the first time in physics, the irreversibility of some processes at the fundamental level.

It was only in 1973 that Kobayashi and Maskawa proposed the  nowadays well-accepted description~\cite{Kobayashi1973} for the observed \CP violation in the kaon system. They included a third  quark family, which extended  the $2\times 2$ Cabibbo mixing matrix \cite{Cabibbo:1963yz} to a $3\times 3$ matrix,  the now called Cabibbo-Kobayashi-Maskawa (CKM) matrix, allowing the presence of a complex phase in the hadronic sector of the weak interaction. The two new quarks, beauty and top, were experimentally observed in 1977 \cite{Herb1977} and 1995 \cite{D01995,CDF1995}, respectively, and finally \CP violation was observed in \B decays, as predicted, in 2001~\cite{Babar2001,Belle2001}. Kobayashi and Maskawa received the Nobel Prize in 2008 for the confirmation of their \CP violation mechanism within the Standard Model (SM).

A crucial question remains unanswered until today: what is the dynamical origin of the quark mixing matrix and, in particular, the weak phase? 
Unlike parity, whose breaking for all fermions comes from the universal vector--axial  nature of the weak currents, \CP breaking in not the same for all quarks.

From its observation, \CP violation represented a  new paradigm, opening the path for the emergence of ideas and concepts. One of them was introduced by Sakharov in 1967 \cite{Sakharov1967}, stating \CP violation as one of the necessary conditions for baryogenesis. Among others, there was also the notion of ``mirror universe'' where \CP violation would be the door between our universe and its \CP mirror version \cite{Okun}, with several implications \cite{Okun2006}. According to some authors, this universe would be the source of the Dark Matter \cite{Ibe2019}.

Even a few years before the CKM ansatz, a different approach  was  proposed by Wolfenstein \cite{Wolfenstein1964} and required a new interaction that was called {\it superweak}. It was able to explain the observed \CP asymmetry associated to oscillation in the neutral-kaon system  but did not allow for \CP violation coming from the decay process itself  -- what is called {\it direct} \CP violation. It was only in the early 2000s  that this idea was definitively ruled out by the observation of  direct \CP violation in kaons by the KTeV \cite{KTeV_1999,KTeV_2003} and NA48 \cite{NA48_2001,NA48_2002} collaborations, through the observable $\Re(\epsilon'/\epsilon)$ which, measured to be $\sim10^{-3}$, implies  a direct \CP asymmetry of $\sim10^{-5}$ \cite{Sozzi2003}.

Given the CKM matrix hierarchy the natural sectors to observe \CP violation are processes involving strange and beauty hadrons. This can be better seen through the Wolfenstein parameterisation~\cite{Wolfenstein_CKMparam_1983}, which describes the CKM elements in powers of $\lambda \approx 0.223$~\cite{PDG2019}. Up to ${\cal O}(\lambda^3)$ it is written as
\begin{equation}
V_{\rm CKM} = \left( \begin{array}{c c c} V_{\uquark\dquark} & V_{\uquark\squark} & V_{\uquark\bquark} \\
                            V_{\cquark\dquark} & V_{\cquark\squark} & V_{\cquark\bquark} \\
                            V_{\tquark\dquark} & V_{\tquark\squark} & V_{\tquark\bquark} \end{array} \right) = \left( \begin{array}{ccc}
1-\lambda^2/2 &  \lambda         &  A\lambda^3(\rho -i\eta) \\
-\lambda      &   1-\lambda^2/2  & A\lambda^2 \\
A\lambda^3(1-\rho -i\eta) & -A\lambda^2 & 1
\end{array}\right) ~. 
\label{CKM}
\end{equation}
The other three parameters are $A$, $\rho$ and $\eta$. The latter is responsible for the complex nature of the CKM matrix and appears at order $\lambda^3$ in the matrix elements $V^{\phantom{*}}_{\tquark\dquark}$ and $V^{\phantom{*}}_{\uquark\bquark}$. The CKM matrix is almost diagonal, breaking at the $\lambda$ level, and almost real, breaking at the $\lambda^3$ level. Being unitary, it leads to orthogonality relations of the type $\sum_{k=u,c,t} V^{\phantom{*}}_{ij}V^*_{ik} = \delta_{jk}$ ($j,k = d,s,b$). The well-known {\it Unitarity Triangle}, constructed from the relation $V^{\phantom{*}}_{\uquark\dquark}V^*_{\uquark\bquark}+ V^{\phantom{*}}_{\cquark\dquark}V^*_{\cquark\bquark}+V^{\phantom{*}}_{\tquark\dquark}V^*_{\tquark\bquark}=0$, defines three angles in the complex plane: $\alpha \equiv \arg\left(-\frac{V^{\phantom{*}}_{\tquark\dquark}V^*_{\tquark\bquark}}{V^{\phantom{*}}_{\uquark\dquark}V^*_{\uquark\bquark}}\right)$, 
 $\beta \equiv \arg\left(-\frac{V^{\phantom{*}}_{\cquark\dquark}V^*_{\cquark\bquark}}{V^{\phantom{*}}_{\tquark\dquark}V^*_{\tquark\bquark}}\right)$ and  $\gamma \equiv \arg\left(-\frac{V^{\phantom{*}}_{\uquark\dquark}V^*_{\uquark\bquark}}{V^{\phantom{*}}_{\cquark\dquark}V^*_{\cquark\bquark}}\right)$. Then, up to ${\cal O}(\lambda^3)$, $V^{\phantom{*}}_{\tquark\dquark} = |V^{\phantom{*}}_{\tquark\dquark}|e^{-i\beta}$ and $V_{\uquark\bquark} = |V^{\phantom{*}}_{\uquark\bquark}|e^{-i\gamma}$.

Since the top quark does not hadronise, the manifestation of $\arg(V_{\tquark\dquark})$ at short distances occurs through box or loop (``penguin'' type) diagrams in $\squark$- and $\bquark$-hadron processes.  Given the top mass is very large, \footnote{The top mass was predicted to be large after the the observation of $\Bz-\Bzb$ oscillations \cite{UA1_Boscill_1987,ARGUS_Boscill_1987,CLEO_Boscill_1989}. Its contribution in the one-loop process is important due to the non-decoupling nature of the SM, manifesting a violation of the Applequist--Carazzone theorem \cite{AppelquistCarazzone}.} its contribution to these diagrams is much more important than those mediated by up and charm quarks, which also almost cancel each other due to the GIM mechanism~\cite{GIM}. From box-type diagrams,  {\it indirect} \CP violation\footnote{The term {\it indirect} \CP violation here refers to processes where \CP violation \cite{PDG2019} requires mixing to occur, and it can arise between the interference of particle and antiparticle decaying to a common final state, or it can appear through the interference between mixing and decay.}   may arise in  neutral meson processes. On the other hand the effect of $\arg(V_{\uquark\bquark})=-\gamma$ can also take place at tree level for $b$-hadron decays. 
It can also be read from Eq.~\ref{CKM} that the possible effects of \CP violation in charm are expected to be tiny, potentially arising due to  $\arg(V_{\uquark\bquark})$ in box or loop diagrams or, at tree level, via a $\lambda^5$ correction in $V_{\cquark\dquark}$.

It is interesting to note the time gap of almost forty years between the discovery of \CP violation in mixing and direct \CP violation for kaon decays --- a consequence of the tiny effect of the latter in the kaon system.  A similar situation, but not so drastic, has occurred in \B decays, where \CP asymmetry in mixing  was observed in $\Bz \to \jpsi\KS$  by the BaBar \cite{Babar2001} and Belle \cite{Belle2001} collaborations while  direct \CP asymmetry was observed (with more than 5 standard deviations) more than ten years later \cite{BaBar2012, LHCb-PAPER-2013-018}.  Interestingly, in the charm sector direct \CP violation has very recently been observed by the LHCb collaboration~\cite{LHCb-PAPER-2019-006} --- and yet there is no sign for indirect \CP violation associated to mixing in \Dz decays \cite{LHCb-PAPER-2017-046,LHCb-PAPER-2019-001,LHCb-PAPER-2019-032}.

From the theory side, there is an even higher challenge regarding estimates for direct \CP violation asymmetries. They depend crucially on the size of the strong phases involved in the decays~\cite{Bigi_Book}.
In fact, since the first theoretical proposal for the study of direct \CP violation in an inclusive way, by Bander, Silverman and Soni~\cite{BSS} and known as the  BSS model, hundreds of papers were published, based on different mechanisms, aiming to  understand the importance of strong interaction phases, mainly in exclusive \B-meson decays. There is fair consensus that the strong phases coming from short-distance effects should be small. If these are assumed to be  the primary source of strong phases, the level of direct \CP violation should be small. On the other hand, if one  considers that long-distance, final-state interaction (FSI) processes in non-leptonic decays introduce significant strong phase shifts, potentially large  direct \CP violation effects should be expected.
         
                  Many good reviews exist addressing  \CP violation in hadrons  (see, for example,   \cite{TimVava, EricaAlberto, Bevan2014,Artuso_CPBs_2016, Hou_Review_2017}). This work is dedicated exclusively to discussing direct \CP violation in $\bquark$-  and $\cquark$-hadron systems. New, interesting, and even unexpected results on direct \CP violation emerged in the recent years which, confronting with the development of new insights from the theory side,  motivate  this review. 
For the \B meson sector, we concentrate the discussion only on well-established results on direct \CP violation and their implications. For $b$ baryons and charm hadrons  the overall \CP violation picture is just in its initial stages, and we address the current state-of-the-art.  BaBar, Belle and a few other experiments have contributed significantly in this area, nevertheless the LHCb collaboration, with unprecedented samples of beauty and charm hadrons, is currently the large source of new experimental results. Belle II  and the LHCb Upgrade I  will soon enter the game. 
                  
                  This review is organised as follows. In Sect.~\ref{chapter2_CP_CPT}, we discuss the fundamental ideas regarding direct \CP violation and call attention on how \CPT conservation enters  as a constraint  for decay modes with the same final-state flavour content. In Sect.~\ref{chapter3}  the major experiments and main strategies  for measuring direct \CP violation observables are discussed. Particular emphasis is given on the importance of multi-body decays, for which \CP violation effects can appear, and  be enhanced, in specific regions of their phase space.  Direct \CP violation in charmless $b$-hadron decays  is presented in Sect.~\ref{chapter4}, followed by Sect.~\ref{chapter5} where a complementary discussion on double-charm contributions to the charmless final states is considered. Section~\ref{chapter6} is dedicated to \B decays to open charm, which represent the ideal environment for the measurement of $\gamma$. Section~\ref{chapter7} addresses the charm sector, contextualising the field after the first observation of direct \CP violation  and discussing the  current status and prospects for many other modes. Finally, we present our conclusions in Sect.~\ref{chapter8}.

 \section{\CP violation in decays and \CPT symmetry}
\label{chapter2_CP_CPT}

\CP violation in decays -- or {\it direct} \CP violation -- is manifested
when the decay width of a particle $P$ to a given final state $f$ differs from the conjugate process:
\begin{equation}
\label{width_difference}
\Delta \Gamma(P\to f) \equiv   \Gamma(P\to f) - \Gamma(\bar P\to \bar f)  \ne 0~.
\end{equation}

In terms of the decay amplitudes, 
\begin{equation}
\label{ampl_difference}
|{\cal A} (P\to f) |^2 - |{\cal A} (\bar P\to \bar f) |^2 \ne 0~.
\end{equation}

The origin of \CP violation relies on the presence of complex terms in the Lagrangian density, which translate to phases in transition amplitudes. If these phases are \CP-odd -- i.e. change sign under a \CP transformation -- \CP violation can appear. Clearly, for phases to become   observables, there should be at least two possible interfering paths producing a final state $f$. 

The usual example \cite{Bigi_Book, Sozzi_Book, Branco_Book} is to consider two decay amplitudes producing a final state $f$ such that
\begin{equation}
{\cal A} (P\to f) = |{\cal A}_1| e^{i\phi_1} + |{\cal A}_2| e^{i\phi_2}~,
\label{eq-amplweakphase}
\end{equation}
where  ${\cal A}_ 1$ and ${\cal A}_2$ have  different phases $\phi_1$ and $\phi_2$. Considering both $\phi_1$ and $\phi_2$ are CP-odd phases, the antiparticle process is given by 
${\cal A} (\bar P\to\bar  f)  = |{\cal A}_1| e^{-i\phi_1} + |{\cal A}_2| e^{-i\phi_2}$. It is easy to 
show that the decay rate difference, Eq.~\ref{ampl_difference},  gives a null result, and \CP violation is not observed. In contrast, if the decay amplitude is given by 
\begin{equation}
{\cal A} (P\to f) = |{\cal A}_1| e^{i\phi_1} e^{i\delta_1}+ |{\cal A}_2| e^{i\phi_2}e^{i\delta_2}~,
\label{eq-amplphases}
\end{equation}
where $\delta_1$ and $\delta_2$ are \CP-even phases, then the decay rate difference is
\begin{equation}
|{\cal A} (P\to f) |^2 - |{\cal A} (\bar P\to \bar f) |^2= 2|{\cal A}_1| |{\cal A}_2| \sin(\phi_1 -\phi_2)\sin(\delta_1-\delta_2).
\label{eq-diffamplsquare}
\end{equation}
Usually the observable is the \CP asymmetry defined as 
\begin{equation}
A_\CP = \frac {|{\cal A} (P\to f) |^2 - |{\cal A} (\bar P\to \bar f) |^2}{|{\cal A} (P\to f) |^2 + |{\cal A} (\bar P\to \bar f) |^2}
\label{CPasym_def}
\end{equation}  that for the amplitude given in Eq.~\ref{eq-amplphases} it is given by 
\begin{equation}
A_\CP = \frac{2|{\cal A}_2/{\cal A}_1|\sin(\delta_1-\delta_2)\sin(\phi_1-\phi_2)}{1 + |{\cal A}_2/{\cal A}_1|^2 + 2 |{\cal A}_2/{\cal A}_1|\cos(\delta_1-\delta_2)\cos(\phi_1-\phi_2)}~.
\label{ACP_sindelta_sinphi}
\end{equation}
Therefore both \CP-odd and \CP-even phases are necessary for \CP violation to be observed. Although this is a simple academic example, the conclusion in quite general.  Moreover, the size of the  \CP asymmetry depends  not only on the phase differences but also on the relative size of the two amplitudes  -- if, for instance, $|{\cal A}_1| \gg |{\cal A}_2|$ the expected \CP asymmetry would be small even with large phase differences.
Within the context of the SM, \CP-odd phases can only come from weak processes, while \CP-even phases may appear through strong (or electromagnetic) interactions. For this reason, \CP-odd and \CP-even phases are usually called {\it weak} and {\it strong}  phases, respectively. 

For a neutral meson decaying to \CP eigenstates like $\pip\pim$, $\Kp\Km$, $\Dz\Dzb$ (or final states accessible for both particle and antiparticle, such as $\KS\piz$) it is necessary to disentangle direct \CP violation and  \CP violation from oscillation and in this case the time-dependent \CP asymmetry, arising from the so-called {\it master equations} \cite{Bigi_Book} need to be used. Assuming \CPT invariance, for a $P^0-\bar P^0 $ system with a mass difference $\Delta M$ between the mass eigenstates (but no width difference), we have
\begin{equation}
A_\CP(t) = \frac {|{\cal A} (P^0(t)\to f) |^2 - |{\cal A}(\bar P^0(t)\to \bar f) |^2}{|{\cal A} (P^0(t)\to f) |^2 + |{\cal A} (\bar P^0(t)\to \bar f) |^2} = {\cal S}_f \sin(\Delta M t) - {\cal C}_f \cos(\Delta M  t)
\label{CPasym_time}
\end{equation} 
where $t$ is the decay time of $P$ decaying to $f$.
The ${\cal C}_f$ and ${\cal S}_f$ observables are sensitive to direct and indirect \CP violation, respectively, with $0 \le {\cal S}_f^2 +{\cal C}_f^2 \le 1$ \cite{Bevan2014}.  Here we are only interested in ${\cal C}_f$, which is given by
\begin{equation}
{\cal C}_f = \frac {|{\cal A} (P^0\to f) |^2 - |{\cal A} (\bar P^0\to f) |^2}{|{\cal A} (P^0\to f) |^2 + |{\cal A} (\bar P^0\to  f) |^2}.
\label{CPasym_C}
\end{equation} 

Generally, while the origin of the weak phase can be pinned down at the quark level to the CKM matrix elements, making it predictable for which decays \CP violation could {\it potentially} be manifested, the size of the direct \CP asymmetry is usually hard to estimate due in part to the rather limited knowledge of the strong phases. 
There are two possible sources of strong phases: one associated with the short-distance penguin contribution at quark level and  another coming from final-state interactions, e.g. hadronic rescatterings in non-leptonic decays. An overview of these mechanisms  is given below.

\subsection{Hadronic strong phases and \CPT constraints on \CP violation}
\label{CPT}

For a particle decaying only weakly, the initial state is stable under strong (and electromagnetic) interactions. On the other hand, following the weak transition the final state is the potential result of strong processes associating states with the same  quantum numbers. Strong phases are the result of these final-state interactions. 
For instance, the total amplitude of the decay of $P$ to a given final state $f$ may include contributions as $P\to f_i\to f$ ($i = 1,2,...,n$) where $P\to f_i$ incorporates the weak transition and $f_i\to f$ the strong processes (elastic or inelastic scatterings):
\begin{equation}
{\cal A} (P\to f) = {\cal A}_1(P\to f_1){\cal A}(f_1\to f) + {\cal A}_2(P\to f_2){\cal A}(f_2\to f)+ .....+ {\cal A}_n(P\to f_n){\cal A}(f_n\to f)~.
\label{eq-f1f2}
\end{equation}
At this point, it is interesting to discuss \CP violation in the context of \CPT constraints, as follows.

The \CPT symmetry, expected to be an exact symmetry in field theories, establishes that the lifetimes -- thus the total decay widths -- of a particle and its antiparticle are the same. On the other hand, \CP violation allows partial decay widths to be different.  Clearly, preserving the equality of total widths for particle and antiparticle while allowing partial widths to be different requires a  ``communication'' between the different decay modes, and this can only happen within modes that have the same flavour quantum numbers.

Thus, and in accordance to what was discussed previously, final-state interactions, connecting states via strong (or electromagnetic) interactions, provide the natural strong phases for \CP violation to be observable, and are also the key ingredient to allow the preservation of  \CPT symmetry \cite{Bigi_Book, Sozzi_Book, Branco_Book,Marshak_Book}.

The constraint from \CPT can become more apparent by considering the arguments from Wolfenstein in \cite{Wolfenstein1990} (see also \cite{Bigi_Book} from which  the notation below is partly adopted). Suppose that particle $P$ can decay to only two final states $f_\alpha$ and $f_\beta$ with the same conserved  quantum numbers. The states $f_\alpha$ and $f_\beta$ can be connected through strong interactions, meaning they can rescatter into each other. This interaction may be described by a scattering matrix ${\cal S}_{\rm strong}$ given by
\begin{equation}
\label{Smatrix_alphabeta}
{\cal S}_{\rm strong} = \left( \begin{array}{cc} e^{2i\delta_\alpha} & i{\cal T}e^{i(\delta_\alpha+ \delta_\beta)} \\  i{\cal T}e^{i(\delta_\alpha+ \delta_\beta)} & e^{2i\delta_\beta} \end{array} \right), 
\end{equation}
where ${\delta_\alpha}$ and ${\delta_\beta}$ are the elastic scattering phases, ${\cal T}$ is assumed to be small so as to be treated only to first order, ${\cal S}_{\rm strong}$ is unitary to that order and symmetrical  to ensure time reversal symmetry. The two states $f_\alpha$ and $f_\beta$ form a complete set of states given their flavour content, and using  ${\cal A}_{\alpha(\beta)}(P\to f_{\alpha(\beta)}) = e^{i\delta_{\alpha(\beta)}} A_{\alpha(\beta)}$ it can be shown that \cite{Bigi_Book} 
\begin{eqnarray}
{\cal A}(P\to f_\alpha)& = & e^{i\delta_\alpha}\left( A_\alpha+ i {\cal T}  A_\beta \right)  ~, \nonumber \\
{\cal A} (\bar P \to \bar f_\alpha) & =&  e^{i\delta_\alpha}\left(  A^*_\alpha+ i {\cal T} A^*_\beta \right) ~, 
\end{eqnarray}
where  $A_{\alpha(\beta)}$ is defined through ${\cal A}_{\alpha(\beta)}(P\to f_{\alpha(\beta)}) = e^{i\delta_{\alpha(\beta)}} A_{\alpha(\beta)}$. The second equation above follows from imposing \CPT invariance. Analogous equations are given for ${\cal A}(P\to f_\beta)$ and $ {\cal A}(\bar P\to \bar f_\beta)$. So, it can be seen that 
\begin{eqnarray}
|{\cal A}(\bar P\to \bar f_\alpha)|^2 - |{\cal A}( P\to f_\alpha)|^2& = & 2{\cal T}\Im ({\cal A}_\alpha^* {\cal A}_\beta ) ~, \nonumber \\
|{\cal A}(\bar P\to \bar f_\beta)|^2 -|{\cal A}( P\to f_\beta)| ^2& = & 2{\cal T}\Im ({\cal A}_\beta^* {\cal A}_\alpha )
\end{eqnarray}
which leads to $\Delta\Gamma(\bar P\to \bar f_\alpha) = - \Delta\Gamma((\bar P\to \bar f_\beta)$ or, alternatively,  
\begin{equation}
\sum_{i = \alpha, \beta} \Delta \Gamma (P \to f_i) = 0~. 
\end{equation}
 
 Even for decays with more channels with same  quantum numbers, the conclusion remains the same \cite{Bigi_Book},
 \begin{equation}
 \sum_{i} \Delta \Gamma (P \to f_i) = 4 \sum_i  \sum_{j\ne i} {\cal T}_{ij} \Im ({\cal A}_i^* {\cal A}_j) = 0~. 
 \end{equation}
 Note that, as made explicit by this Wolfenstein mechanism \cite{Wolfenstein1990}, it is the interference terms between  (at least) two hadronic rescattering amplitudes that generate an observable \CP asymmetry.
 
  The main lesson here is that \CPT constrains not only the total widths of particle and antiparticle to be the same, but also the sum of partial widths to final states with the same quantum numbers.  For instance, if one finds a sizeable positive \CP asymmetry in a given decay mode, there should be strongly-coupled final states for which the \CP asymmetry has negative sign to compensate. 
  One can also immediately expect,  as a consequence, that if a given final state is stable under strong interactions, e.g. without possible rescattering process, direct \CP violation  would not be observed, even if the weak transition involves a weak phase. This is  the case of a semi-leptonic decay of a charged hadron such as $B^+\to \pi^0\mu^+\nu_\mu$.\footnote{Unless explicitly stated otherwise, charge conjugation is implied throughout the text.}   Another example is the decay $K^+\to \pi^+\pi^0$ which,  due to G-parity, cannot rescatter to any other final state,  implying  $\Gamma(\Kp\to\pip\piz)=\Gamma(\Km\to \pim\piz)$. Analogously,  the sum of the partial widths of the two 3-pion final states for $K^+$ and $K^-$ has to be the same~\cite{Sozzi_Book, Marshak_Book}, or equivalently 
\begin{equation}
\Gamma(K^+\to \pi^+\pi^-\pi^+)+\Gamma(K^+\to\pi^+\pi^0\pi^0) = \Gamma(K^-\to \pi^-\pi^+\pi^-)+\Gamma(K^-\to\pi^-\pi^0\pi^0)~.
\end{equation}

Clearly the \CPT constraint gains flexibility when the number of possible connected final states increase. For non-leptonic charm decays, in principle the families can be divided in terms of the strangeness quantum number $S$: Cabibbo-favoured transitions, $c\to su\bar d$ , with $\Delta S =+1$; Cabibbo-suppressed transitions, $c\to du\bar d$ or $c\to su\bar s$ , with $\Delta S = 0$; and doubly-Cabibbo suppressed decays, $c\to du\bar s$, with $\Delta S =-1$.   For beauty decays, there are in principle eight different families, considering the combinations of transitions with $\Delta S =0,\pm 1$ and $\Delta C = 0, \pm 1$ ($C$ being the charm quantum number). Since the b-hadron masses are higher than those of charm hadrons,  the multiplicity of decay channels for each family is larger, making it difficult to foresee where and how the compensation of an eventual \CP asymmetry in a particular channel would appear in the other strongly-coupled channels. 

\subsection{Short-distance effects and inclusive \B decays}

\begin{figure}[htb]
\centering
\includegraphics[width=.6\columnwidth,angle=0]{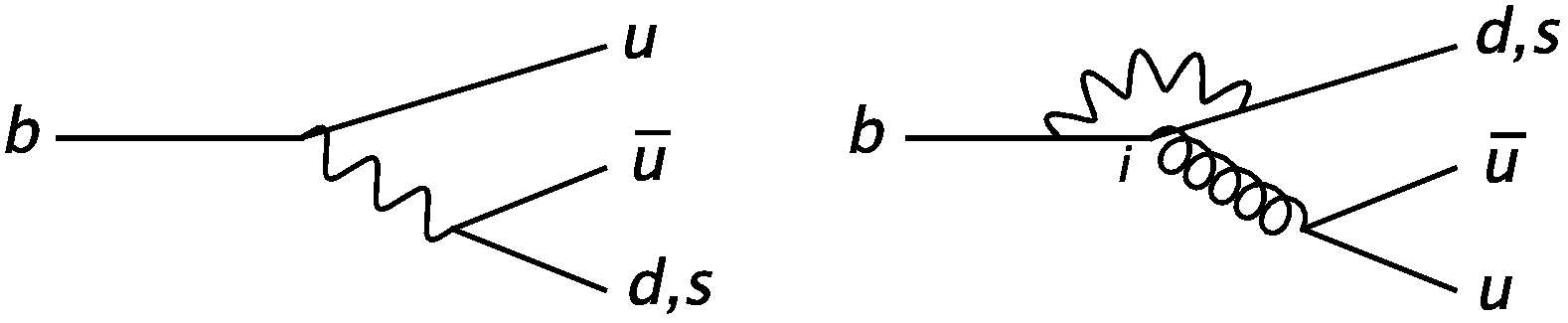}
\caption{ (Left) Tree and (right) penguin diagram contributions to inclusive charmless $\bquark \to \dquark, \squark$ transition. The index $i$ on the right plot  sums over internal quarks $\uquark, \cquark, \tquark$.  }
\label{treepenguin}
\end{figure}

As discussed above, the presence of at least two amplitudes, with different weak and strong phases, is a requirement for the manifestation of \CP violation in decays. While weak phases in the SM appear from specific CKM matrix elements, strong phases can appear from different sources. At short distances, a strong phase can be produced from loop (second-order) processes. This effect is described for inclusive charmless  \B decays by the BSS model \cite{BSS}. The tree amplitude with $ \bquark \to \uquark $ transition has the CKM weak phase $\gamma$, and a strong, short-distance phase may arise from a penguin amplitude.  Both contributions are shown in Fig.~\ref{treepenguin}. 

\begin{figure}[htb]
\centering
\includegraphics[width=.5\columnwidth,angle=0]{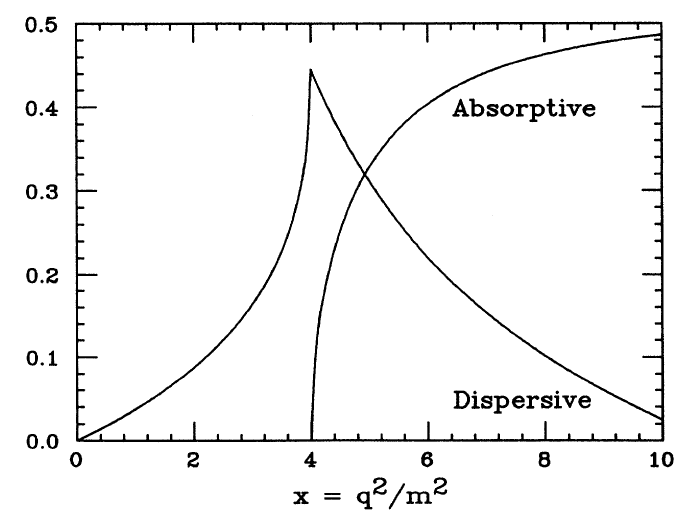}
\caption{Absorptive and dispersive penguin amplitudes, in arbitrary units, as a function of the ratio of the gluon momentum transfer  and the internal quark mass \cite{gerardhou1991}. }
\label{penguinphase}
\end{figure}

For the penguin operator to produce a strong phase, the intermediate particles need to be on-shell \cite{Sozzi_Book,Wolfenstein1990}. With an internal charm line, this can take place  in a small part of the phase space, as discussed in Ref.~\cite{gerardhou1991} and shown in Fig.~\ref{penguinphase}. For charmless $b \to s u \bar u $ it occurs  only for $q^2 > 4m_c^2$ ($q$ being the gluon momentum and $m_c$ the charm quark mass), while, on the other hand,  the tree amplitude is real. 

For a given inclusive flavour content of a \B decay, for example $\Delta S=1$ and  $\Delta C=0$,  quark-hadron duality \cite{Bigi1987,Uraltsev1992} involves the summation over all channels with final states with the same quantum numbers, as
\begin{equation}
\Delta\Gamma = \Gamma(\bquarkbar \to \squarkbar\uquark\uquarkbar) - \Gamma( \bquark \to \squark\uquarkbar\uquark) = -4\Im(\lambda_i\lambda^*_j)
\sum_{{f}} \Im\left[A_i(b \to f)\cdot A^*_j(b \to f)\right]~,
\label{inclusive}
\end{equation}
where $f$ are the possible final states for the process, $A_i$ and $A_j$ are the decay amplitudes for each  final state $f$, and  $\lambda_i, \lambda^*_j$  are the combinations of the relevant CKM matrix elements associated to the transition and are factored out. The assumption of quark-hadron duality in principle allows the quoted \CP asymmetries to be free of  hadronic uncertainties.  

However, as discussed previously, \CPT conservation demands  the sum of partial width differences for particle and antiparticle to be zero within a family of final states with the same quantum numbers. In particular for the example being treated,  both charmless \B decays with $\Delta S=1$ and  $\Delta S=0$ must each have the total $A_{\CP} = 0$. Gérard and Hou \cite{gerardhou1991, gerardhou1989} noticed this inconsistency in the BSS mechanism and presented a new proposal considering the \CPT constraint, mostly including ${\cal O}(\alpha_S ^2) $ contributions.\footnote{An interesting relationship between these ${\cal O}(\alpha_S ^2) $ contributions and the recent direct \CP symmetry distributions observed in charmless three-body \B decays can be found in \cite{Hou_2019}. Such calculations can also be found in  \cite{Lenz1999}. } 
In any case, for a large number of decay channels within a flavour family, the concept of an  inclusive \CP asymmetry has no practical experimental usage other than to give an estimate of the order of magnitude expected for \CP violation.

\subsection{Exclusive \B decays and final-state interaction phases}

The situation is quite different from the above when considering  exclusive decays, where important contributions may come from hadronic rescattering, as presented in Eq.~\ref{eq-f1f2}. In principle, these FSI  can introduce strong phase shifts which are not restricted to be small according to  some authors \cite {Bigi1987,Suzuki2007,Soni2005,Donoghue1996,Smith2003}.

But this is far from being a consensus.  From a different standpoint, the authors of Ref. \cite{Beneke1999,Beneke2000} propose a factorisation based on Quantum Chromodynamics (QCD) that assumes that,  in the limit of the beauty mass $m_b \to  \infty$, the number of intermediate physical states is arbitrarily large and the soft rescattering phases vanish at this limit.\footnote{The heavy-quark limit  is also used in a different approach known as soft-collinear effective theory (SCET) \cite{Zupan2006}.} 
Thus the complications associated to the non-perturbative nature of   hadronic \B decays  can be treated systematically as $1/m_b$  power corrections, which vanish at the heavy quark limit through systematic cancellations \cite{Neubert2000}.  Moreover, the large number of intermediate states is used as an argument for not taking the \CPT constraint into  account. Very recently the calculation has been extended to next-to-next-leading-order  (NNLO) in QCD factorisation with a sizeable impact  \cite{Bell2020}.

 This long-term discussion has been using the $\B \to \PK \pi$  system as the preferred place for confrontation. These charmless \B decays have branching fractions at ${\cal O}(10^{-5})$  and clean experimental signatures. Indeed the $\Bz \to \Km\pip$   decay was the first charmless \B decay to be observed, and the first that had its direct \CP asymmetry well determined experimentally.  The theoretical interest comes from the dominance of the penguin contribution in all   $\B \to \PK \pi$ decays,  and  the  contribution of a tree component in most of them, with the $ \bquark \to \uquark$  process providing the weak phase $\gamma$. 
      
                   Besides the short-distance contribution mentioned above, there are two possible classes of hadronic rescattering  contributions for $\B \to \PK \pi$ decays, one coming from a process involving charm hadronic intermediate states, the other involving only  light-meson final states. A suitable division for these contributions  is proposed in Ref.~\cite{Buras1997}: the first group is defined as $ \B \to  f_c  \to \PK\pi$, with $f_c = \D^{(*)} \D_s^{(*)}$ and also called  hard hadronic rescattering  \cite{Suzuki2007,Soni2005,Smith2003}. The second group involves   $\B \to  f_s \to \PK\pi$, with $f_s = \Kstar \pi, \Kstar\rho, \Kstar\eta$, among others, and is  called soft rescattering \cite{Soni2005,Donoghue1996,Buras1997,AtwoodSoni1997}.  Although the rescattering involving the $f_s$ family of intermediate states is more likely than that of the $f_c$ family, the two orders of the magnitude larger branching fractions of the latter family relative to the former  may allow both possibilities to be competitive.
 
           The advantage of the short-distance QCD factorisation approach  is the possibility to predict the level of direct  \CP asymmetry for several \B exclusive final states  ~\cite{Beneke1999,Beneke2000} to be compared to those from experimental data. On the other hand, the hadronic rescattering approaches to $\B \to \PK \pi$ (or two-body decays in general) do not allow a similar direct comparison. However, as discussed in Sect.~\ref{chapter4} and \ref{chapter5}, multi-body \B decays can shed light on this debate.
           
\subsection{What to expect in charm decays}
\label{charm_cap2}

\begin{figure}
\centering
\includegraphics[width = 0.9\textwidth]{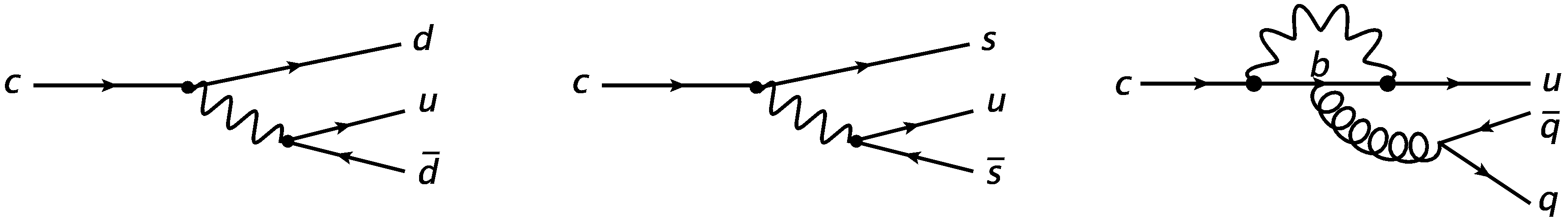}
\caption{Tree and penguin topologies for a charm-hadron CS decay, where $q$ can be an $s$ or a $d$ quark.}
\label{CS-diagrams}
\end{figure}

At a mass scale around $2\gev$,\footnote{Natural units are used.} charm hadrons lie in a region of semi-perturbative QCD, at most, for which the heavy quark limit mentioned above is not such a good tool.  For general reviews on charm physics, see for instance Refs.~\cite{Bianco_2003,Artuso_Charm_2008}. Strong interactions at low energy play a major role which makes it a challenge, from the theoretical point of view, to describe the dynamics of their hadronic decays. 

Charm transitions at tree level are driven by the 2$\times$2 Cabibbo ``sub-space'' of the CKM matrix, which is  unitary and real up to ${\cal O}(\lambda^4)$. Loop and box processes allow, however, the phenomena of $\Dz-\Dzb$ mixing and \CP violation to occur at very low rates. Oscillations in the neutral charm system have been observed for the first time in 2007~\cite{BaBar_DOscill_2007,Belle_DOscill_2007,CDF_DOscill_2008} and is now very well established \cite{LHCb-PAPER-2017-046,LHCb-PAPER-2019-001} but so far there is no evidence for \CP violation associated to mixing~\cite{LHCb-PAPER-2019-032}. 

Within the CKM mechanism, direct \CP violation in charm is only possible in singly Cabibbo-suppressed (CS) decays. As for the case of \B decays, both tree and penguin amplitudes can produce final states with the same flavour content, as depicted in Fig.~\ref{CS-diagrams}. 
Contrary to \B decays, though, the penguin amplitude is expected to be very small.  Within the Wolfenstein parameterisation~\cite{Wolfenstein_CKMparam_1983}, the tree topology amplitude is proportional to $\lambda$ and for the penguin topology the relevant contribution is with the $\bquark$ quark in the loop, which carries the weak phase $\gamma$ at $\lambda^5$ level, so the interference term appears with a $\lambda^4$ dependence relative to the leading term. No weak phases are expected from penguin contributions with $\dquark$ and $\squark$ quarks in the loop, which moreover almost cancel due to the GIM mechanism.
A tree-level interference  between $\cquark\to \dquark\dquarkbar\uquark$ and  $\cquark\to \squark\squarkbar\uquark$ is also sensitive to the complex nature of the CKM matrix since $V_{cd}$ gets an imaginary term at order ${\cal O}(\lambda^5)$. The imaginary part of the interference term is then $\frac{V^{\phantom{*}}_{ud}V^*_{cd}}{V^{\phantom{*}}_{us}V^*_{cs}}$ which, by the CKM orthogonality relations,  is also at the $\lambda^4$ level relative to the leading term.  This tree-level effect can appear for certain final states such as $\Dz\to \KS\KS,\eta'\piz, \Kp\Km\pip\pim$, and $\Dp\to \pim\pip\pip, \Km\Kp\pip$ and can also play a role through rescattering for $\Dz\to\pim\pip, \Km\Kp$ \cite{Grossman_2019,Bianco_Bigi_2020}. 

A general form for the amplitude of a CS decay can be written as \cite{Lenz_CHARM2013}
\begin{equation}
{\cal A}_{\rm CS} = \frac{G_F}{\sqrt{2}}\lambda_d \,T \left[1+\frac{\lambda_b}{\lambda_d}\frac{P}{T}\right],
\end{equation}
where $G_F$ is the Fermi weak coupling constant, $\lambda_q = V^*_{cq}V_{uq}$ ($q=d,s,b$) is the product of the relevant CKM matrix elements (and by unitarity $\lambda_s = -\lambda_d - \lambda_b$) and $T$ and $P$ are the contributions  driven by $\lambda_d$ and $\lambda_b$, respectively, thus the latter carrying the weak phase. By computing the \CP asymmetry, 
as in Eq.~\ref{ACP_sindelta_sinphi} (and neglecting the $|P/T|$ terms in the denominator when compared to 1) it is easy to see that
\begin{equation}
A_\CP^{\rm (CS)} = 2 \frac{\lambda_b}{\lambda_d}\sin\gamma\sin\delta_{\rm s}\left|\frac{P}{T}\right|
\label{ACP_CS}
\end{equation}
where the weak ($\gamma$) and strong ($\delta_s$) phases are shown explicitly. The ratio $\left|\frac{P}{T}\right|$ is unknown, as well as the strong phase. While this expression is rather general, it refers to the direct \CP violation contribution. For  \Dz decays, the observed time-integrated asymmetry also receives a contribution from indirect \CP violation (as discussed  in Sect.~\ref{Dto2body}). For multibody decays the asymmetries can also be accessed throughout the phase space while Eq.~\ref{ACP_CS} refers to the phase-space integrated asymmetry. With the current values for $\lambda_b$, $\lambda_d$ and $\gamma$ ~\cite{CKMfitter2015}, $A_\CP^{\rm (CS)} \approx 1.4\times 10^{-3} \sin\delta_{\rm s}\left|\frac{P}{T}\right|$ is obtained.

An estimate of the ratio $\left|\frac{P}{T}\right|$ is challenging in the SM, and many authors arrive to different results, depending on the assumptions made. Short-distance QCD gives $\left|\frac{P}{T}\right| \approx \frac{\alpha_s}{\pi} \approx 0.1$ which leads to  $A_\CP^{\rm (CS)} \approx 10^{-4}$ (assuming $\sin\delta_{\rm s}\approx 1$) or even less \cite{Grossman_CPV_CS_2007,Khodjamirian_Petrov_2017}. Yet non-perturbative effects may give rise to penguin enhancements, flavour SU(3) symmetry breaking and rescattering effects which could push asymmetries to as high as a few per mille (see for instance \cite{Golden_Grinstein_1989,Cheng_Chiang_2012_1,Brod_ConsistentPenguins_2012,Grossman_2019}). This is further discussed in Sect.~\ref{chapter7}. 

Direct \CP violation has now been observed  through  the difference in \CP asymmetries between the decays $\Dz\to\Kp\Km$ and $\Dz\to\pip\pim$~\cite{LHCb-PAPER-2019-006} at the level of ${\cal O}(10^{-3})$. This is an impressive achievement  and further stimulates the search for signs of \CP violation in other decays. The measured value is higher than those from short-distance expectations and in principle favours  models with hadronic enhancements. It also give some fuel for new physics (NP) models \cite{Nir_Implications_CharmCPV_2019,Chala_2019,Calibbi_2019}.

Indeed the search for \CP violation in charm is an excellent place to look for new dynamics due to the low ``background'' from the SM. There are many channels for which the experimental uncertainties are still above a few percent and can be tested for sources of \CP violation beyond the SM.

In any case, SM or beyond, the interplay of weak and hadronic amplitudes still dictates the level of \CP violation to be observed. Moreover, as discussed previously, for particle and antiparticle decays the \CPT invariance constrains the sum of their partial widths to final states with same flavour content to be the same. The multiplicity of final states here, when compared to \B decays, is not so high and thus the  \CPT constraint can turn out to be more enlightening and helpful than that in the $b$-quark sector. 

   \section{Experimental facilities and analysis tools} 
 \label{chapter3}
 
 As discussed previously, it took a long time since the first observation of \CP violation in the neutral kaon system to the start of a rich programme of \CP violation measurements, occurring at the beginning of the $21^{\rm st}$ century with the measurement of $\sin{2\beta}$  in  $\Bz\to \jpsi\KS$ decays by the Babar~\cite{Babar2001} and Belle collaborations~\cite{Belle2001}. Now, two decades has passed and we have witness how a coherent picture of the \CP violation phenomena has emerged from a wide range of results, with a remarkable consistency with the CKM description~\cite{CKMfitter2015,UTfit-UT}. 
 Within the last 10 years or so, many experiments, such as D$\emptyset$, CDF, CLEO, and BESIII,  have contributed to  direct \CP violation measurements, but largely the bulk of results comes from  BaBar, Belle and LHCb. Here we briefly describe some of their main features,  and then discuss Belle II and the Upgrade LHCb, which will lead the field for the next decade. The principal aspects and techniques for the measurement of direct \CP violation observables are presented afterwards.

 \subsection{Main facilities for \CP violation in beauty and charm hadrons}
 
 The fundamental aspects of a detector dedicated to beauty and charm physics are: excellent vertexing and tracking capabilities, in order to provide a good reconstruction of charged particles and a precise determination of the $b$- or $c$-hadron  decay vertex; a reliable particle identification system, fundamental to distinguish the final-state particles (protons, kaons, pions, muons, etc); good momentum and time resolutions. For the study of neutral \B and \D mesons it is also fundamental a trustful {\it tagging} system, {\it i.e.} the knowledge of the flavour of the neutral meson at production (or at a given time), to allow the study of its time evolution and its decay to self-\CP-conjugated final states. 
 
 In the quest for the first signs of \CP violation in the beauty sector, the $B$ factories, BaBar~\cite{Aubert:2001tu,TheBABAR:2013jta} and Belle~\cite{Abashian:2000cg}, took the leading role for many years. They both relied on a copious production of $\Bp\Bm$ and $\Bz\Bzb$ pairs at the $\FourS$ bottonium resonance, corresponding to a centre-of-mass energy of $10.58\gev$. The PEPII machine  provided $\en$ and $\ep$ beams to BaBar at energies of 9.0 and 3.1\,\gev, respectively, while KEKB machine, for Belle,  had $\en$ and $\ep$ energies at 8.0 and 3.5\,\gev, respectively. The energy-asymmetric beams were designed to provide the $\B\Bb$ pair a boost in the laboratory, in order for the \B-decay vertices to be well separated from the production vertex. The $\Bz\Bzb$ pair were produced as an entangled quantum system, and by measuring the flavour of one \B meson at its decay vertex through a flavour-specific mode it was possible to determine the flavour of the other (conjugate) \B meson. Since charm mesons are common decay products of \B mesons, BaBar and Belle had also a significant impact on charm physics. With $\en\ep$ collisions producing low hadronic background, and an almost $4\pi$ of geometrical coverage,  the experiments were very efficient for reconstructing both charged and neutral final-state particles. 
 The BaBar experiment ran from 1999 to 2008 and cumulated about $424\invfb$ of data at the $\FourS$ resonance while Belle ran from 1999 to 2010, cumulating $711\invfb$ at the same energy. \footnote{Although the majority of the data taking occurred at the  $\FourS$  resonance, both BaBar and Belle took data off-resonance to study background. BaBar also had runs at $\TwoS$ and$\ThreeS$, while Belle had a dedicated $\FiveS$ run which allowed the production of $\Bs\Bsb$ pairs.}  For further details, a nice review can be found in Ref.~\cite{Bevan2014}.
  
\begin{figure}[!htbp]
 \centering
 \includegraphics[align=c, width=0.4\textwidth]{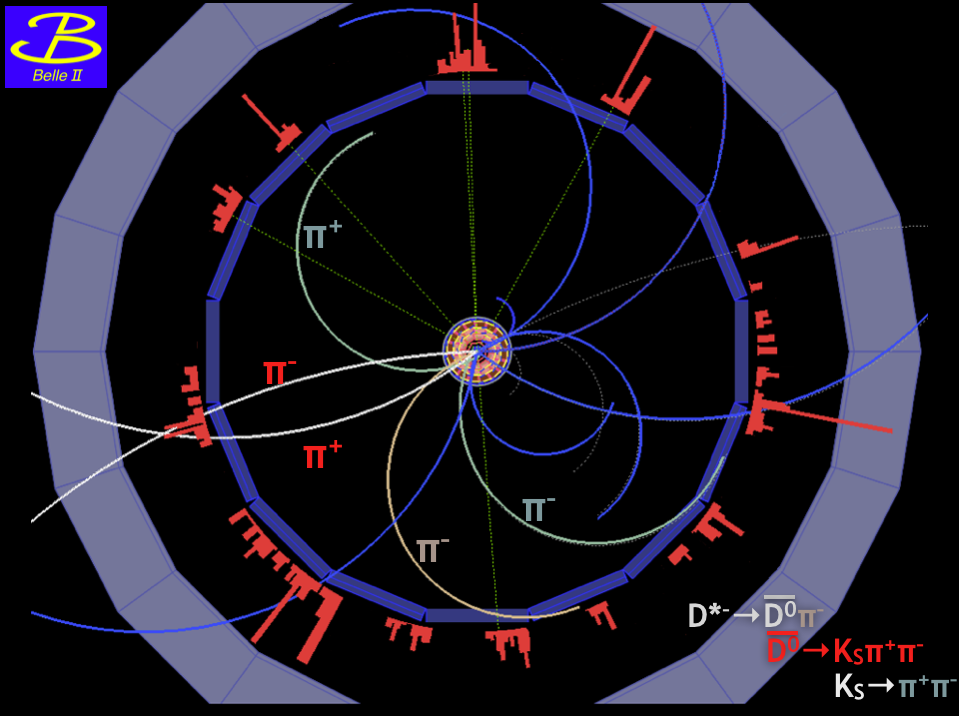}\hspace*{0.05\textwidth}
  \includegraphics[align=c,width=0.4\textwidth]{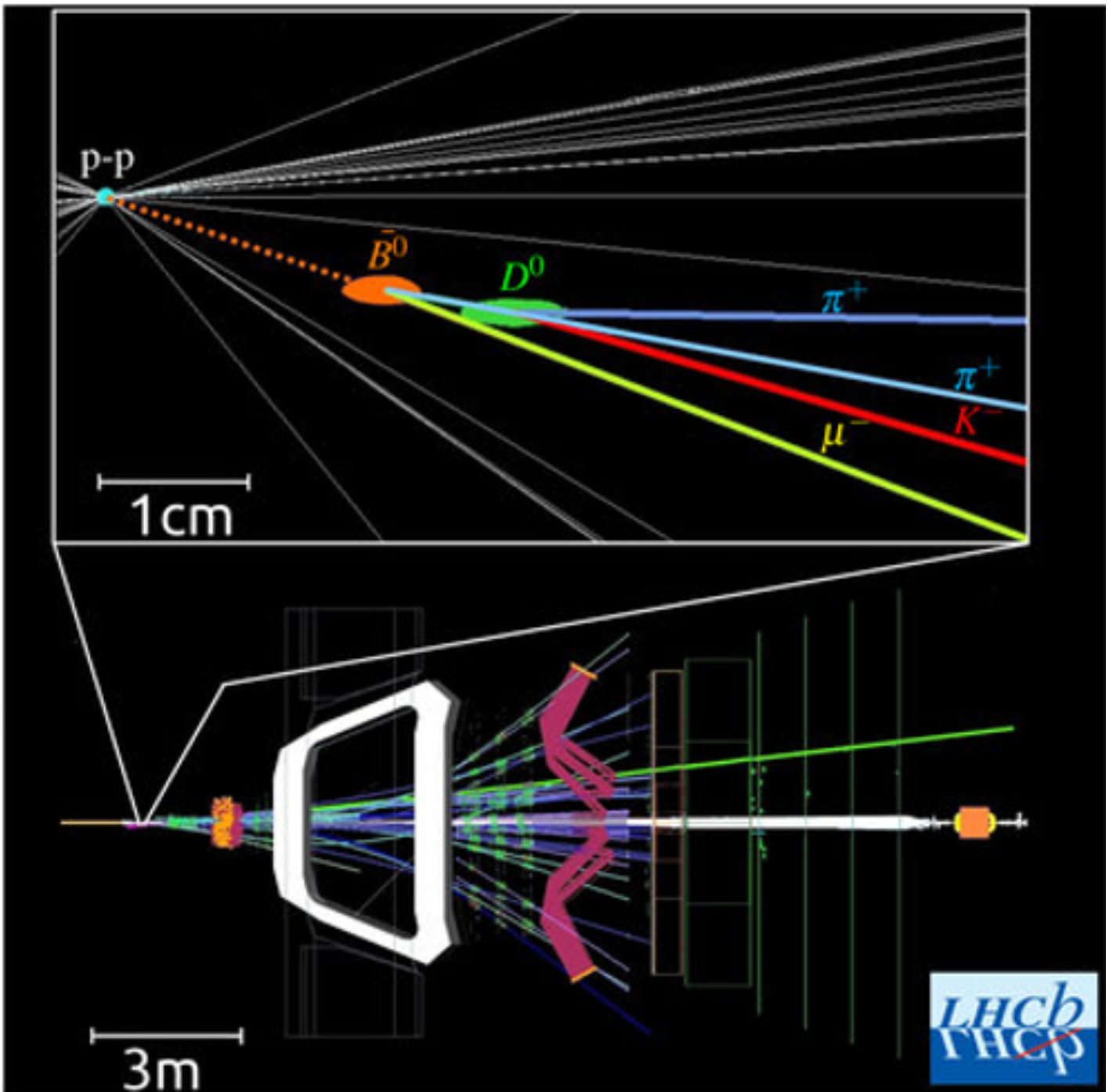}
 \caption{(left) A typical event from $\en\ep$ collisions at Belle II, as obtained from simulation (source: image courtesy of Belle II/KEK).  (right) A $\proton\proton$-collision event recorded by the LHCb experiment (source: taken from ~\cite{Ciezarek2017}).}
 \label{BellevsLHCb_eventdisplay}
 \end{figure}

 The LHCb experiment~\cite{LHCb-DP-2008-001,LHCb-DP-2014-002} has entered the game with the start of the LHC in 2010 with proton-proton collisions at $7\Tev$, and since the last 5 years is leading the experimental field of  \CP violation.   The LHCb is a single-arm forward detector with an angular coverage corresponding to  the pseudorapidity in the range $2<\eta<5$,  taking advantage that $b$ and $c$ hadrons are produced at relatively small angles with respect to the $\proton\proton$ beam direction. In contrast to  $\en\ep$ machines, the initial state  is not known for $\proton\proton$ collisions at the LHC; the main mechanism for heavy-flavour production is  gluon-gluon fusion, and a typical event may produce a hundred charged particles. For illustration, a  comparison of the event profiles at  Belle and LHCb is shown in Fig.~\ref{BellevsLHCb_eventdisplay}.  
 To distinguish the $b$- and $c$-hadron decay products from a large background, LHCb relies on an efficient and reliable trigger system, capable of full online reconstruction, exploiting  the topological characteristics of the desired decay processes. The reconstruction of neutrals, nevertheless, are not very efficient when compared to that from low-occupation $\en\ep$ collisions. Another important difference, very relevant for \CP violation studies, is that the initial state is not \CP symmetric and this leads to particle-antiparticle production asymmetries that have to be considered (subtracted) for the study of (or search for)  \CP asymmetries. 
The data was taken in two periods: run I with collisions at a centre-of-mass energy of $7\Tev$ in 2011 and at $8\Tev$ in 2012, with integrated luminosities of $1\invfb$ and $2\invfb$, respectively; and run II, from 2015 to 2018, at a centre-of-mass energy of $13\Tev$ with an integrated luminosity of $6\invfb$. 
 While the great majority of LHCb analyses based on run I data are completed, many studies based on run II data are still under progress, and we shall expect new results throughout the next few years. 
 
 Meanwhile, the Belle II experiment~\cite{BelleII_TDR,BelleII_PhysicsBook_2018} has started its physics run in 2019 and will continue until 2025. All components of the detector are new or highly upgraded with respect to the former Belle detector to be able to operate at an event rate 40 times higher than before, now provided by the SuperKEKB accelerator \cite{SuperKEKB}. As KEKB, the  $\en\ep$ collisions occur at centre-of-mass energies in the region of the  $\PUpsilon$ resonances, mainly the $\FourS$. The expected integrated luminosity is $50 \rm ab^{-1}$. 
 
 The  LHCb Upgrade I~\cite{LHCb-TDR-012} is currently being commissioned for the start of run III of LHC in 2021. It is a major upgrade, with a factor of five higher luminosity as compared to run II,  with the goal to collect $50\invfb$ by the end of run IV in 2028. The LHCb Upgrade II~\cite{LHCb-PII-EoI,LHCb-PII-Physics} is planned from 2031 onwards (run V) during the High Luminosity LHC (HL-HLT) operation period.  
 
 LHCb and Belle II will allow comprehensive and complementary programmes on quark-flavour physics in the next decade. While Belle II will be able to efficiently study  final states with neutral particles and benefit from a very efficient tagging system, LHCb will have much larger yields for final states with charged particles and the capability to produce all $b$-hadron species, specially $\Bs$, $\Bc$  and $b$-baryons.

  \subsection{Direct \CP violation in the decay rate}
As discussed in Sect.~\ref{chapter2_CP_CPT}, the usual observable when measuring or searching for \CP violation in decays is the charge asymmetry, as defined in Eq.~\ref{CPasym_def}. For two-body decays of \D and \B mesons, this is the only \CP violation observable. Experimentally, though, one measures the particle and antiparticle {\it yields},  $N(P\to f)$ and $N(\bar P\to \bar f)$, respectively,  which are affected by a chain of effects from production, reconstruction, and final selection of events. The observed  {\it raw} asymmetry is obtained,
\begin{equation}
A_{\rm raw} = \frac{N(P\to f) - N(\bar P\to \bar f)}{N(P\to f) + N(\bar P\to \bar f)}~,
\label{Araw_def}
\end{equation}
which needs then to be corrected for eventual production, $A_{\rm P}$, and detection, $A_{\rm D}$, asymmetries. Usually these constitute small effects, of a few percent at most, which allow the approximation \cite{Gersabeck_Interplay_2012}
\begin{equation}
A_\CP \approx A_{\rm raw} - A_{\rm P} - A_{\rm D}~.
\end{equation}
The precise measurement of these nuisance asymmetries is thus fundamental. Usually they are measured through control channels for which no \CP violation effects are expected.

For baryons and for multi-body decays other measurements can also be accessed, due to  polarisation and phase space, as  discussed in the following.

 \subsection{Direct \CP violation in multi-body decays}
 \label{multibody}

For three or more particles in the final state, besides the measurement of the (total) charge asymmetry, \CP violation can be studied through the phase space distribution of the decay. Usually in \B and \D mesons decays, the process is dominated by the formation of resonances as intermediate states, which then decay strongly or electromagnetically to form the detectable final state. The distribution of events throughout the phase space is the result of the superposition of the various amplitudes, and the interference pattern depends directly on the strong and weak phases involved. The rich dynamics potentially allow different sources of strong phases to appear.  It is natural to expect localised \CP asymmetries to be stronger than phase-space integrated ones, and can even change sign. Altogether these features make multi-body decays an excellent tool for studying  \CP violation in the hadronic sector. 

To measure or to search for \CP violation over the phase space, two approaches are pursued: model-dependent and model-independent analyses.  The first is based on an amplitude analysis fit;  the decay amplitude is modelled as a coherent sum of intermediate states such that their relative contributions can differ for particle and antiparticle decays. In  model-independent strategies the phase space distributions for particle and antiparticle decays are directly compared to look for regions where there are statistically significant differences, and localised asymmetries can be measured. The two approaches are complementary: while model-independent techniques can pin down the regions of the phase space where \CP violation is manifest, amplitude analysis can identify its dynamical source. 

\subsubsection{Amplitude analysis}
\label{AmAn}

Decays of \D and \B mesons to three pseudo-scalar mesons in the final state have their dynamics fully described in terms of only  two invariants.  By writing a general decay as $P\to h_1h_2h_3$ ($P=\B, \D$ and $h_{1,2,3}$ light mesons)  this two-dimensional phase space can be constructed in terms of two-body squared invariant masses such as $s_{12} = m^2(h_1h_2)$ and $s_{23} = m^2(h_2h_3)$, the so-called Dalitz plot. The distribution of events in the Dalitz plot is directly proportional to the total amplitude squared, with no phase-space factors involved. 

The simplest and most common approach for an amplitude analysis is the so-called isobar model where the total amplitude is described as a coherent sum of resonant amplitudes,  for instance $P\to R_{12} h_3$, $R_{12}$ being a resonance decaying to $h_1h_2$; the ``companion'' hadron $h_3$ is assumed to not interact with the $h_1h_2$ system. 

The intermediate resonance is described by a propagator, usually a relativistic Breit--Wigner, which is a good representation only for narrow and non-overlapping resonances, having the form
\begin{equation}
{\it BW_{\! R} }=\frac{1}{m_R^2 - s_{ij} -i m_R \Gamma_R(s_{ij})}~,
\label{BW}
\end{equation}
where $\Gamma_R$ is the energy-dependent width, which introduces a strong phase variation of $180^\circ$ around the nominal resonance mass $m_R$, and $i,j$ represent the label of the hadron pair it decays to. Each resonant amplitude is described as 
\begin{equation}
{\cal A}_R= {\it  {^J\!F}_{\! P}  {\it ^J\!F}_{\! R}  \times {^J\!M}_{\! R} \times  BW_{\! R} }~,
\label{AM}
\end{equation}
where $J$ is the spin of the resonance, the factors ${\it ^J\!F}_{\! P,R}$  are the Blatt--Weisskopf damping factors \cite{Blatt:1952ije}, accounting for the finite size of the $P$ and $R$ mesons, and ${\it ^J\!M}_{\! R}$ is an angular factor, ensuring angular-momentum conservation \cite{Argus}. Explicitly, ${^0\!M}_{\! R} = 1$, ${^1\!M}_{\! R} =- 2|p_k||p_i|\cos\theta_{ki}^{R_{ij}}$ and \mbox{${^2\!M}_{\! R} =4(|p_3||p_2|)^2(3\cos^2\theta_{ki}^{R_{ij}} -1)$} for resonances of spin 0, 1 and 2, respectively, where $p_{i(k)}$ is the modulus of the momentum of particle $h_{i(k)}$ and $\theta_{ki}^{R_{ij}}$ is the angle between the momenta of particles $h_k$ and $h_i$ in the $h_ih_j$ rest frame.  

The total amplitude is then written as a superposition of individual amplitudes
\begin{equation}
{\cal A}(s_{12}, s_{23}) = a_{\rm nr}e^{i\delta_{\rm nr}} {\cal A}_{\rm nr}(s_{12},s_{23})+ \sum_{{n =1}}^{N} a_n e^{i\delta_n}{\cal A}_{R_n}(s_{12},s_{23})~,
\label{Coherent}
\end{equation}
where a magnitude $a_n$ and a phase $\delta_n$ are associated to each resonant contribution $n$; these are parameters to be determined by the amplitude fit,  the former giving the relative strength  and the latter representing an attributed phase incorporating potentially the effects of both  weak and strong phases. The first term allows for a possible non-resonant contribution, with magnitude $a_{\rm nr}$, phase $\delta_{\rm nr}$ and the shape ${\cal A}_{\rm nr}$ usually assumed to be a  constant for \D decays and an exponential or a single-pole function for \B decays \cite{Argus,Belle_bkpipi,Tobias_CPT2,LHCb-PAPER-2018-051}. The Dalitz plot is fitted allowing for \CP violation, that is, $a_n$ and $\delta_n$, for the decay of $P$ meson,  and $\bar a_n$ and $\bar \delta_n$, for the decay of $\bar P$ meson, are independent fit parameters.

The isobar approach, while simple, has limitations. One of them is the assumption that the companion hadron (that not resulting from the decaying resonance) does not take part in the dynamics. This is the $2$+$1$ or quasi-two-body approximation and is discussed below in Sect.~\ref{quasi2body}. The other is that it is known to be fairly good only for narrow, non-overlapping resonances with the same quantum numbers. There are examples of  three- or four-body \B and \D decays where low-mass ($\lesssim 2 \gev$) broad scalar resonances are very important or even dominant. Assuming the  $2$+$1$ approximation to be exact, the simple sum of relativistic Breit--Wigners would not preserve two-body unitarity.\footnote{However the validity of the isobar model finds support  with arguments  in Ref.~\cite{Svec2000}.} To overcome such a problem, some analyses have been using alternative approaches for the parameterisation of the S-wave. One is a quasi-model-independent approach, where the S-wave amplitude is binned throughout the two-body invariant mass range to fit for its amplitude and phase in each bin, as in Refs.~\cite{E791_Brian,Focus_PWA,BaBar_ds3pi,LHCb-PAPER-2019-017}. Another one is the K-matrix approach~\cite{Aitchison_Kmatrix}, which imposes two-body unitarity by using external data from scattering experiments, as in Refs.~\cite{LHCb-PAPER-2019-017,FOCUS_Kmatrix_3pi}. 
   
Yet, while approximate, the simple and elegant isobar model can reveal important features related to what has been discussed in Sect.~\ref{chapter2_CP_CPT}. To illustrate this, we show a simple example with only two resonant amplitudes appearing in the $h_1h_2$ channel, interfering in a particular region of the Dalitz plot. Suppose one is a vector resonance, for example $\rho(770)^0$ and the other a scalar,  $f_0(980)$.\footnote{Scalar plus vector amplitudes are the dominant configuration in the low-mass regions in three-body heavy meson decays.} For simplicity and without loss of generality, we represent each resonant amplitude by a Breit--Wigner, and consider only the relevant $\cos\theta$ factor for the $\rho(770)^0$ amplitude ($\cos\theta^{R_{12}}_{13}$ in this case).\footnote{Unless otherwise stated, wherever $\cos\theta$ appears throughout the text, $\theta$ refers to the angle $\theta_{ki}^{R_{ij}}$ defined earlier, that is, that of the direction of the companion hadron $h_k$ with $h_i$ in the $h_ih_j$ rest frame.} The  two charge-conjugate amplitudes are written  as follows:
\begin{eqnarray}
\Ap = \arp e^{i \drp} \BWr \cos\theta + \afp e^{i \dfp} \BWf  \nonumber \\
\Am = \arm e^{i \drm} \BWr \cos\theta + \afm e^{i \dfm} \BWf~.
\label{Arho_f0}
\end{eqnarray}
Besides the inherent (and energy-dependent) Breit--Wigner phases, the phases $\delta$ and $\bar\delta$ represent, as mentioned before,  the net effect of contributions from the strong phase due to on-shell penguin diagrams, final-state (soft) hadron-hadron interactions, and the underlying weak phase from the CKM matrix.  
The difference of the total amplitude squared for particle and antiparticle is  given by
\begin{eqnarray}
\Ad= \Apsq - \Amsq = [\arp^2 - \arm^2]\BWrsq\cos^2\theta + [\afp^2 - \afm^2]\BWfsq + \nonumber \\ 
2 \BWrsq \BWfsq \cos\theta \cdot   \nonumber  \\
\Big\{ \left[(\mrr^2 -s_{12})(\mf^2 -s_{12})+  (\mrr \gr)(\mf \gf)\right]  \left[ \arp \afp \cos(\drp-\dfp) - \arm \afm \cos(\drm-\dfm)\right] \nonumber \\
- \left[\mrr \gr(\mf^2 -s_{12}) - \mf \gf(\mrr^2 -s_{12})\right] \left[\arp \afp \sin(\drp-\dfp) - \arm \afm \sin(\drm-\dfm)\right]  \Big\}~.
\label{AcpAmplitude}
\end{eqnarray}

The first two terms correspond to a \CP asymmetry associated to each of the resonances individually. A short-distance tree-penguin interference would manifest through them (assuming the resonance can be formed by both processes).  It would be the analogous case of a two-body asymmetry, but here as a quasi-two-body state. Due to its vector nature, the signature from the first term is the $\cos^2 \theta$  distribution, and this feature shall be better discussed in Sect.~\ref{quasi2body}.  
From the amplitude fit, a \CP asymmetry for each individual amplitude can be defined in terms of the magnitudes. For instance, for the $\rho(770)^0$ amplitude
\begin{equation}
A_{\CP}^\rho = \frac{a_\rho^2 - \bar a_\rho^2}{a_\rho^2 + \bar a_\rho^2}~.
\label{ACP_r}
\end{equation}

 The interference term between the two resonant amplitudes gives rise to the third term of Eq.~\ref{AcpAmplitude}, and is proportional to $\cos\theta$. As such, this term drives a change of sign of the asymmetry throughout the phase space, the inflexion region being around $\cos\theta \approx 0$.  Yet, this factor multiplies terms related to the complex nature of the Breit--Wigner. When  $s_{12}$ crosses the resonance nominal mass $m_n^2$ (either $\rho(770)^0$ or $f_0(980)$), the \CP asymmetry associated with this contribution tends to change sign. This effect has been seen on data in the decays  $ \B^\pm \to \pi^\pm \pi^+\pi^+$ and $ \B^\pm \to K^\pm \pi^+\pi^+$ as discussed later in Sect.~\ref{CPfromcostheta}. 
This S- and P-wave interference term, integrated in the whole phase space, must be equal to zero due to orthogonality, and consequently the net \CP asymmetry coming from the third term should be equal to zero. 
The integrated asymmetry may still receive contributions from the first and second terms though. In Sect.~\ref{quasi2body}, we discuss that, by  strictly  assuming the $2$+$1$ approximation,  \CP violation involving light vector mesons should be very small.

\subsubsection{The quasi-two-body approximation}
\label{quasi2body}

 Any approximation neglecting hadron interactions  in multi-hadron decays, such as  the $2$+$1$ approximation,  has to be taken with parsimony. Weak decays are point-like interactions, and strong forces mediate the final state  formation. The validity of the  approximation may depend on specific cases but in general one would expect to be better for \B than for \D decays, and with less hadrons in the final state. 
       
\begin{figure}[!htb]
\centering
\includegraphics[width=.45\columnwidth,angle=0]{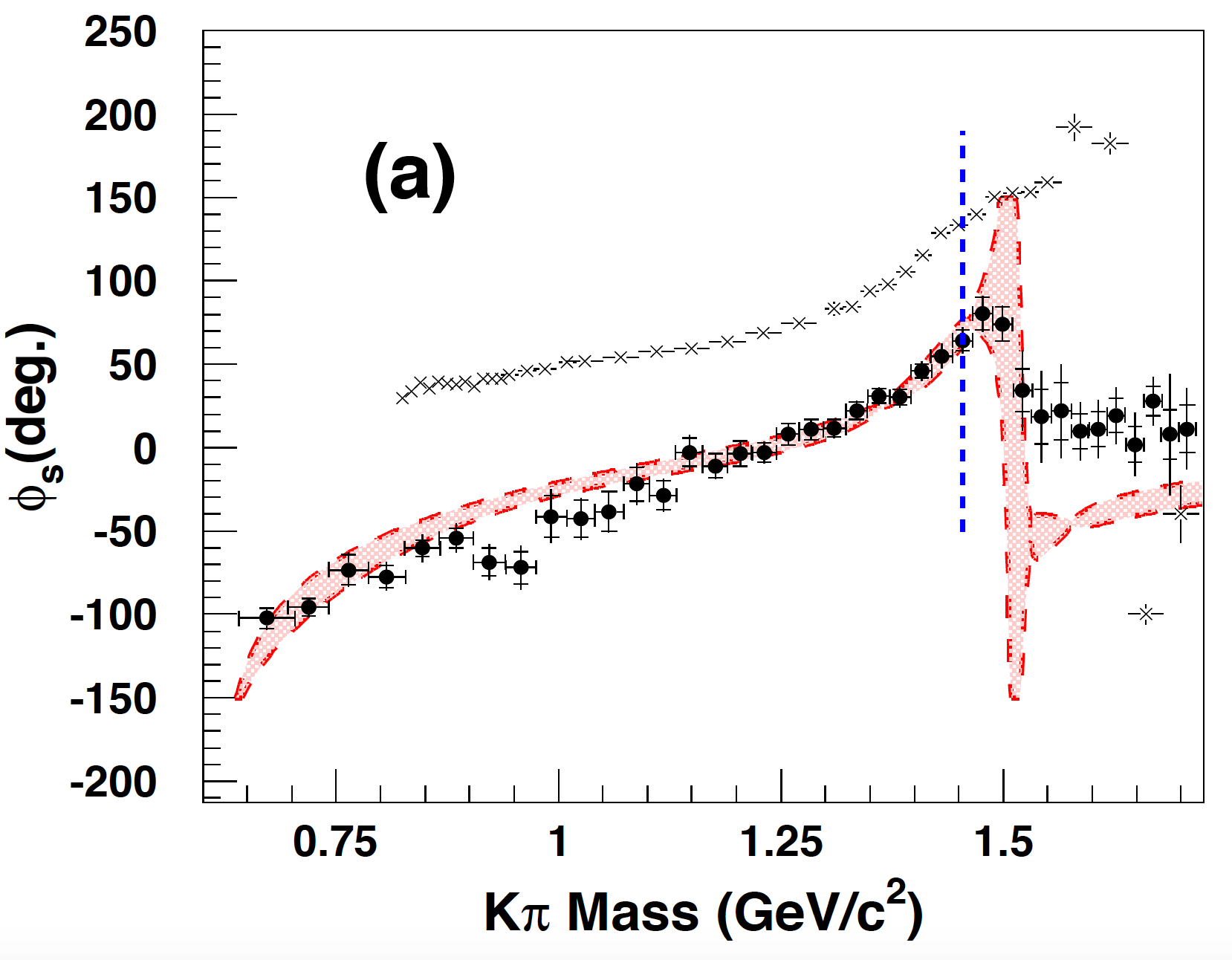}
\caption{The S-wave phase as a function of the $\Km\pip$  mass from the QMI-PWA
 $ \Dp\to\Km\pip\pip$ Dalitz plot fit, taken from \cite{E791_Brian}. The band shows the result of the Isobar model fit in a previous work \cite{E791_kappa}. The hashed vertical line shows the elastic range according
to LASS \cite{LASS}. }
\label{e791PWA}
\end{figure}

In fact, despite the large use of the $2$+$1$ approximation in three-body \D amplitude analyses, some problems have been observed with this approach in experimental studies. The Fermilab E791  \cite{E791_Brian} and FOCUS~\cite{Focus_PWA} experiments have performed a quasi-model-independent partial wave amplitude (QMI-PWA) approach to measure the $\Km\pip$ S-wave in the $\Dp\to\Km\pip\pip$ decays.  Both results  deviate from the $\Km\pip$  S-wave measured by the LASS scattering experiment \cite{LASS}, as shown in Fig.~\ref{e791PWA} for the E791 result. The  $\Km\pip$  system, if isolated, should present the same phase behaviour independently of the production process, according to  the Watson theorem~\cite{Watson}. A plausible explanation for the difference in the $\Km\pip$  behaviour is the interaction with the companion \pip in the $\Dp\to\Km\pip\pip$ decay. 

The QMI-PWA of the $ \Dsp\to\pim\pip\pip$~\cite{BaBar_ds3pi} and  $\Dsp\to \Km\Kp\pip$~\cite{BaBar_dskkpi} decays have also presented a  different behaviour for the $\pim\pip$ and $\Km\Kp$ S-wave, respectively,  when compared to those from two-bodies elastic scattering~\cite{CERN-Munich,Hadron2017}. Altogether these results reinforce the idea that the companion particle in three-body \D decays takes a role in the final state. 

\begin{figure}[!htb]
\centering
\includegraphics[width=.4\columnwidth,angle=0]{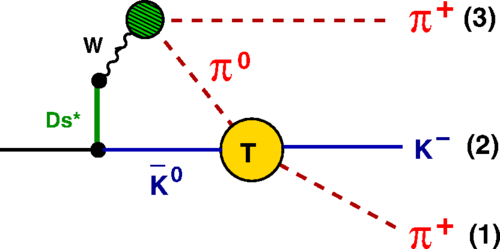}
\caption{Leading vector-current contribution to the $ \Dp\to\Km\pip\pip$ decay~\cite{Pat_Rob2015}.}
\label{pat2015}
\end{figure}

\begin{figure}[htb]
\centering
\includegraphics[width=.85
\columnwidth,angle=0]{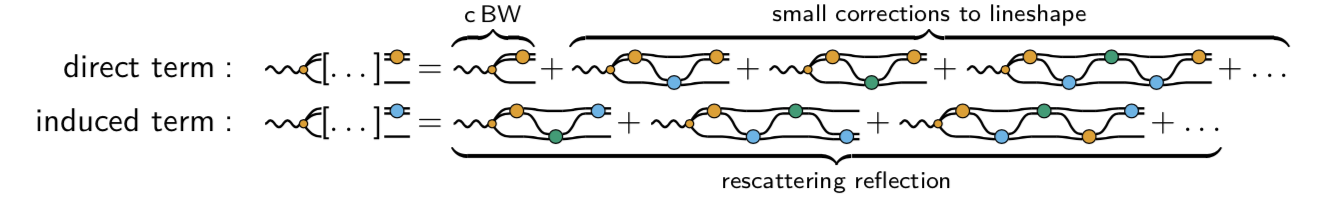}
\caption{ Direct and Induced terms produced in a three-body  rescattering \cite{Mikhasenko_private}. }
\label{Mikhashenko1}
\end{figure}

\begin{figure}[htb]
\centering
\includegraphics[width=.85
\columnwidth,angle=0]{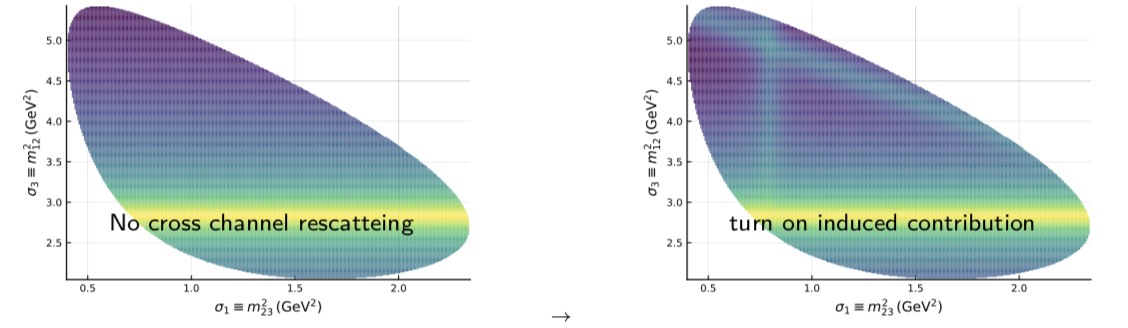}
\caption{ Signatures in the Dalitz plot  for the direct and Induced terms in re-scattering \cite{Mikhasenko_private}.}
\label{Mikhashenko2}
\end{figure}

The importance of three-body final-state interactions (FSI) in heavy-meson decays has been investigated by different theoretical groups~\cite{Boito, Pat_Rob2011,Pat_Rob2015,Nakamura,Guimaraes2014}. They use a simple model implementing three-body unitarity, solving the Faddeev equation. The first-order rescattering  interaction is represented through a vector-resonance intermediate state as shown in Fig.~\ref{pat2015}. Additional contributions  can be added by including higher-order loops in a  transition matrix that sums the connected scattering series from  a ladder graph~\cite{Guimaraes2014,Mikhasenko_private}, as shown in Fig.~\ref{Mikhashenko1}.
In this figure, three-body rescattering effects in these decays have been classified as a direct term, with a small correction to the Breit--Wigner lineshape~\cite{Pat_Ig}, and an induced term, with a possible  presence of a resonant term to  the same three-body final state \cite{Mikhasenko}. Figure~\ref{Mikhashenko2} presents the Dalitz plot distribution for these two terms.

\begin{figure}[htb]
\centering
\includegraphics[width=.25\columnwidth,angle=0]{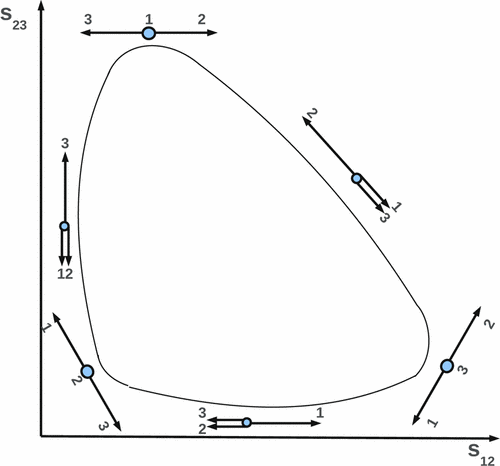}
\caption{A sketch of the relative momenta  for particles 1, 2 and 3 in the Dalitz plot constructed by $s_{12}$ and $s_{23}$ invariants. Taken from~\cite{Bediaga:2012up}.}
\label{Dalitz_momentum}
\end{figure}

For three-body charmless \B decays, with a much larger phase space, the $ 2\!+\!1$ approximation would be expected to be less problematic. However, depending on the momentum distribution of the final-state particles in the Dalitz plot, the validity of the approximation is better  in some regions than in others. 
For instance, in the \B-meson rest frame, the corners of the Dalitz plot (see Fig.~\ref{Dalitz_momentum}) correspond to a configuration in which one particle is at rest. The midway between the corners corresponds to a configuration where two of the particles are colinear, thus having the same scalar momentum as the third particle. The centre of the Dalitz plot has all three particles with the same scalar momentum. Thus the validity of the $2$+$1$ approximation may be  better in some regions than in others. The participation of the third particle in the two-body interaction can change the two-body strong phase over the Dalitz plot and should be taken into account in \CP violation studies.

The characteristics of the three-body momentum distribution over the phase space for the specific process of $ \Bp\to\pim\pip\pip$ have been addressed in Refs.~\cite{Mannel1,Mannel2}. The authors show that the peripheral regions of the Dalitz plot, where light resonances appear, are essentially dominated by non-perturbative effects. On the other hand, the central region of the Dalitz is dominated by large  momentum transfer, allowing for a quasi-perturbative treatment of QCD. 

\subsubsection{The {\it Miranda} techniques}    
\label{miranda}

The use of amplitude analyses allows to access amplitude differences -- both their magnitudes and phases -- between particle and antiparticle decays, but relies on an adequate modelling of the possible intermediate states and, as discussed in Sect.~\ref{AmAn},  the description of three- and also four-body decays is a complicated task with non-perturbative QCD effects possibly playing a major role depending on the region of phase space.  
The study of \CP violation in multi-decays can also be carried out by model-independent techniques by comparing the distribution of events for particle and antiparticle over the phase space. 

\begin{figure}[htb]
\centering
\includegraphics[width=.6
\columnwidth,angle=0]{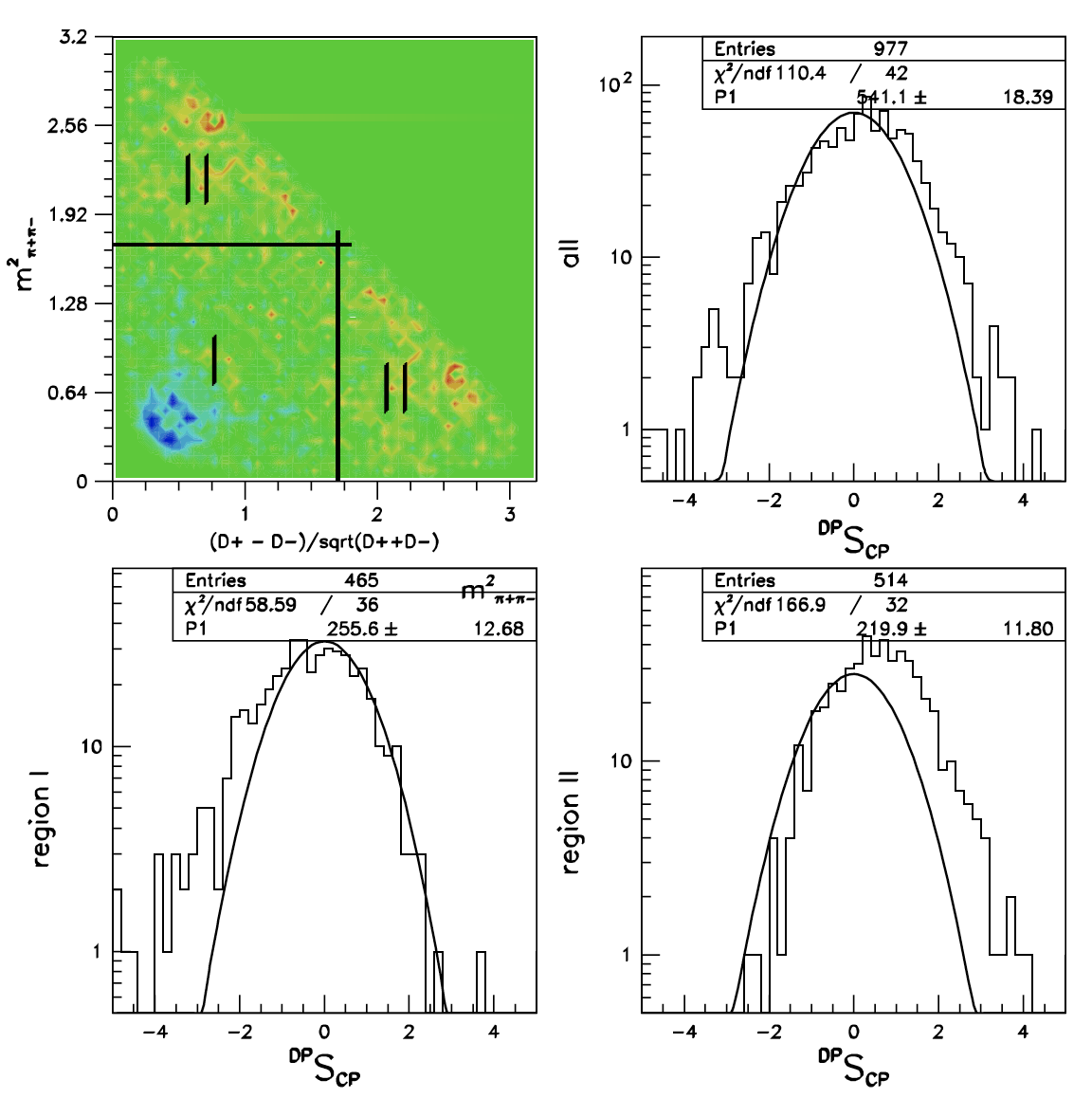}
\caption{Example of $S_\CP$ distribution for simulated $\D^\pm \to\pi^\pm\pim\pip$ decays where a 1\% phase difference was introduced between $\rho(770)^0\pi^+$ and $\rho(770)^0\pim$ amplitudes. From the top-left plot clockwise: $S_\CP$ across the Dalitz plot, distribution of $S_\CP$ for the whole Dalitz plot, $S_\CP$ for regions II, and I, as taken from Ref.~\cite{bigi1}. }
\label{mirandaD3pi_original}
\end{figure}

As a search tool for \CP violation, which is particularly relevant for \D decays, the {\it Miranda} technique~\cite{bigi1} consists in dividing the phase space in bins of the relevant variables (for instance, the two Dalitz plot variables in a three-body decay) and for each bin $i$ the number of observed events $N_i(P)$ and $N_i(\bar P)$ for the decay of particle $P$ and its \CP conjugate process, respectively,  are obtained. The bin sizes can be uniform across the phase space, or adaptively defined according to a given criterion (for example, number of events in each bin). An asymmetry significance is calculated for each bin
\begin{equation}
S_\CP(i) = \frac{N^i(P) -\alpha N^i(\bar P)}{\sqrt{\alpha(N^i(P) +N^i(\bar P))}}~,
\label{SCP}
\end{equation}
where  the factor \mbox{$\alpha = \sum_i N^i(P)/\sum_i N^i (\bar P)$} is used to normalise the total sample size of $\bar P$ decays to that of $P$. It serves to correct for any global asymmetries,  such as those coming from production and detection effects, but also includes the net effect of the integrated \CP asymmetry.  The method is highly sensitive to {\it local} asymmetries, expected to be more pronounced that the integrated one. Under  \CP conservation, the distribution of $S_\CP(i)$ follows a Gaussian distribution, with mean and width equal to zero and one, respectively.\footnote{The number of entries in each bin is ensured to be large enough to guarantee Gaussian uncertainties.} Any deviation is indication of \CP violation. In this case, the distribution pattern of $S_\CP(i)$ through the phase space evidences the regions where the effect appears. A p-value for the \CP conservation hypothesis can be obtained by constructing a $\chi^2 \equiv \sum_i S^2_\CP(i)$ and the number of degrees of freedom, which is equal to the number of bins minus one (due to the $\alpha$ normalisation). Figure~\ref{mirandaD3pi_original}(a) shows a typical Miranda distribution where a simulated sample of about 1 million $\D^\pm \to\pi^\pm\pim\pip$  decays with a 1\% ($3.6^\circ$) phase difference is included in the $\rho(770)^0\pip$ component with respect to that of $\rho(770)^0\pim$~\cite{bigi1}. 

\begin{figure}[htb]
\centering
\includegraphics[width=.45\columnwidth, height =.3\columnwidth, angle=0]{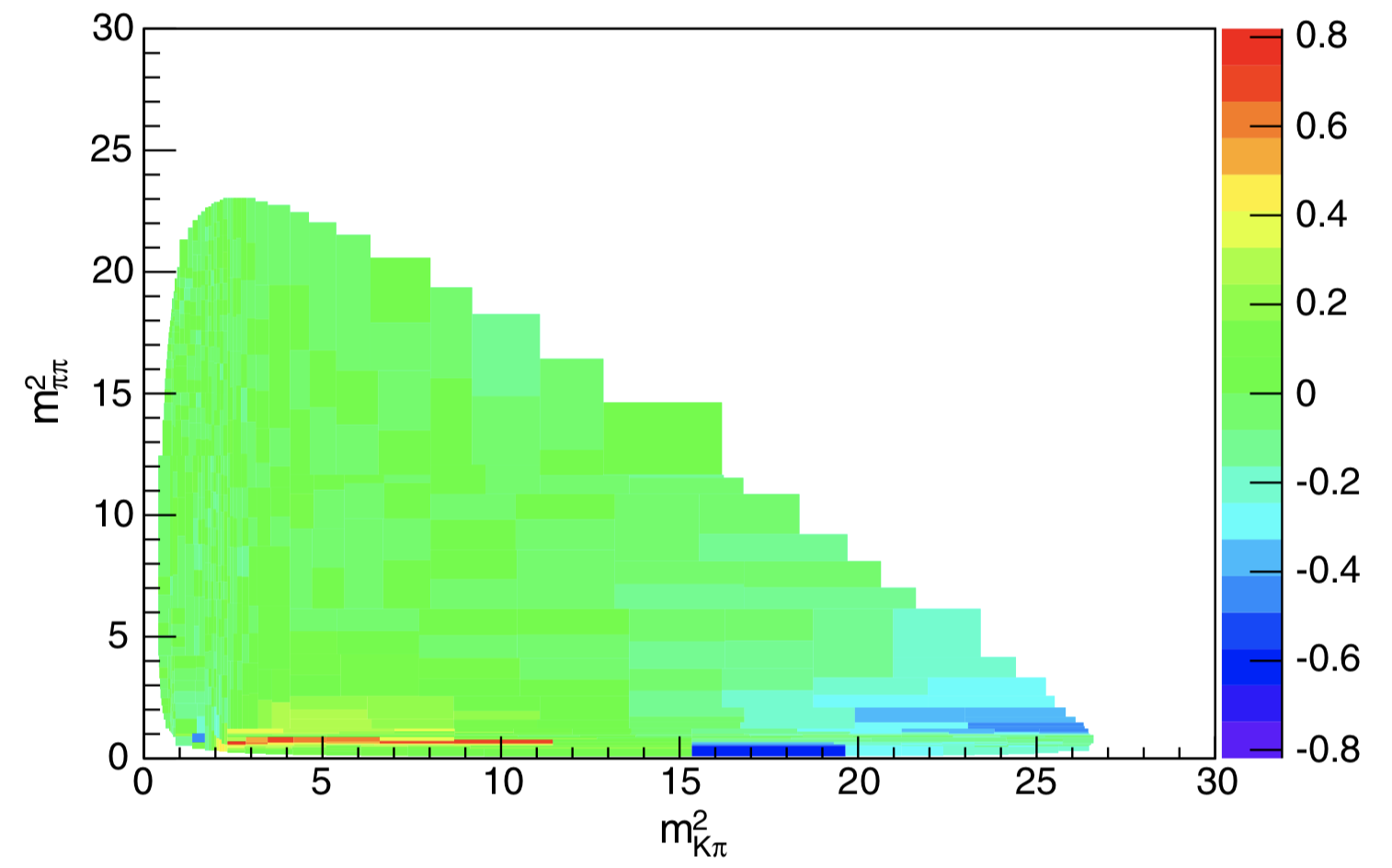}\includegraphics[width=.45\columnwidth,height =.3\columnwidth,angle=0]{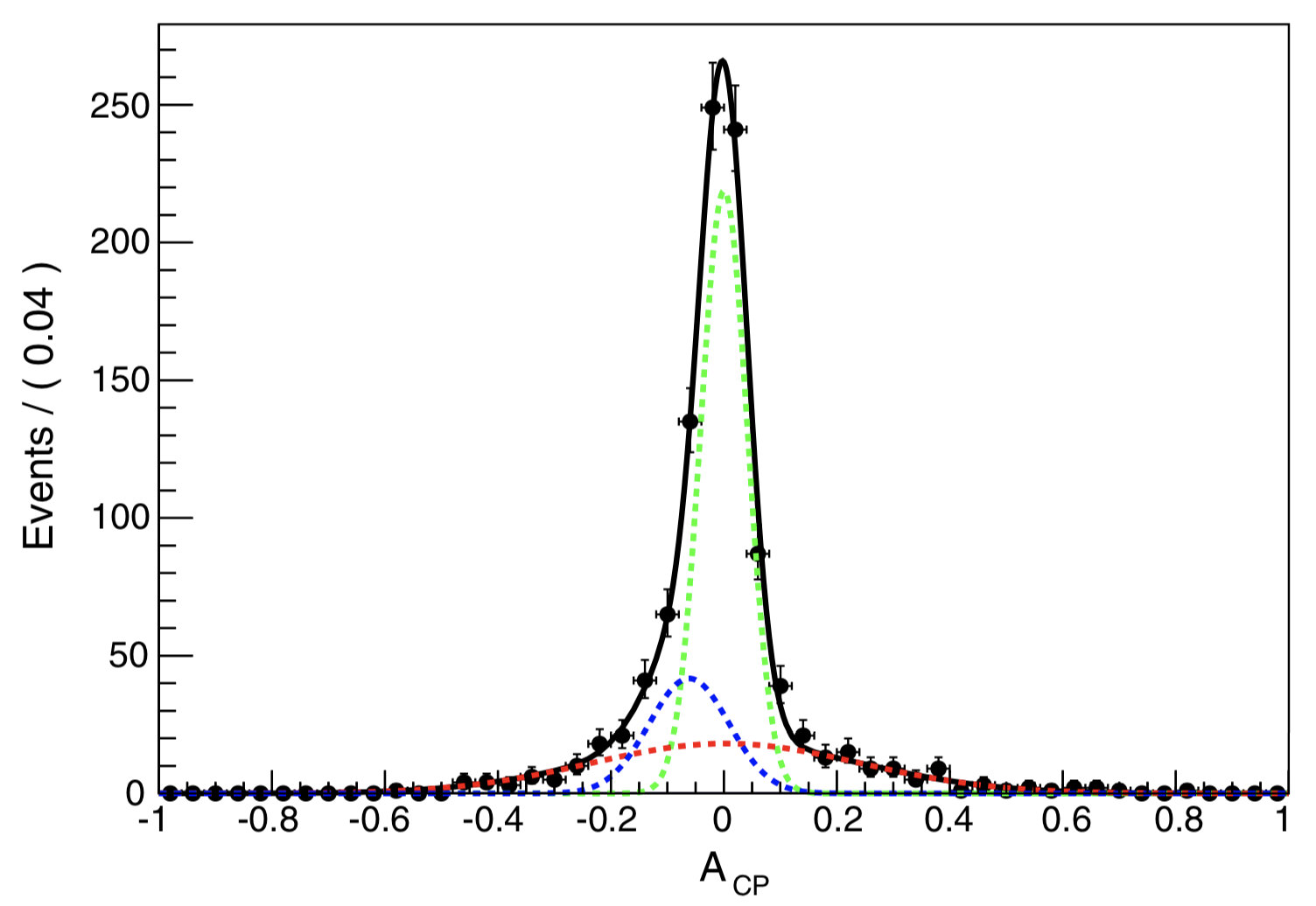}
\caption{(left) Charge asymmetries across the Dalitz plot from  simulated  $\Bz\to \KS\pip\pim$ decays; (right) distribution of the charge asymmetries in different regions of the phase space~\cite{bigi2}.}
\label{2ndMiranda}
\end{figure}

In the case of presence of a measurable \CP violation effect, the method above is very sensitive to it, giving p-values incompatible with the null hypothesis, and it is possible to distinguish the regions in the Dalitz plot with substantial $S_\CP$ values, but otherwise gives no quantification. A ``second generation'' Miranda technique~\cite{bigi2} is dedicated to the measurement of \CP asymmetries across the phase space. The method strengthens local effects by obtaining asymmetries through the division of the Dalitz plot in bins of equal population for the combined samples of particle and antiparticle decays: each bin has $ N_i^{\rm tot} = N_i(P) + N_i (\bar P)$ events. In Fig.~\ref{2ndMiranda} a simulation of $\Bz\to\KS\pip\pim$ decays is shown, where \CP violation was introduced as a $60^\circ$ phase difference in the $\rho(770)^0\KS$ mode. In the left plot, localised positive and negative asymmetries are clearly seen, reaching values as high as 80\%, although the total integrated asymmetry is only about 1\%. In the right plot, the distribution of asymmetries is shown. This is a clear example evidencing how \CP violation effects can be substantially stronger in regions of the phase space when compared to the total, integrated, \CP effect. 

We discuss the use of these techniques in Sects.~\ref{chapter4} and \ref{chapter6}.

\subsubsection{Unbinned techniques}
\label{unbinned}

The search for \CP violation within the phase space in a model-independent way can also be performed through unbinned, event-by-event techniques. One such technique is the  so-called {\it energy test} \cite{Aslan_energytest1_2005,Aslan_energytest2_2005,Williams_energytest_2011}. The method relies on a test statistic, $T$,  for a comparison of distances of pairs of events in the phase space,
\begin{equation}
T = \sum_{i,j>i}^N \frac{\psi_{ij}}{N(N-1)} + \sum_{i,j>i}^{\bar N} \frac{\psi_{ij}}{\bar N(\bar N-1)} - \sum_{i}^N \sum_j^{\bar N}\frac{\psi_{ij}}{N\bar N}. 
\label{T_energytest}
\end{equation}
The first and second terms correspond to the average weighted-distance between events in the sample of particle $P$ (with $N$ total events) and antiparticle $\bar P$ ($\bar N$ events), respectively. The third term measures the same quantity but now between $P$ and $\bar P$ events. The metric function $\psi_{ij}$ is usually chosen as a Gaussian, $\psi_{ij} = e^{-d^2_{ij}/2\delta^2}$ where $d_{ij}$ is the ``distance'' (measured in \gevgevcccc) between the $i$ and $j$ events in the phase space. The parameter $\delta$ is an effective radius which defines the region where the distributions for $P$ and $\bar P$  are being compared. If localised \CP asymmetries are present, the last term in Eq.~\ref{T_energytest} is smaller, since the average distances are larger, when compared to a scenario with no asymmetries, and the $T$ value will be larger. To do this comparison and access a p-value for the hypothesis of \CP  conservation, the nominal $T$ value is compared to a distribution of $T$ values obtained from permutation samples (associating randomly the ``flavour'' -- $P$ or $\bar P$ -- to all events for each permutation sample). When compared to the binned technique, the energy test has been shown to have equivalent or higher sensitivity to \CP violation effects~\cite{Williams_energytest_2011}.

An alternative unbinned technique is the k-nearest neighbour (kNN) method~\cite{Henze_kNN_1988,Schilling_kNN_1986}. Instead of using a metric function to define an effective region around each event, a given number of  nearest neighbours is defined and among them  the number of particle and antiparticle events are evaluated. Similar to  the energy-test method, a test statistic $T$ is constructed for the whole phase space (or for phase-space regions) 
\begin{equation}
T = \frac{1}{n_k(N+\bar N)}\sum\limits_{i=1}^{N+\bar N}\sum\limits_{k=1}^{n_k}I(i,k),
\end{equation}
where  $I(i,k)=1$ if the $i^{\rm th} $event and its $k^{\rm th}$nearest neighbour have
the same charge and $I(i,k)=0$ otherwise;  $N$ ($\bar N$) is the number of $P$ ($\bar P$) events. The test statistic $T$ is the mean fraction of like-charged neighbour pairs
in the combined $P$ and $\bar P$  samples. The advantage of 
the kNN method, in comparison with energy-test, is that the calculation of $T$ is simple and fast 
and the expected distribution of $T$ is well known: for the null hypothesis 
it follows a Gaussian distribution with mean $\mu_T$ and 
variance $\sigma^2_T$  calculated from known parameters of the distributions. A p-value for the null hypothesis can then be obtained. 

Both unbinned methods have been used for searches of \CP violation in three- and four-body charm decays, as discussed in Sect.~\ref{chapter6}. As in the case of the binned Miranda method given by Eq.~\ref{SCP}, these methods are search tools and  rely on the assumption that potential nuisance asymmetries, those coming from production or detection effects, do not vary across the phase space. Typically control modes, for which \CP violation signals are not expected, are studied prior to the modes of interest, to guarantee this premise.  
\subsubsection{Triple products}
\label{tripleproducts}

For four-body decays or for baryon three-body decays (due to the baryon polarisation), other \CP-violating observables can be accessed. A triple-product in terms of the final-state particle momenta can be constructed,  
\begin{equation}
C_T \equiv \vec{p}_1\cdot {\vec p}_2 \times {\vec p}_3, 
\label{C_T}
\end{equation}
where subscripts $1$, $2$ and $3$ represent the final-state particles (three out of three for baryon decays, or three out of four for meson decays). The $C_T$ quantity constitutes a T-odd observable ---  which reverses sign under reversal of momenta and spin of all involved particles (it is not the same as a true time reversal odd variable). For a particle $P$ (meson or baryon) decaying to a final state $f$, the triple product T-odd asymmetry  is defined as 
\begin{equation}
A_T = \frac{\Gamma(P\to f; C_T>0) - \Gamma(P\to f;C_T<0)}{\Gamma(P\to f;C_T>0) + \Gamma(P\to f;C_T<0)}
\label{A_T}
\end{equation}
and analogously for the $\bar P$ process, 
\begin{equation}
{\bar A}_T = \frac{\Gamma(\bar P\to \bar f; -C_T>0) - \Gamma(\bar P\to \bar f;-C_T<0)}{\Gamma(\bar P\to \bar f;-C_T>0) + \Gamma(\bar P\to \bar f;-C_T<0)}.
\label{ATbar}
\end{equation}
In the presence of final-state interactions, $A_T$ and ${\bar A}_T$ are generally different from zero even in the absence of \CP violation, but a  \CP violation observable can be constructed through the  asymmetry 
\begin{equation}
a_\CP^{T-\rm odd} \equiv \frac{1}{2}\left(A_T - {\bar A}_T\right), 
\label{a_CP_Todd}
\end{equation}
which, under \CPT conservation,  is different from zero if \CP is violated --- even in the absence of strong phases. Indeed, not relying on strong phases constitutes a clear advantage of this \CP observable, making this kind of measurement complementary to the usual asymmetries. Moreover, by construction $a_\CP^{T-\rm odd}$ as well as $A_T$ and ${\bar A}_T$ are insensitive to production and detection asymmetries.

In Sect.~\ref{b-baryons} and \ref{Dto4body} results of searches for \CP violation using triple products are presented.  
  \section{Direct \CP violation in charmless \B decays}     
\label{chapter4}

For $b$-hadron decays, $b\to u$ transitions -- leading to charmless final states --   are highly suppressed due to the small value of $|V_{ub}|$, when compared to $b\to c$ processes. The weak phase $\gamma$ enters at the tree-level amplitude. Penguin transitions  leading to the same final states can be of comparable size as the tree amplitude due to the  top quark in the loop, thus interference terms for which \CP violation could be assessed are naively expected to be large, depending on the size of the strong phases.

There are a variety of two-, three- and multi-body charmless final states from $b$-hadron decays. The measurements of \CP violation effects for these channels make use of some of the techniques described in the previous section. For two-body decays, $A_{\CP}$ is obtained  directly from the event yields from particle and antiparticle decays, correcting for nuisance asymmetries when applicable. The same goes for phase-space integrated measurements of multi-body decays, while for resonant intermediate states or for asymmetries in localised regions of the phase space the use of amplitude analyses and model-independent techniques is required.

\subsection{Two-body decays}

\begin{figure}[!htb]
\centering
\includegraphics[width=.65\columnwidth]{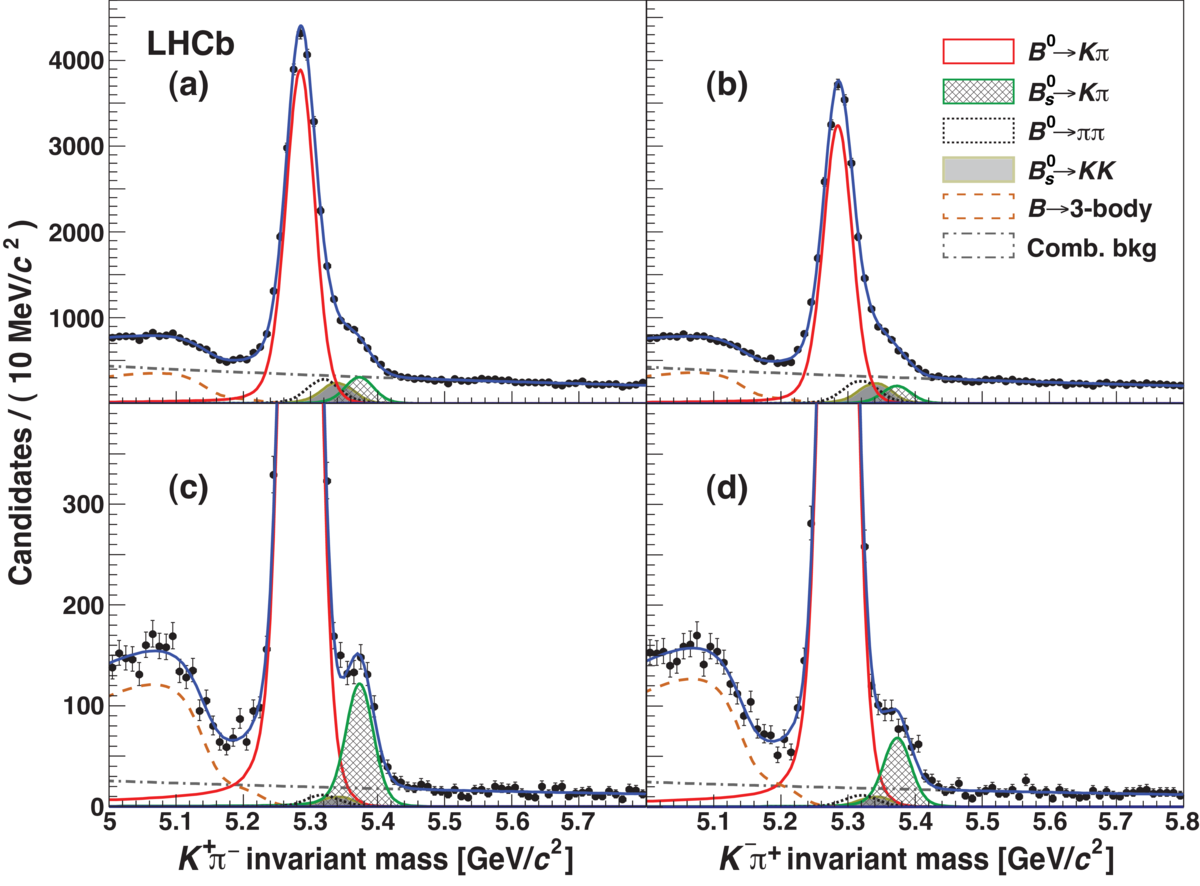}
\caption{Invariant-mass distributions for (a) $\Bz\to\Kp \pim$, (b) $\Bzb\to\Km \pip$, (c) $\Bs\to\Km\pim$ and (d) $\Bsb\to\Km\pip$. The legend describes the fit components \cite{LHCb-PAPER-2013-018}. }
\label{LHCB_CP_BsKpi}
\end{figure}

Although there are many two-body charmless \B decays, for only a very few of them the measured \CP asymmetry is statistically significant. In 2004,  BaBar \cite{BaBar_CPKpi} and Belle \cite{Belle_CPKpi} collaborations presented the first experimental evidence for \CP violation in $  \Bz \to \Kp \pim$  decay,  with a statistical significance around $4 \sigma$ (where $\sigma$ is one standard deviation). It was only in 2012 that  the BaBar collaboration \cite{BaBar2012}  observed this decay channel with more than $5 \sigma$, followed by direct \CP asymmetry observation of  \Bs to the same final state from the LHCb collaboration \cite{LHCb-PAPER-2013-018}.  The $\Kp\pim$ and $\Km\pip$ mass spectra obtained are shown in Fig.~\ref{LHCB_CP_BsKpi} where one can clearly see that the yields for both \Bz and \Bs are higher that those for \Bzb and \Bsb.     

The best measurement for these decays, with \CP violation fully established, comes from the recent update from the LHCb experiment \cite{LHCb-PAPER-2018-006}:
\begin{eqnarray}
A_{\CP}(  \Bz \to \Kp \pim)= -0.084 \pm 0.004 \pm 0.003~,\nonumber \\
A_{\CP}(  \Bs \to \Kp \pim )= 0.213 \pm 0.015\pm 0.007~,
\label{ACP_Bd}
\end{eqnarray} 
where the first uncertainty is statistical and the second systematic.\footnote{Unless otherwise stated, whenever measurements appear with two uncertainties, the first is statistical and the second systematic.}

The theoretical prediction from  NNLO approach  of the \CP asymmetry for the  $\Bz \to \Kp \pim$  decay is $(8.08 \pm1.52 \pm 2.60)\%$ (not including long-distance effects)\cite{Beneke2015}.\footnote{Curiously enough, this prediction   is very close to one made 30 years ago in Ref.~\cite{Bigi1987}.} The experimental value above, with opposite sign with respect to the theory prediction, opened  speculations about the possibility of new physics (NP) effects.  An interesting and robust relationship between the $ A_{\CP}$ of this decay and the $ A_{\CP}$ of $\Bs \to \Kp \pim $ decay was proposed by Lipkin~\cite{Lipkin2005} to test  whether the result was consistent with the SM.  Using U-spin symmetry \cite{Gronau2000}, the interference terms from these two decays are expected to be very similar, producing the approximate equality  \mbox{$ | A(\Bs \to \pi^+ K^-)|^2 - | A(\bar \Bs \to \pi^- K^+)|^2 \approx | A(\bar \Bz \to \pi^+ K^-)|^2 - | A(  \Bz \to \pi^- K^+)|^2 $}, which can be rewritten as 
\begin{eqnarray}
\Delta =  \frac{A_{\CP}(  \Bz \to \Kp \pim)}{A_{\CP}(  \Bs \to \Kp \pim )} +  \frac{ {\cal B} (  \Bs \to \Kp \pim ) \tau_d}{ {\cal B} (  \Bz \to \Kp \pim) \tau_s} \approx 0, 
\label{Delta}
\end{eqnarray} 
where $\tau_d$ and $\tau_s$ are the \Bz and \Bs mean lifetimes, respectively. Thus, from the measurements in  Eq.~\ref{ACP_Bd},  the LHCb collaboration quotes $ \Delta = 0.11\pm 0.04 \pm 0.03$ \cite{LHCb-PAPER-2018-006}, which is consistent with Eq.~\ref{Delta} within experimental uncertainties and favouring the SM nature of \CP violation in these decays.   

A third charmless two-body \B decay with well established  direct \CP violation is the $\Bz\to\pip\pim$ mode. Since this final state is  self-\CP conjugate, the measurement was performed with the time-dependent formalism presented in Eq.~\ref{CPasym_time}. The parameters ${\cal C}_{\pip\pim}$ and ${\cal S}_{\pip\pim}$  were measured by Belle \cite{Belle_pipi}, BaBar \cite{BaBar2012} and more recently by LHCb \cite{LHCb-PAPER-2018-006} collaborations. 

\begin{figure}[htbp]
\centering
\includegraphics[width=.4\columnwidth,angle=0]{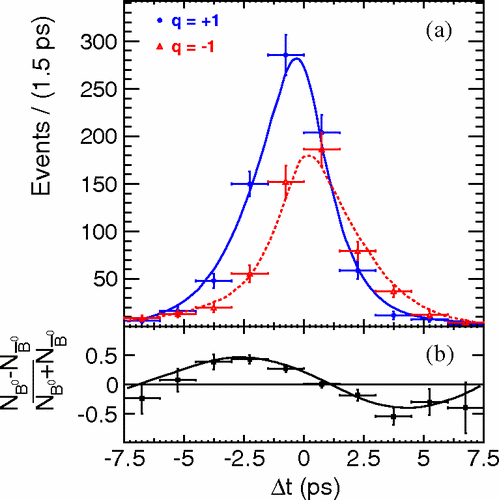}
\caption{ Background subtracted time-dependent fit results for $B^0 \to \pi^+\pi^ -$  \cite{Belle_pipi}. (a) $\Delta t$ distribution for \Bz (solid blue line) and \Bzb (dashed red line) (b) Time-dependent asymmetry  distribution.}
\label{Belle_time}
\end{figure}

The time-dependent distribution for $A_\CP$ obtained by the Belle collaboration is presented in Fig.~\ref{Belle_time}, from where the parameters ${\cal C}_{\pip\pim}$ and ${\cal S}_{\pip\pim}$ were extracted.\footnote{$\Bz$ and $\Bzb$ are produced as entangled pairs at Belle, so $t$ in Eq.~\ref{CPasym_time} is replaced by $\Delta t$, the time elapsed since the other \B (used for tagging) decayed.}

\begin{figure}[htbp]
\centering
\includegraphics[width=.4\columnwidth,angle=0]{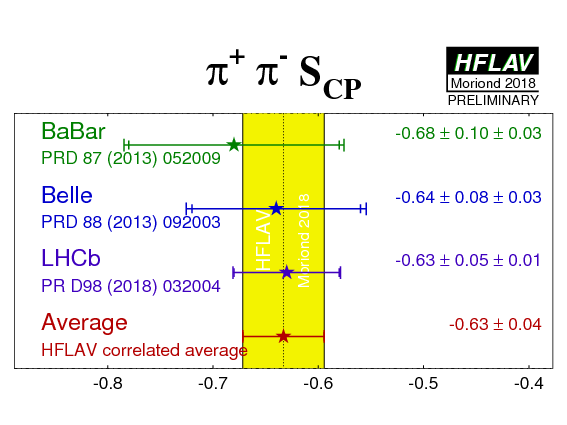}
\includegraphics[width=.4\columnwidth,angle=0]{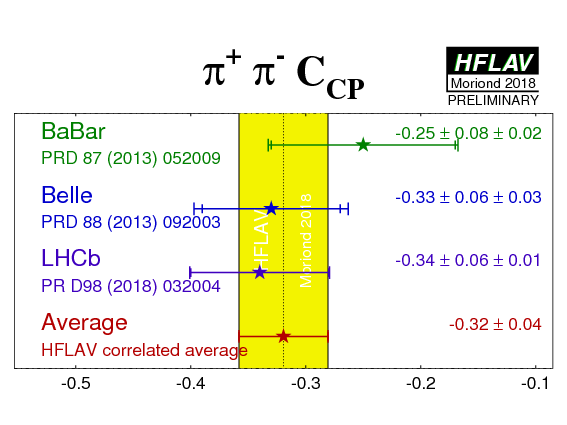}
\caption{ Averages of (left)   ${\cal S}_{\pip\pim}$ and (right) ${\cal C}_{\pip\pim}$ for $\Bz\to\pip\pim$ decay mode \cite{HFLAV19}.}
\label{Bpipi_HFLAV}
\end{figure}

The combined results from HFLAV \cite{HFLAV19} are shown in Fig.~\ref{Bpipi_HFLAV}  with the average result for direct \CP violation  ${\cal C}_{\pip\pim}=-0.32 \pm 0.04$ .

Many other two-body channels have been studied, with just one of them presenting  a \CP asymmetry with more than $3\sigma$ significance: $A_{\CP}(\Bp\to\Kp\eta)$ has been measured by Belle  \cite{Belle_CPKeta} and BaBar \cite{BaBar_CPKeta}, with consistent values. The current average values for $A_\CP$ from the PDG~\cite{PDG2019} for different channels are reproduced in Table~\ref{ACP_2body}.

\begin{table}[htpb]
\begin{center}
\caption{Average $A_\CP$ values for two-body charmless \B decays \cite{PDG2019}.}
\begin{tabular}{| l c |}\hline
Channel & $A_\CP$ \\ \hline \hline
$\Bp \to \KS \pip$ &  $-0.017 \pm 0.016$ \\ \hline
$\Bp \to \Kp \piz$   & $ -0.037 \pm 0.021$\\ \hline
$\Bp \to \Kp \eta'$   &  $ 0.004 \pm 0.011$\\ \hline
$\Bp \to \Kp \eta$    & $ -0.37 \pm 0.08$ \\ \hline
$\Bp \to \Kp \Kz$    & $ -0.04 \pm 0.14$\\ \hline
$\Bp \to \Kp \KS$ &  $-0.21 \pm 0.14$\\ \hline
$\Bp \to \pip \piz$  &  $0.03 \pm 0.04$\\ \hline
$\Bp \to  \pip\eta$  &  $-0.14 \pm 0.07$\\ \hline
$\Bp\to \pip\eta'$  &   $ 0.06 \pm 0.16$\\ \hline \hline
 $ \Bz \to \Kzb\Kz$ & $-0.6 \pm 0.7$ \\ \hline
$ \Bz \to \Kp \pim$  & $-0.082 \pm 0.006$ \\ \hline \hline
$ \Bs \to \Kp \pim $ & $ 0.26 \pm 0.04$ \\ \hline \hline
\end{tabular}
\label{ACP_2body}
\end{center}
\end{table}

From the theoretical point of view, branching fraction and direct \CP asymmetry estimates for charmless two-body \B decays are challenging.  For branching fractions, there is quite  reasonable agreement between calculations based on factorisation techniques and the experimental results, as can be found in Ref.~\cite{ChengChiangKuo2014}. Yet, for  \CP asymmetries the  consistency between calculations and measured values is currently not so good. Despite the theoretical developments and the experimental improvements over the last decade, the origin of the strong phase leading to \CP asymmetries is not clear yet~\cite{Beneke2015}.

One crucial example is the difference between the well-measured \CP asymmetry in the  $ \Bzb \to \Km\pip$  and  $ \Bp \to \Kp\piz$ channels.   From the short-distance point of view, these decay channels are equivalent --- the dominant quark-level diagrams coincide with the only difference being the  spectator quark, $\dquarkbar$ quark for \Bzb and $\uquarkbar$ quark for \Bp and the colour-suppressed tree contribution. As such,  in principle they would have similar behaviour with respect to \CP violation, in particular one should expect $\Delta A_\CP(\B\to \PK\pi) \equiv  A_{\CP}( \Bz \to \Km\pip) -A_{\CP}(\Bp \to \Kp\piz)\approx 0$. Instead, it is measured to be $\Delta A_{\CP}(\B\to \PK\pi)= -0.122 \pm 0.022$ ~\cite{PDG2019}. This result also seems to point towards a large influence of long-distance, final-state interaction effects, with hadronic re-scattering processes taking a leading role.                     
                       
                        Invoking the \CPT constraint, as highlighted by Wolfenstein \cite{Wolfenstein1990} and discussed in Sect.~\ref{chapter2_CP_CPT},  the significant negative \CP asymmetry observed in $ \Bz \to \Km\pip$ decays has to be compensated by other channels in the same flavour family.  If one also assumes that  a two-body final state   would preferably rescatter to other two-body final states \cite{Smith2003}, the most probable candidates to present a positive \CP asymmetry are the channels  $\Kz\piz$, $\Kz\eta$ and $\Kz\eta'$. The \Bz decay to these final states also has a penguin contribution plus a colour-suppressed tree component carrying the weak phase $\gamma$. For the decay $\Bp \to \Kp\piz$, the most probable rescattered two-body final states would be $\Kz\pip, \Kp\eta$ and $\Kp\eta'$. For these, however, the \Bp decay has contribution from the penguin diagram only, carrying no weak phase.  This means that, if \CP violation is observed in any of these final states, it can have its origin from rescattering effects, while preserving \CPT.

 It is not trivial, however, to estimate the impact of  rescattering effects in the aforementioned processes and consequently in the value of $\Delta A_{\CP}(\B \to K\pi)$.  But, in any case, it seems plausible that the presence of strong hadronic phases, appearing through final-state interactions and not taken into account in short-distance calculations, are key ingredients to understand the  discrepancies between experimental results and theory predictions --- although there are no available theory nor experimental results to support how important these FSI phases are, especially above 2 GeV.  The experimental measurement of \CP asymmetries for $ \Bz \to \Kz\piz$, $ \Bz \to \Kz\eta $, and  $ \Bz \to \Kz\eta'$ can shed  light on this puzzle.  In this respect, Belle II experiment~\cite{BelleII_PhysicsBook_2018} will be the main actor.

   \subsection{Three-body decays}
 \label{Bhhh}
 
 As discussed in Sect.~\ref{multibody}, decays of three or more particles in the final state offer a richer environment for  the study of \CP violation, since the phase space can be scanned and observables other than the total \CP asymmetry exist.

\CP violation was first observed in 2013 by the LHCb collaboration in local regions of the Dalitz plot for the charmless decays $\Bp\to \Kp\pip\pim$, $ \Bp\to \Kp\Kp\Km$, $\Bp\to \pip\pip\pim$ and  $\Bp\to \pip\Kp\Km$ ~\cite{LHCb-PAPER-2013-027,LHCb-PAPER-2013-051}. Integrated \CP asymmetries were then measured by the same collaboration in 2014,  using the total available sample of $3\invfb$ from run I. 
The studies have shown sizeable integrated \CP asymmetries but also confirmed rich structures of \CP asymmetries across the phase space~\cite{LHCb-PAPER-2014-044}. The results for the integrated \CP asymmetries were
\begin{eqnarray}
A_{\CP}(\Bp\to \Kp\pip\pim) & = +0.025 \pm 0.004 \pm 0.004 \pm 0.007,  \nonumber  \\
A_{\CP}( \Bp\to \Kp\Kp\Km) & = -0.036 \pm 0.004 \pm 0.002 \pm 0.007, \nonumber\\
A_{\CP}( \Bp\to \pip\pip\pim) & = +0.058 \pm 0.008 \pm 0.009 \pm 0.007, \nonumber\\
A_{\CP}( \Bp\to \pip\Kp\Km) & = -0.123 \pm 0.017 \pm 0.012 \pm 0.007, 
\label{B3h_ACP}
\end{eqnarray}
where the first uncertainties are statistical, the second systematic, and the third due to the limited knowledge of the \CP asymmetry in the $\Bp\to \Kp\jpsi$, used as a reference mode. The statistical significances of the \CP asymmetries were respectively 2.8,  4.3, 4.2, and 5.6$\sigma$.   The sample sizes were about 180 thousand decays for $\Bp\to \Kp\pip\pim $, 110 thousand for $\Bp\to \Kp\Kp\Km  $, 25 thousand for $ \Bp\to \pip\pip\pim $ and 6 thousand for $ \Bp\to \pip\Kp\Km $. 

\begin{figure}[!htb]
\centering
 \includegraphics[width=.4\columnwidth,angle=0]{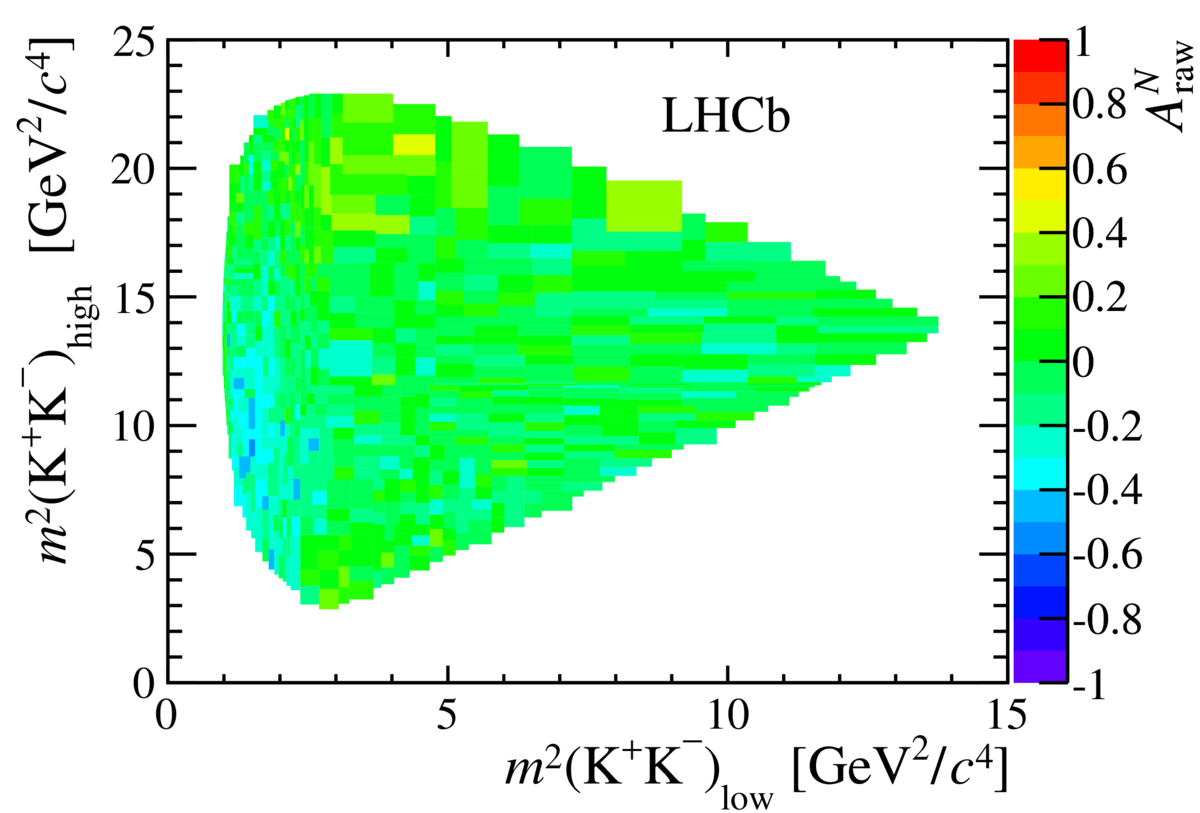} 
\includegraphics[width=.4\columnwidth,angle=0]{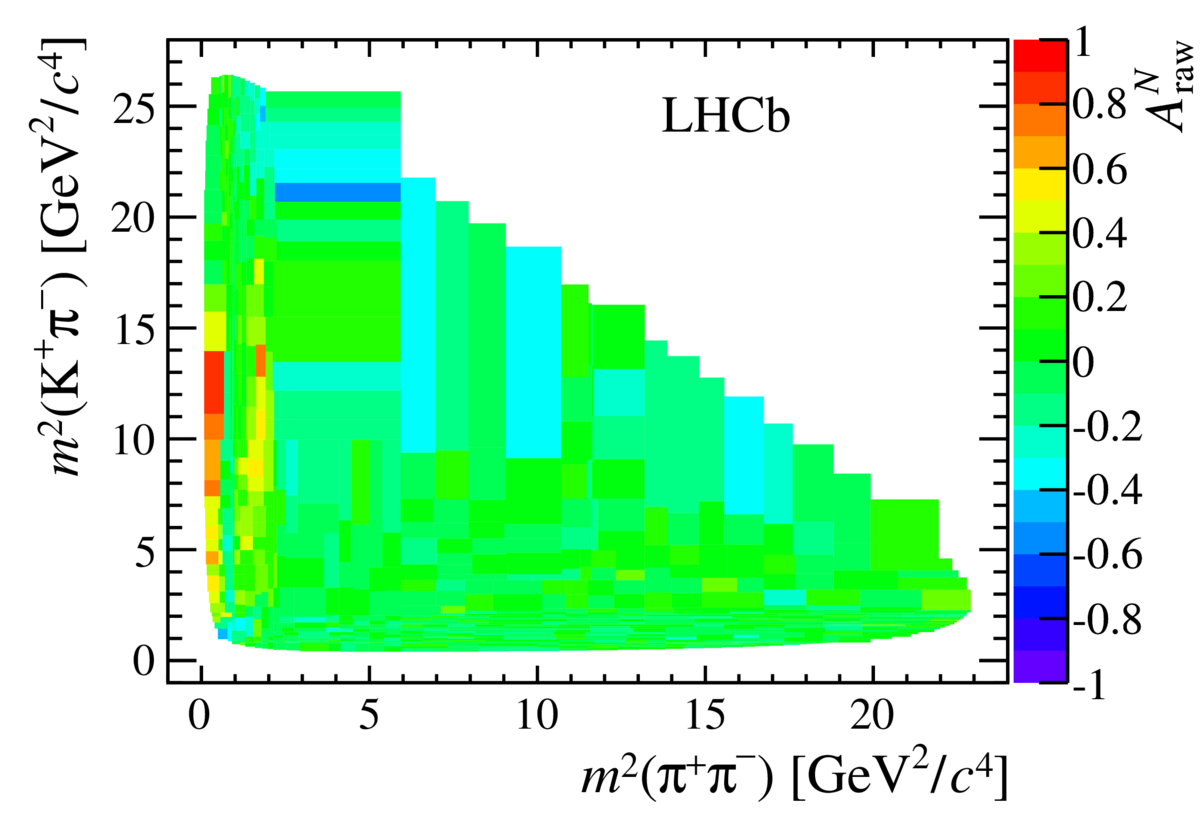}  \\
 \includegraphics[width=.4\columnwidth,angle=0]{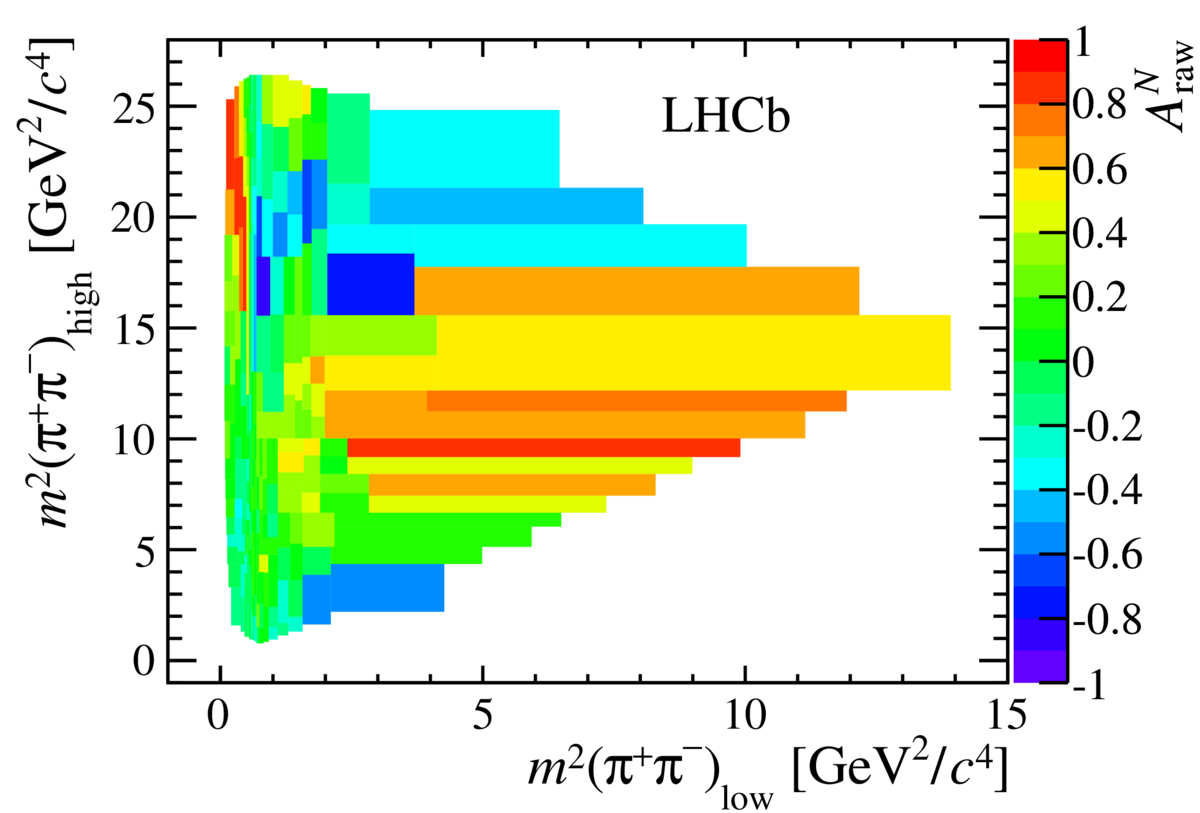} 
 \includegraphics[width=.4\columnwidth,angle=0]{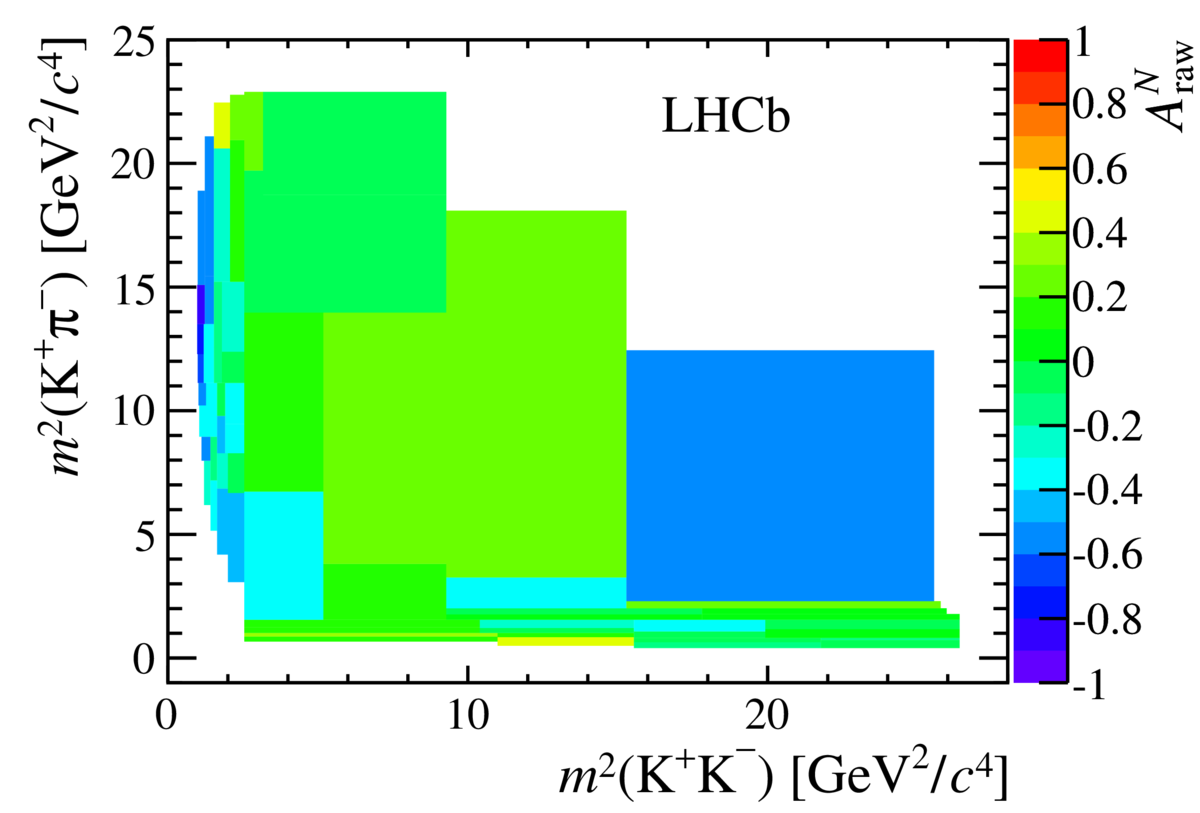} 
\caption{\CP asymmetry distributions with the Miranda technique, with  background-subtracted and acceptance-corrected events for: (top left) $\Bp\to\Kp\Kp\Km$, (top right) $\Bp\to\Kp\pip\pim$, (bottom left)  $\Bp\to\pip\pip\pim$ and (bottom right)   $\Bp\to\pip\Kp\Km$ \cite{LHCb-PAPER-2014-044}. }
\label{Mirandizing4channels}
\end{figure}

The \CP asymmetries across the phase space for the four channels were obtained through the second Miranda technique~\cite{bigi2}, as presented in Sect.~\ref{miranda}, and are reproduced here in Fig.~\ref{Mirandizing4channels}. Rich \CP violation patterns can be seen along the Dalitz plots, with positive (red) and negative (blue) \CP asymmetries coexisting in the same final state. Since the source of weak phase for all these charmless \B decays is the CKM angle $\gamma$, the \CP asymmetry variations over the Dalitz plot have to come from the variation of the strong phase. The Miranda distributions evidence the richness involved in the \CP violation dynamics of these decays.   A discussion on these features follows. 

\subsubsection{Interference between different angular-momentum waves}     
\label{CPfromcostheta}

\begin{figure}[!htb]
\centering
 \includegraphics[width=.4\columnwidth,angle=0]{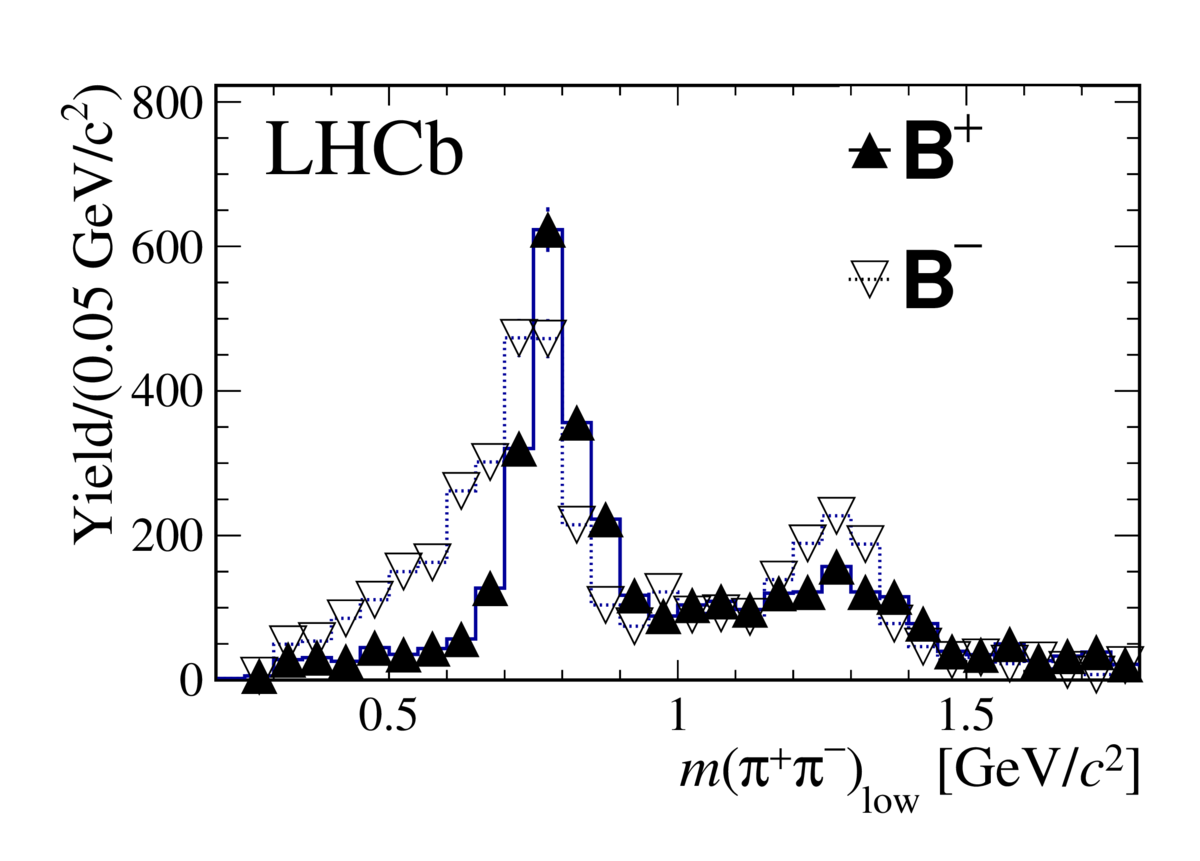} 
\includegraphics[width=.4\columnwidth,angle=0]{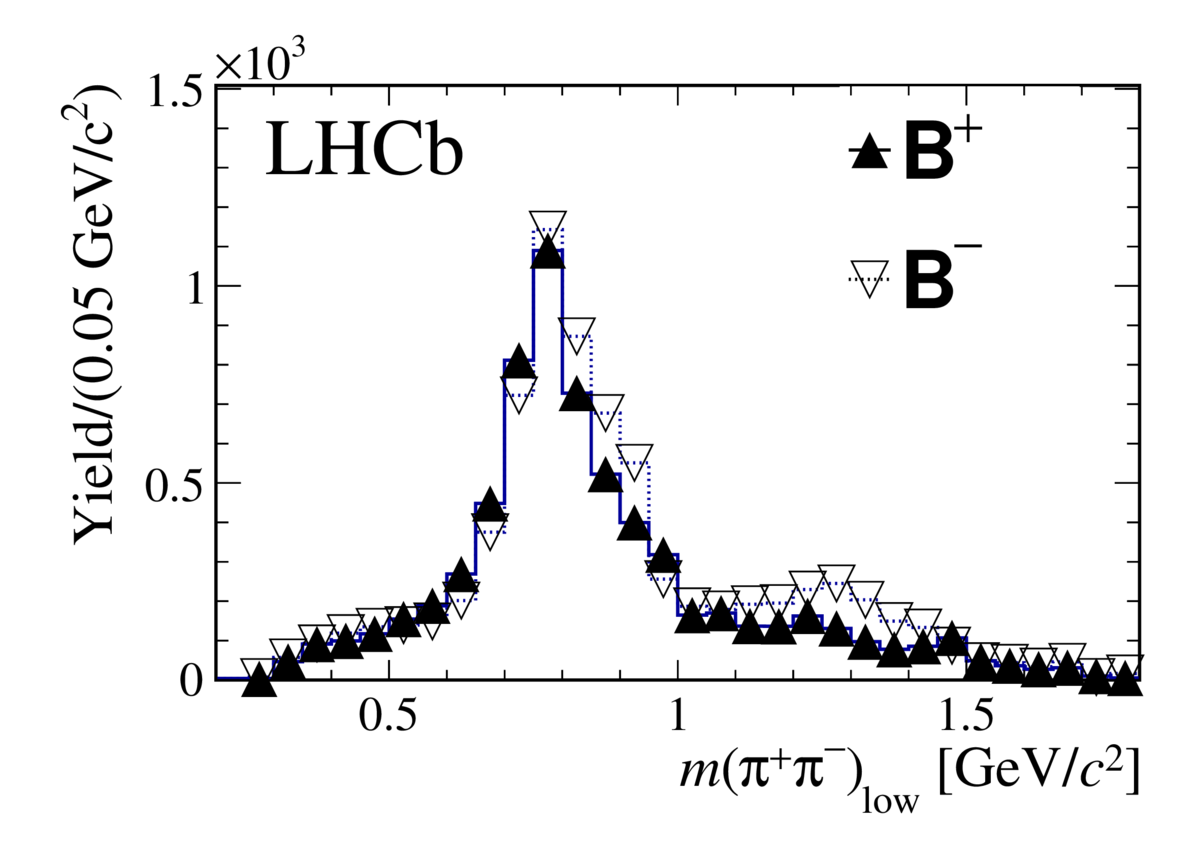}  \\
 \includegraphics[width=.4\columnwidth,angle=0]{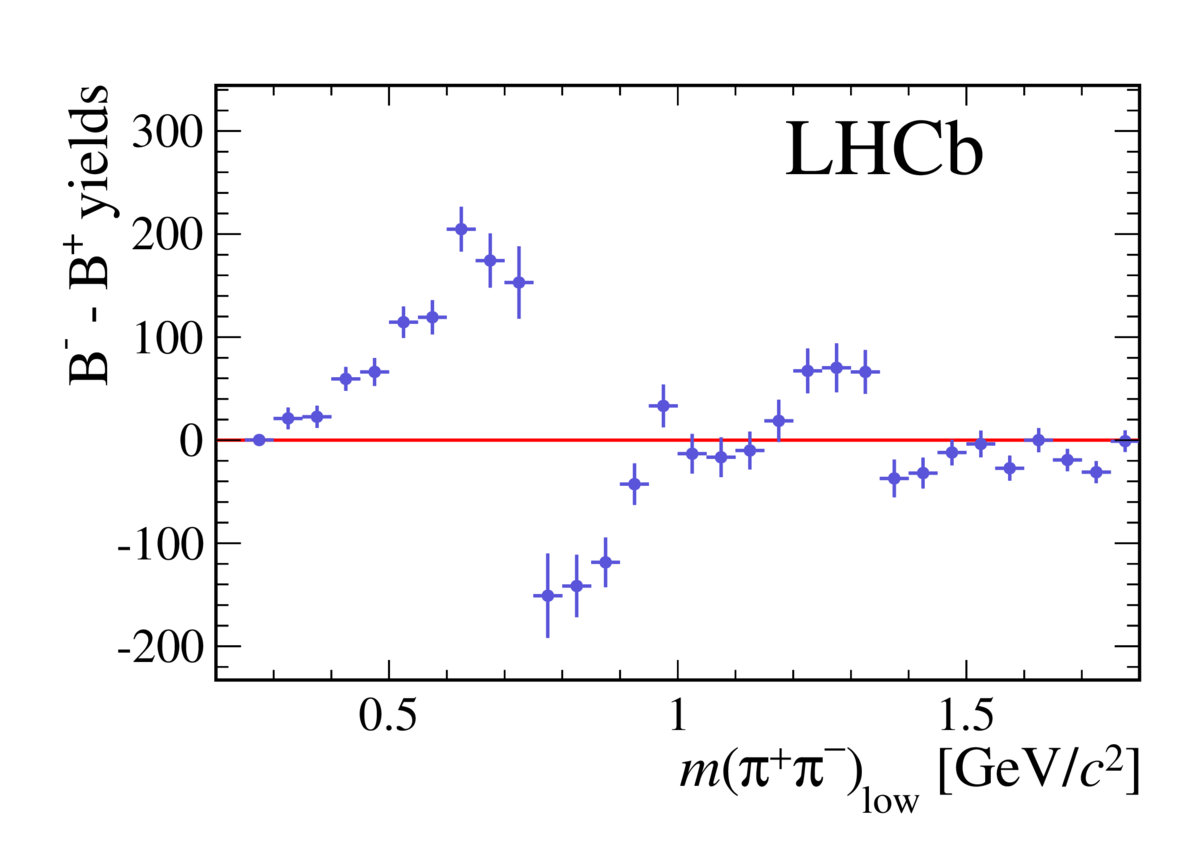} 
 \includegraphics[width=.4\columnwidth,angle=0]{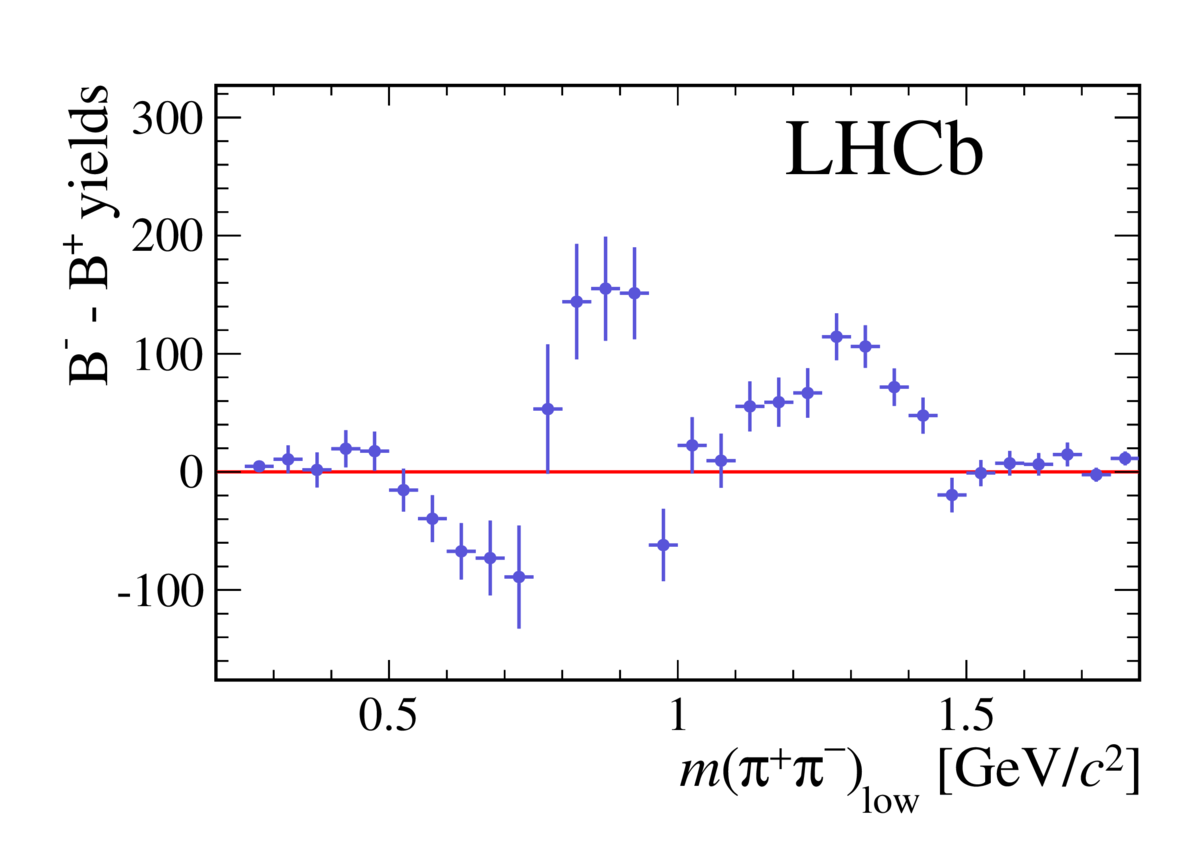} 
\caption{Projections of the $m(\pip \pim)_{\rm low}$  variable  for  $B^\pm \to \pi^\pm \pip\pim  $  signal events  separated by charge for (top left) $\cos \theta < 0 $ and (top right)  $\cos \theta > 0$. The respective subtracted  \Bp and \Bm distributions are shown in the bottom plots.\ \cite{LHCb-PAPER-2014-044}.}
\label{SP_CPdistribution}
\end{figure}

By inspecting the low $\pip\pim$ invariant-mass region in both $\Bp\to\Kp\pip\pim$ and  $\Bp\to\pip\pip\pim$ Dalitz plots, the results from the LHCb experiment \cite{LHCb-PAPER-2014-044} show a peculiar \CP asymmetry distribution, with  positive and negative $A_{\CP}$ around the $\rho(770)^0$ mass. As discussed in Sect.~\ref{AmAn} (see Eq.~\ref{AcpAmplitude}), this is a signature from  the interference term between scalar and vector amplitudes, proportional to $\cos \theta $. This can be better seen in Fig.~\ref{SP_CPdistribution}, where the  $\pip\pim$ invariant-mass combination\footnote{In $\Bp\to\pip\pip\pim$, with two identical pions, $m(\pip \pim)_{\rm low\,(high)}$  is the lower (higher) $\pip\pim$ mass among the two possible combinations. The same subscript $\rm low\, (high)$ is used for other decays with identical particles.} for $\cos\theta<0$ and $\cos\theta>0$ are shown for both $\Bp\to\pip\pip\pim$ and $\Bm\to\pim\pim\pip$ decays --- the subtraction of the histograms is also shown, evidencing the charge distribution difference. 
Besides, the observed zeros in the bottom plots in Fig.\ref{SP_CPdistribution} around the $\rho(770)^0$ nominal mass indicate a dominance of the real part of the Breit--Wigner in this process. 
       
The LHCb collaboration very recently has extended the study by performing a full amplitude analysis of the decay $\Bp\to\pip\pip\pim$ to understand the features being observed in data \cite{LHCb-PAPER-2019-017,LHCb-PAPER-2019-018}. Three different models for the scalar amplitude were used: the isobar model with $ \sigma(500)$ pole and a $ \pi\pi-\PK\PK$ rescattering amplitude, the K-matrix formalism, and the QMI-PWA method. Resonant amplitudes for spin-1 and spin-2 were included through the isobar model for the three approaches. The general agreement between these three approaches was good. The dominant amplitudes for this decay were found to be the $\rho(770)^0\pip$ amplitude, with about 50\% fit fraction, followed by the S-wave amplitude with about 25\% fit fraction.            
           Within the isobar model, the phase difference between the charged conjugated decays $\Bm \to \sigma(500)\pim $ and $\Bp \to\sigma(500)\pip $, measured with respect to that of the $\rho(770)^0 \pi^\pm$ components, was measured to be about  $60^\circ$ \cite{LHCb-PAPER-2019-017}, confirming the interference trend  presented in Fig.~\ref{SP_CPdistribution},  with zeros at  the $\rho(770)^0$ nominal mass. The conservation of \CP symmetry  in the interference between S- and P-wave was excluded with a significance of $25\sigma$.

\begin{figure}[!htb]
\centering
 \includegraphics[width=.4\columnwidth,angle=0]{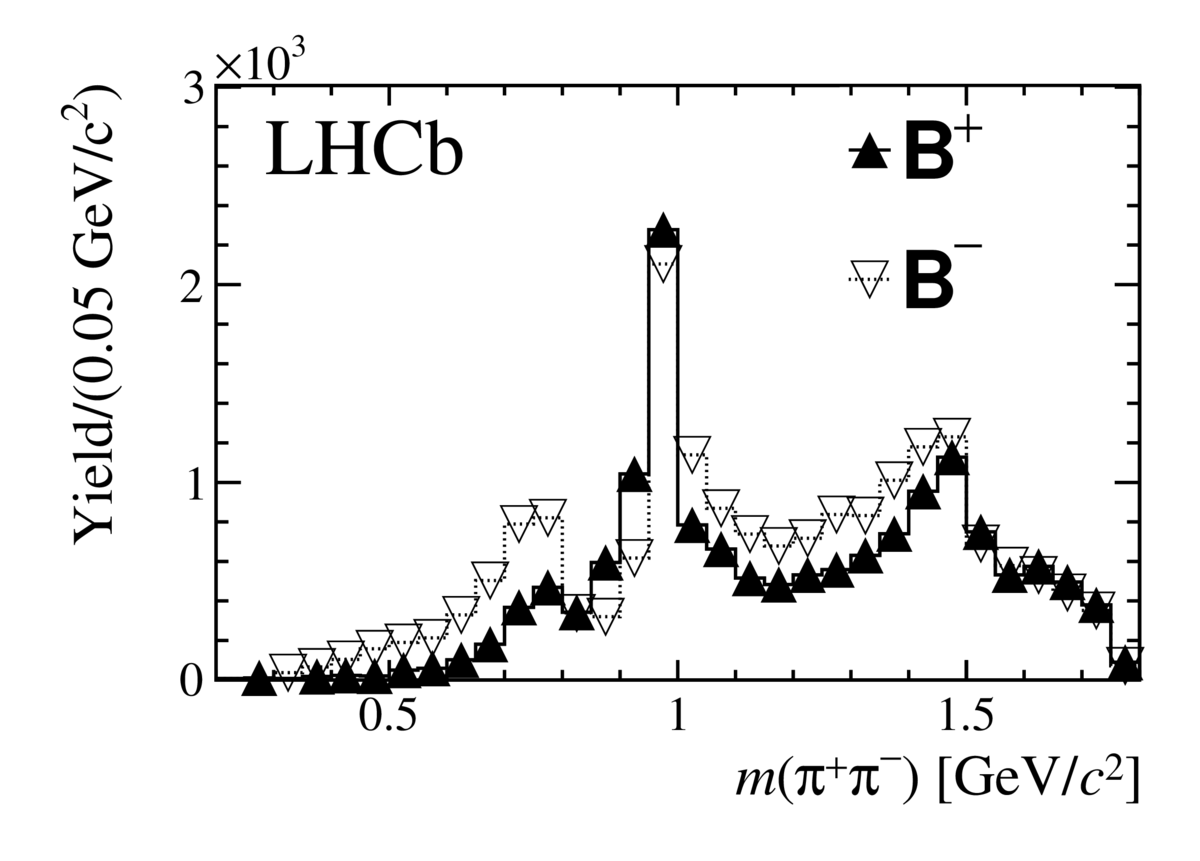} 
\includegraphics[width=.4\columnwidth,angle=0]{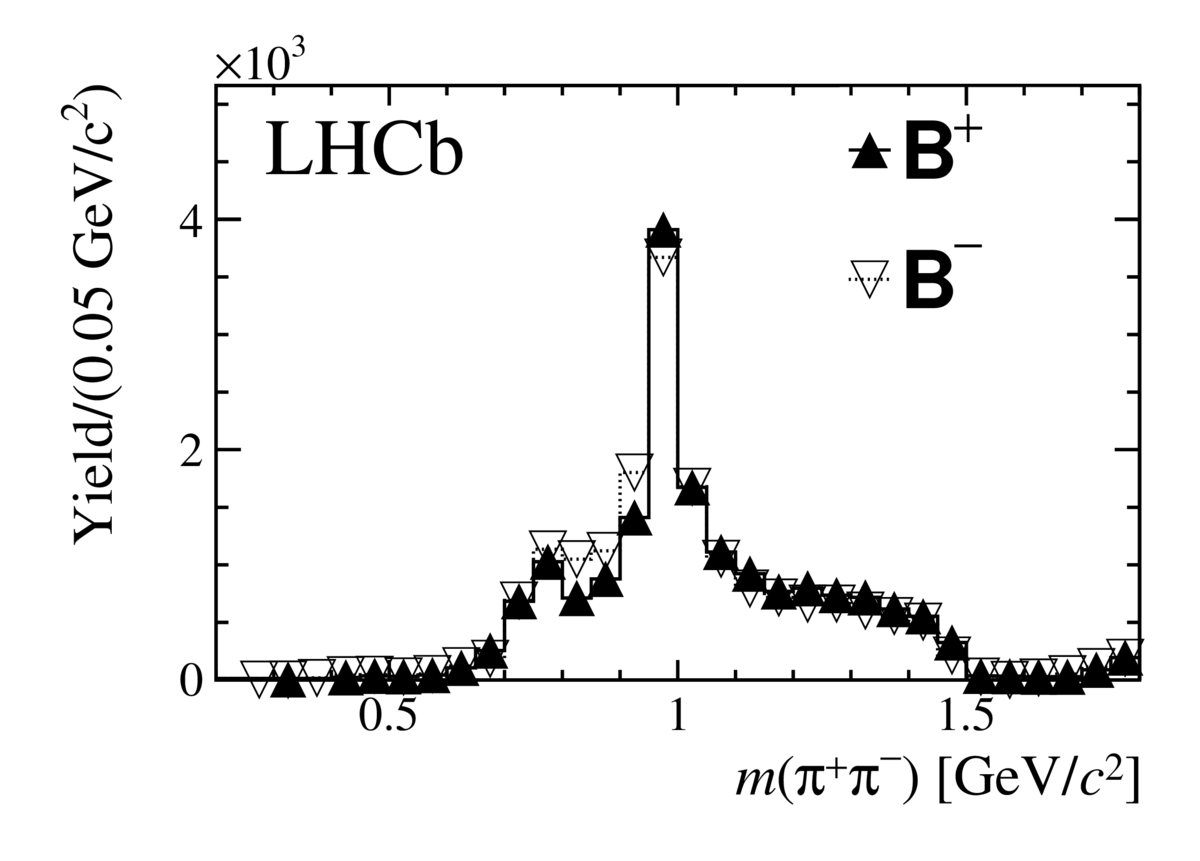}  \\
 \includegraphics[width=.4\columnwidth,angle=0]{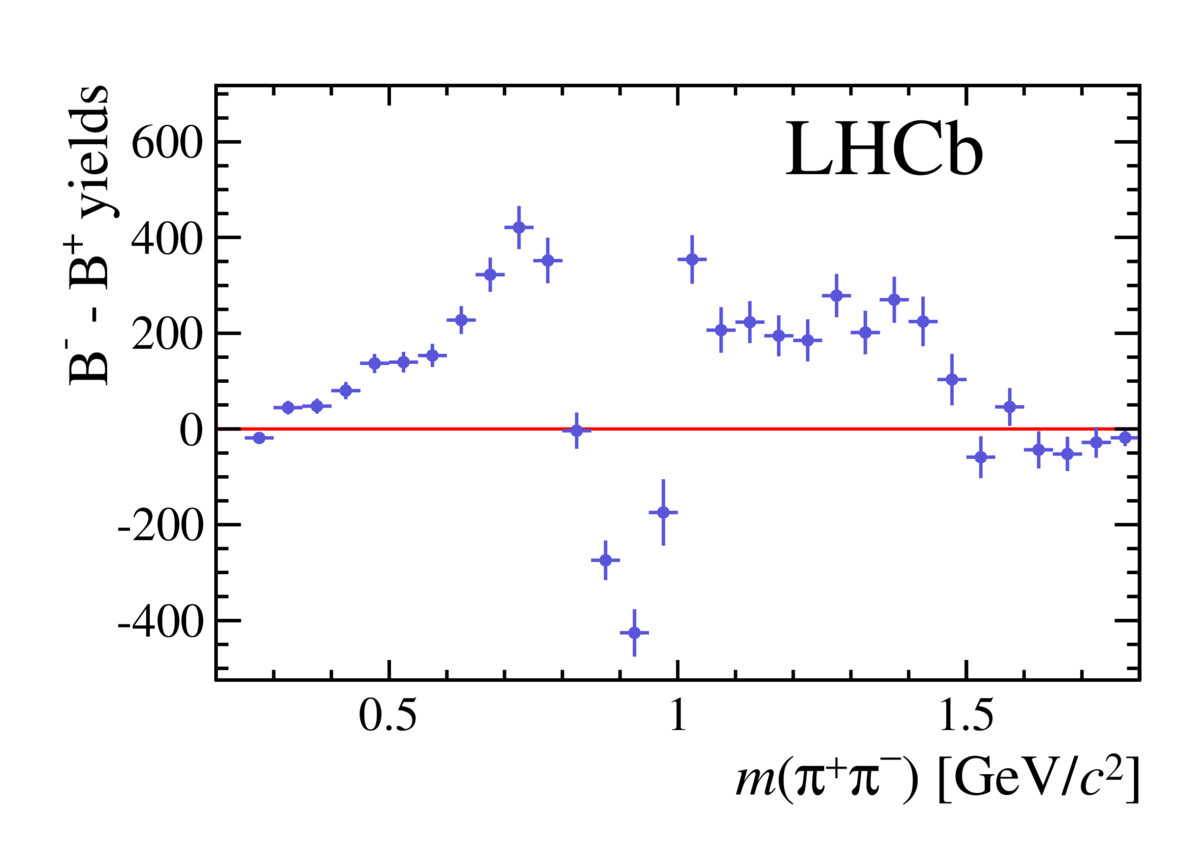} 
 \includegraphics[width=.4\columnwidth,angle=0]{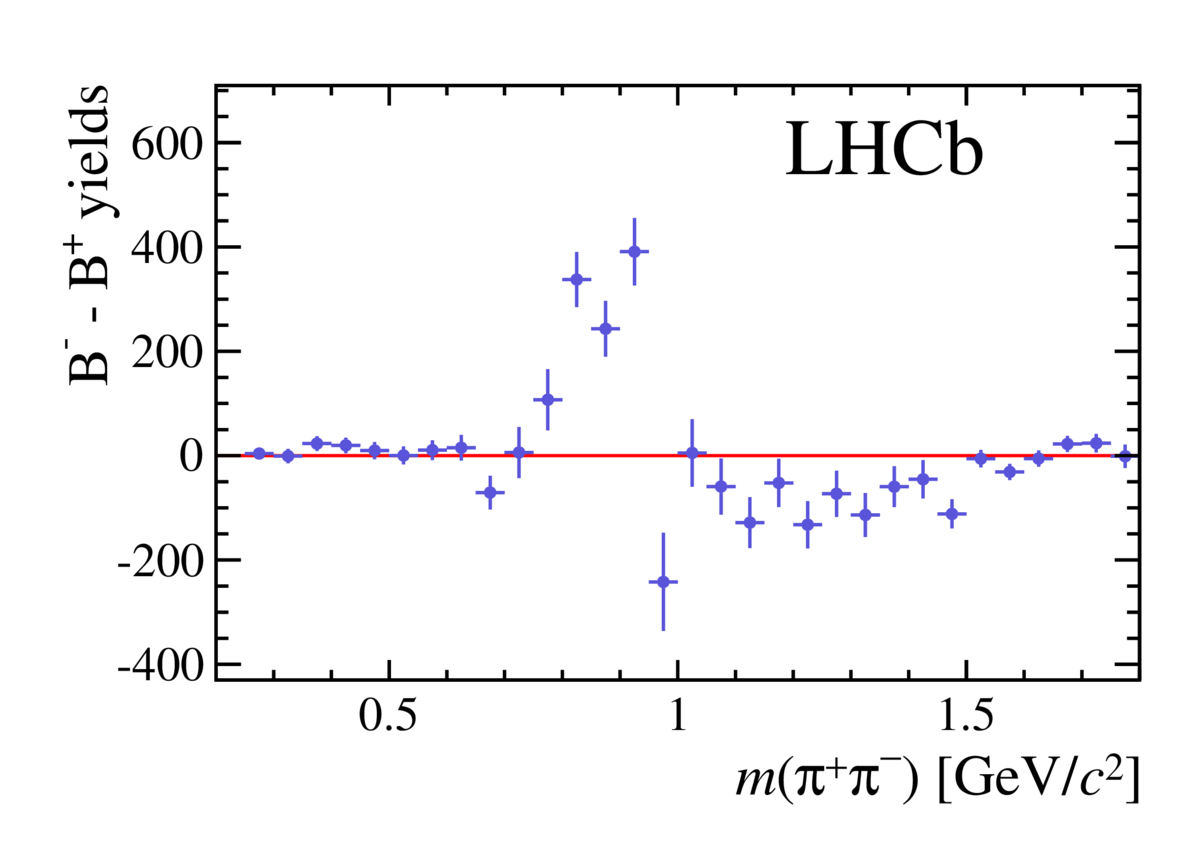} 
\caption{Projections of the $m(\pip \pim)$  variable  for  $B^\pm \to K^\pm \pip\pim  $ signal events  separated by charge for (top left) $\cos \theta < 0 $ and (top right)  $\cos \theta > 0$. The respective subtracted  \Bp and \Bm distributions are shown in the bottom plots \cite{LHCb-PAPER-2014-044}.}               
\label{SP_CPinterferenceKpipi}
\end{figure}

        A similar result for the interference distribution was observed in  the $\pim \pip$ invariant mass for the   \mbox{$ \Bp\to \Kp \pip\pim$} decay \cite{LHCb-PAPER-2014-044} as can be seen in Fig.~\ref{SP_CPinterferenceKpipi}. However, there are  important differences between these two decays. In $\Bp\to\pip\pip\pim$ decay, the $\rho(770)^0$ amplitude is dominant, with a smaller  scalar amplitude \cite{LHCb-PAPER-2019-017}. For the $ \Bp\to \Kp \pip\pim$  decay, instead,  the scalar resonance $f_0(980)$ is dominant, with a low  $\rho(770)^0$ amplitude. In the bottom right plot in Fig.~\ref{SP_CPinterferenceKpipi}, the presence of two zeros is clearly seen, one around the $\rho(770)^0$ mass and another near the  scalar resonance $f_0(980)$.  Another remarkable difference is the apparent absence of \CP asymmetry below the $\rho(770)^0$ mass for $\cos \theta > 0$ (see bottom right plot in Fig.~\ref{SP_CPinterferenceKpipi}), as observed for the equivalent region in  $\Bp\to\pip\pip\pim$ decay.  There is yet no amplitude analysis with this data set from LHCb.  BaBar \cite{BaBar_bkpipi} and Belle   \cite{Belle_bkpipi} collaborations   presented  results for the $ \Bp \to \Kp \pim \pip$  with large statistical uncertainties in the phase for both  $\Kp\rho(770)$    and $\Kp f_0(980) $ amplitudes. In any case, larger data samples coming from LHCb run II may shed light on these differences. 
                
\subsubsection{$\pim \pip  \to \Km \Kp$ re-scattering}
\label{kkpipiresc}

\begin{figure}[htb]
\centering
 \includegraphics[width=.4\columnwidth,angle=0]{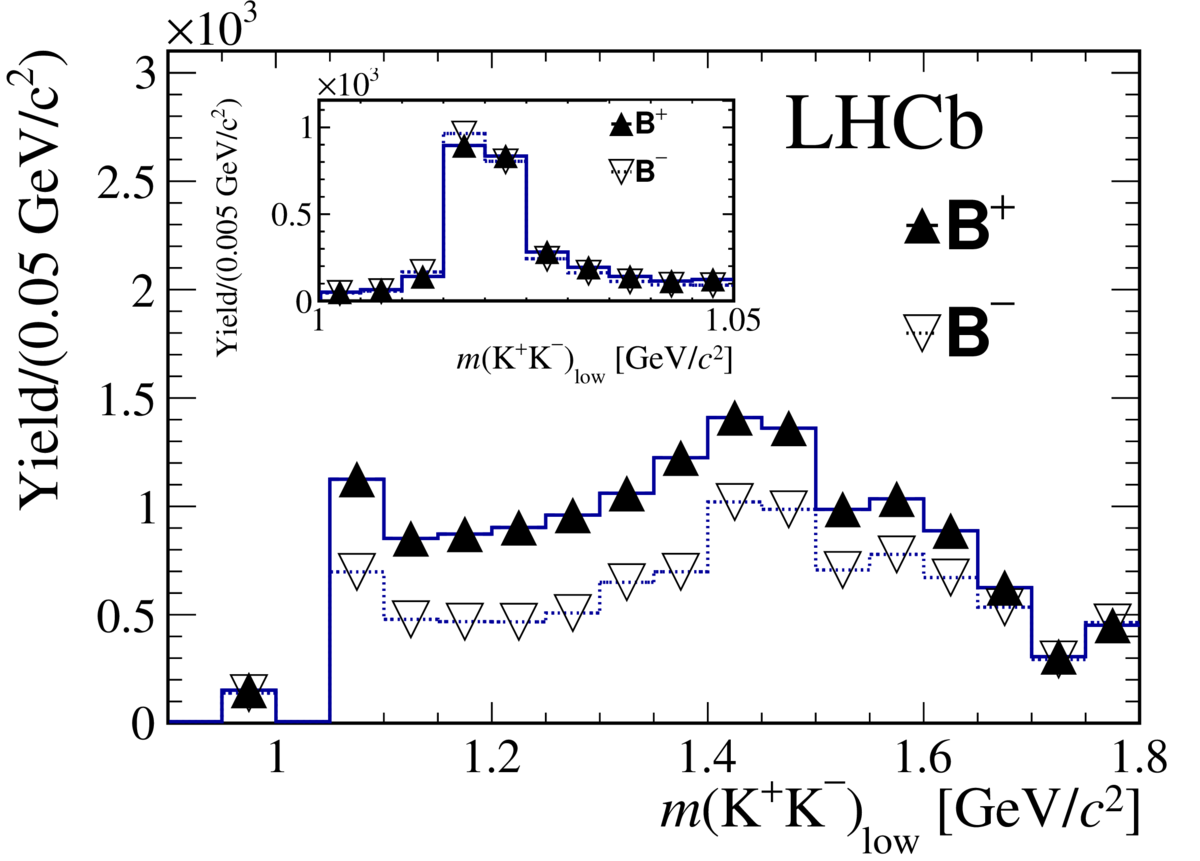} 
\includegraphics[width=.4\columnwidth,angle=0]{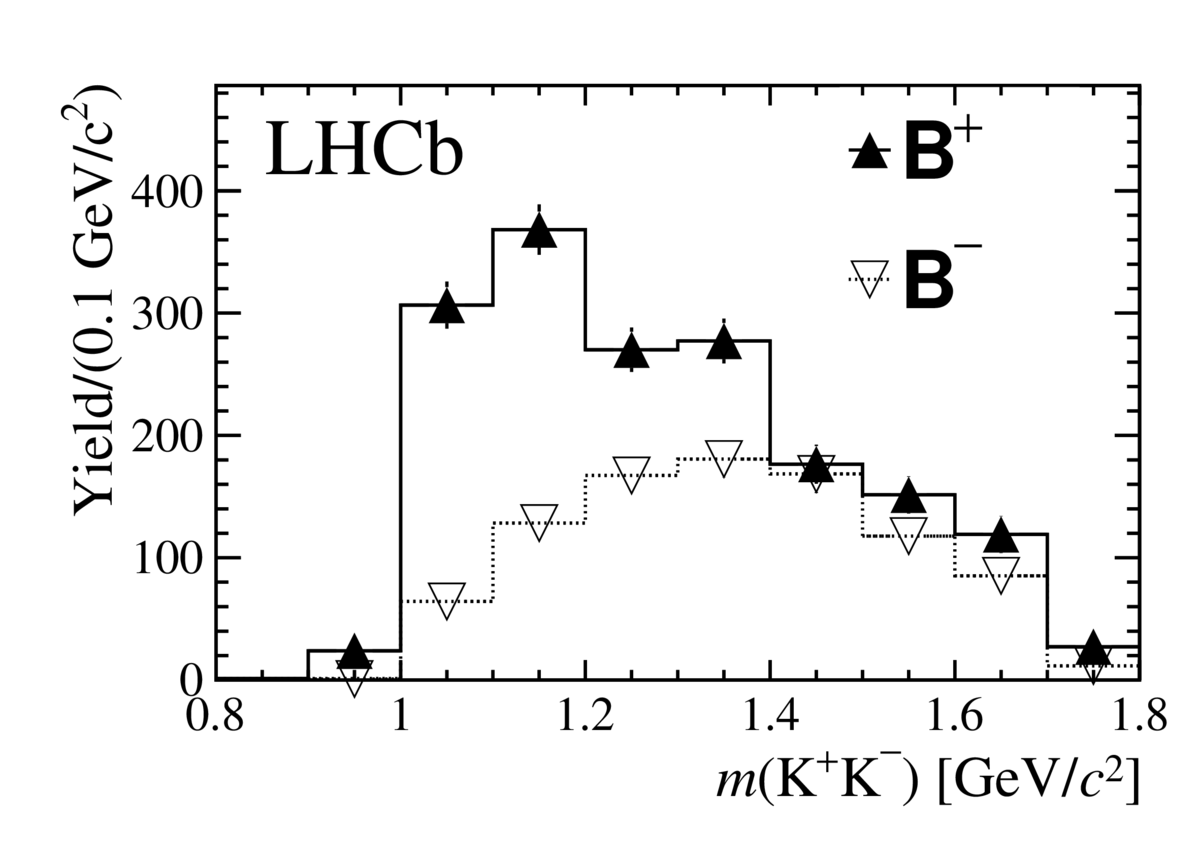}  \\
\caption{$\Kp\Km$ invariant mass distributions for (left) $\B^\pm\to\PK^\pm\Kp\Km$ and (right) $\B^\pm\to\pi^\pm\Kp\Km$ decays, separated by charge. The inset in the left plot magnifies the $\phi(1020)$ mass region \cite{LHCb-PAPER-2014-044}.}
\label{B-KK}
\end{figure}

       A very interesting result in charmless $\Bp$ decays obtained from LHCb   \cite{LHCb-PAPER-2013-027,LHCb-PAPER-2013-051,LHCb-PAPER-2014-044} was the pattern of \CP asymmetries observed in the $\pim \pip $  and $ \Km \Kp$ mass regions between 1.0 and 1.5 \gev for the flavour coupled-channels $\Bp\to \pip\Kp\Km$ and $\Bp\to \pip\pip\pim$,  and also  $\Bp\to \Kp\Kp\Km$ and $\Bp\to \Kp\pip\pim$. Figure~\ref{B-KK} shows the distributions for $\Bp\to \Kp\Kp\Km$  and $\Bp\to \pip\Kp\Km$, separated by charge.

       In Table~\ref{ACP_rescattering}, the resulting \CP asymmetries for the $\pip\pim$ and $\Kp\Km$ invariant-mass  region between $1.0$ and $1.5 \gev$ are shown. The values are larger than those integrated in the phase space (Eq.~\ref{B3h_ACP}) and evidence the change in sign among the pairs of final states with same flavour content. 
\begingroup
\begin{table}[httb] 
\caption{Charge asymmetries in the rescattering $\pim \pip  \to \Km \Kp$ region --- between $1.0$ and $1.5 \gev$. Uncertainties are first statistical, second systematic, and the third is due to the \CP asymmetry of the  $\Bp\to \jpsi \Kp$  reference mode\cite{LHCb-PAPER-2014-044}.}
\begin{center} 
\begin{tabular}{lc}
\hline 
 Decay & $A_{\CP}$ \\ 
\hline 
$ \Bp\to \Kp\pip\pim$  &  $+0.123 \pm 0.012 \pm 0.017 \pm 0.006$ \\ 
$ \Bp\to \Kp\Kp\Km$  &   $-0.209 \pm 0.011 \pm 0.004 \pm 0.006$ \\ 
$ \Bp\to \pip\pip\pim$ & $+0.173 \pm 0.021 \pm 0.015 \pm 0.006$ \\ 
$ \Bp\to \pip\Kp\Km$  & $-0.326 \pm 0.028 \pm 0.029 \pm 0.006$ \\ 
\hline 
\end{tabular} \end{center} 
\label{ACP_rescattering}
\end{table} 
\endgroup

\begin{figure}[htb]
 \centering
\includegraphics[width=.35\columnwidth,angle=0]{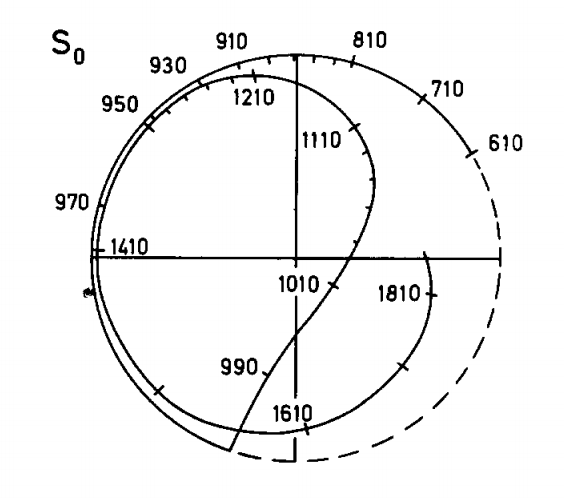}
\includegraphics[width=.43\columnwidth,angle=0]{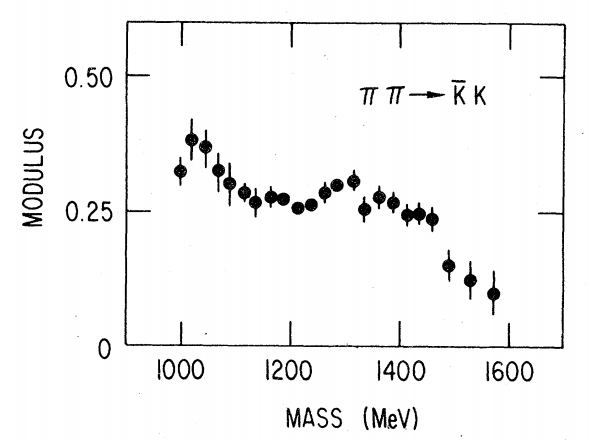}
\caption{  (left) Argand circle of the scalar $\pim \pip$ elastic scattering  amplitude \cite{CERN-Munich}. (right) Modulos of the scalar $\pi \pi  \to \PK\PK$ amplitude  distribution \cite{Cohen1980}.}
\label{CERN-Munich}
\end{figure}

A possible explanation for this effect is $\pi\pi-KK$ rescattering.  A well-known  $\pim \pip $   elastic-scattering data  result \cite{CERN-Munich} shows that the Argand plot for the scalar amplitude departs from the unitary circle at about $1\gev$ and returns at around $1.5\gev$, as shown in Fig.~\ref{CERN-Munich}(a). Also in the region $1.0-1.5\gev$ a bump is observed in the inelastic interaction  $\pim \pip  \to \Km \Kp$, as shown in  Fig.~\ref{CERN-Munich}(b) \cite{Cohen1980}. 
In other words, $\pi\pi-KK$ rescattering may be at least partially responsible for the opposite \CP asymmetries found above, and would be in line with the Wolfenstein mechanism for \CP violation \cite{Wolfenstein1990} through coupled channels with the same quantum number \cite{Bediaga2013,Nogueira2015}. 
    
\begin{figure}[htb]
\centering
\includegraphics[width=.65\columnwidth,angle=0]{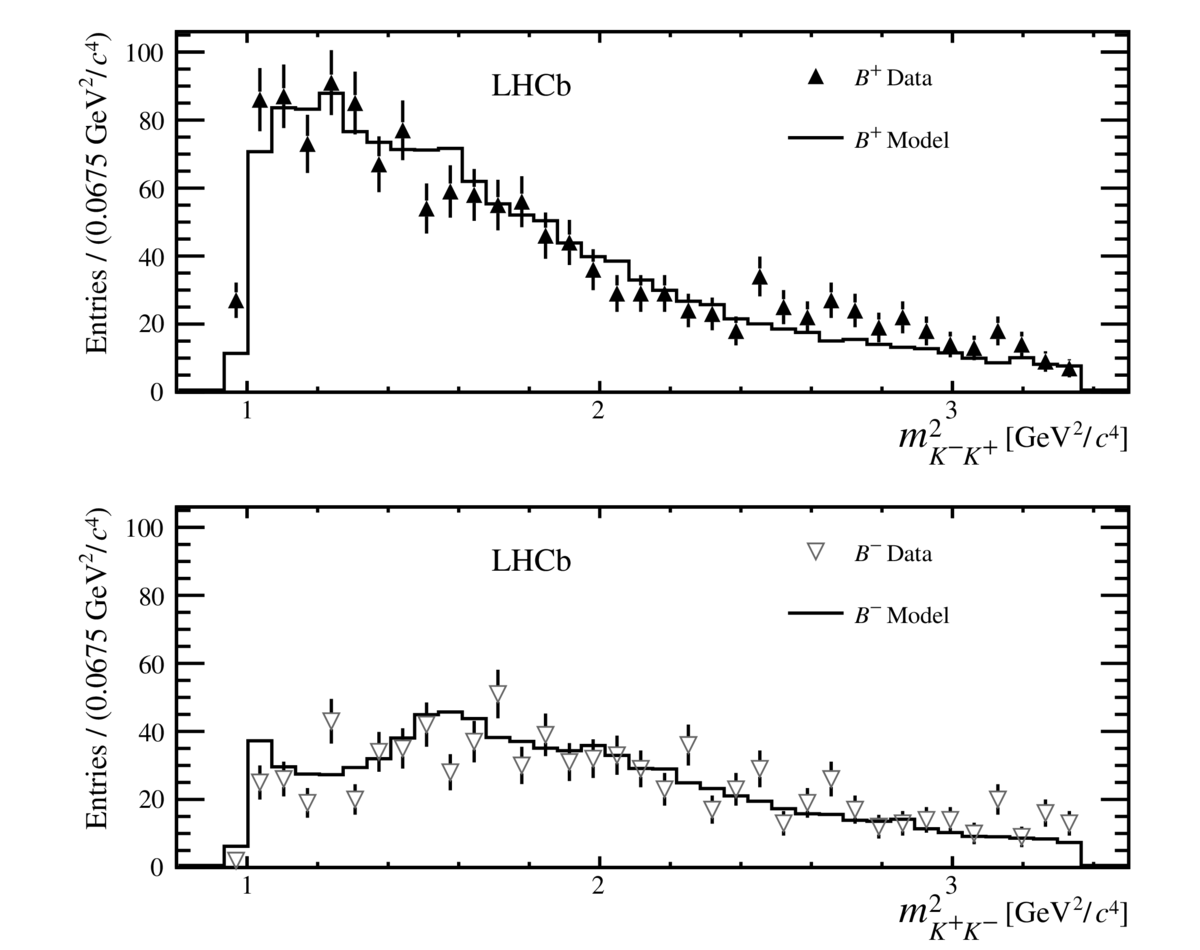}
\caption{$\Kp \Km$ squared mass distribution for (top) $\Bp\to \pip \Kp \Km$  and (bottom) $\Bm\to\pim \Km \Kp$ decays.  The total amplitude fit model, including a term for the $\pi\pi-KK$ rescattering, is overlaid \cite{LHCb-PAPER-2018-051}.}
\label{KKPI_AmplitudeAnalysis}
\end{figure}

Recently, the LHCb collaboration published the first amplitude analysis for the decay $\Bp\to \pip\Kp \Km$ with run I data \cite{LHCb-PAPER-2018-051}. Apart from a dominant non-resonant component and a few resonant contributions, an explicit amplitude parameterisation for the $\pi\pi-KK$ rescattering has been included, following Ref.~\cite{Pelaez2004}. The fit model shows good description of the data, as can be seen in Fig.~\ref{KKPI_AmplitudeAnalysis}. The $\B^\pm$ average fit fraction for the rescattering contribution  is found to be about 23$\%$  while presenting the largest \CP asymmetry observed for a single amplitude, with a measured value of  $A_{\CP} =( -66 \pm 4\pm 2)$. 

          The same amplitude is also included in the amplitude analysis for the  $\Bp\to \pip\pip\pim$ decay \cite{LHCb-PAPER-2019-017,LHCb-PAPER-2019-018}. The  total contribution for this channel is much lower than that obtained in $\Bp\to  \pip \Kp \Km$. However, when compared to the  $\Kp \Km$  mass spectrum in the rescattering region, the $\pip\pim$ mass spectrum has different resonant contributions, including part of the broad and scalar $\sigma(500)$ and the tensor $f_2(1270)$, both with significant important positive \CP asymmetries, as outlined in the next sub-section. It would be important to perform a coupled-channel amplitude analysis -- in particular a simultaneous fit for $\Bp\to \pip\pip\pim$ and $\Bp\to  \pip \Kp \Km$ decays -- to clarify better a possible correlation between $\pi\pi$ and $\PK\PK$ amplitudes through rescattering as a way to understand the \CP violation mechanisms involved.

\subsubsection{Quasi-two-body decays}
\label{Bquasi2body}

As discussed in Sects.~\ref{AmAn} and \ref{quasi2body}, the dynamics of a three-body decay are usually dominated by the formation of resonances. For a \B-meson decay these are quasi-two-body processes with a scalar, vector or tensor resonance  plus a pseudo-scalar. Although the resonance formation is an intermediate {\it amplitude} path -- and as such one cannot strictly measure a decay rate or a charge asymmetry associated to it -- it is still useful to define such quantities in terms of its squared amplitude. A \CP asymmetry can be defined as presented in Eq.~\ref{ACP_r}, exemplified for a  $\rho(770)^0$ intermediate resonance. 

A few values of \CP asymmetries for  quasi-two body modes are presented in Table~\ref{ACP_quasi2body}, as quoted by the PDG~\cite{PDG2019}. 

\begin{table}[htbp]
\centering
\caption{\CP asymmetries for a few quasi two-body $B$ decays~\cite{PDG2019}.}
\begin{tabular}{lc} \hline
Mode & $A_\CP$  \\ \hline
$\Bp\to \Kp f_2(1270)$ & $ -0.68 \pm 0.19$ \\
$\Bp\to \Kp\rho(770)^0$ & $\phantom{-}0.37 \pm 0.10$ \\ 
$\Bp\to \Kstarp \eta$ & $ \phantom{-}0.19 \pm 0.05$ \\ \hline
\end{tabular}
\label{ACP_quasi2body}
\end{table}

Recent amplitude analyses performed by the  LHCb collaboration using run I data   were able to further  determine the following \CP asymmetries associated to resonant intermediate decays:

\begin{enumerate}
\item $A_{\CP}(  \Bz \to \Kstar(892)^+ \pim)= -0.30 \pm 0.06$ in $\Bz\to \KS\pip\pim$ decays\cite{LHCb-PAPER-2017-033};
\item $A_{\CP}(\Bp \to \sigma(500)\pip)= -0.16 \pm 0.02 \pm 0.03$   in $\Bp\to\pip\pip\pim$ decays \cite{LHCb-PAPER-2019-017};
\item $A_{\CP}( \Bp \to f_2(1270)\pip)= -0.47 \pm 0.06 \pm 0.6$  in $\Bp\to\pip\pip\pim$ decays  \cite{LHCb-PAPER-2019-017}.
\item $A_{\CP}^0(  \Bz \to \rho(770)^0\Kstar(892)^0) = -0.62\pm 0.09 \pm 0.09$, for the longitudinal amplitude component in $\Bz\to (\pip\pim)(\Kp\Km)$ decays \cite{LHCb-PAPER-2018-042}.
\end{enumerate}

Although with the mentioned caveat on the interpretation of \CP asymmetries in quasi two-body decays, it is very interesting to note that the values of asymmetries here are considerably larger than those of two-body \B decays presented in Table~\ref{ACP_2body} (not considering those with large uncertainties). This is probably related to the richness of the hadronic processes in a multi-body decay, where different sources of strong phases, necessary to observe \CP violation, are possible. The \CPT  constraint has to enter somehow connecting final states with the same flavour content: resonant states usually have more than one decay channel, so a natural interconnection of \CP asymmetries may arise. Moreover, possible rescattering involving the companion pseudoscalar, and as such breaking the $2\!+1\!$ approximation,   can be of some importance, at least in a part of the phase space.  

Interestingly enough, the quasi-two-body decay $\Bp\to \rho(770)^0\pip$, which is the dominant resonant contribution of the decay $\Bp\to\pip\pip\pim$, has a \CP asymmetry compatible with zero: within the isobar model in Ref.~\cite{LHCb-PAPER-2019-017} it was measured to be $A_{\CP}= (0.7 \pm 1.1 \pm 2.2) \%$. This value contradicts some theoretical predictions with a high value for this quantity \cite{ChengChiangKuo2014,Zhou2017}.\footnote{A very recent work though, using QCD factorisation, presents a result consistent with no \CP violation in this intermediate mode ~\cite{Cheng:2020hyj}.} Yet, the $\rho(770)^0$ resonance has only one possible decay,   $\rho(770)^0 \to \pim \pip$,  since it cannot decay to two neutral pions.  So, if one takes the $2$+$1$ approximation at its limit of validity, the \CP asymmetry should indeed be zero, since there is no possible re-scattering to interconnect to another channel to produce an opposite \CP asymmetry \cite{Josue}. 

In this sense, within the $2$+$1$ approximation, the LHCb observation of a large negative \CP asymmetry  \cite{LHCb-PAPER-2017-033} for the   $  \Bz \to \Kstar(892)^+ \pim$ decay, with $\Kstar(892)^+\to \Kz \pip $,  should be compensated by a similarly large positive \CP asymmetry in the  channel $  \Bz \to \Kstar(892)^+ \pim$ with  $ \Kstar(892)^+ \to \Kp \piz $. There is yet no analysis of the final state $\Bz\to \Kp\pim\piz$ where this \CP asymmetry could be measured. With two neutral pions in the final state, this study is challenging for LHCb but less so for Belle II.

Two- and quasi two-body modes have been given quite some attention from the theory side.  In general, estimates for  \CP asymmetries involve final states with  two pseudoscalars ($\B\to PP$) or one pseudoscalar and one low-mass vector resonance ($\B\to PV$). Yet and again recalling the caveat on quasi-two-body decays, to measure the \CP asymmetry  in a $\B\to PV$ process  it is necessary to manage  the complexity of three-body decay, with interferences of different resonant amplitudes, with different spins,  across the phase space.

In certain situations, to avoid much of this complexity, a model-independent analysis method \cite{Josue} has been proposed, taking into account the angular momentum distribution for  $\B\to PV$, the relatively low invariant mass (assuming $V$ to be a resonance below $\sim 2 \gev$) where this amplitude is dominant and the proximity with a possible light scalar resonance. The method takes a slice around the nominal  mass of the light vector resonance including the region of interference with the  light scalar resonance. The potential of this method was tested successfully with fast Monte Carlo~\cite{Josue}.

\subsubsection{\B decays involving baryons} 

With a mass exceeding 5 \gev, \B mesons can  also decay to final states involving a nucleon and an antinucleon. Three-body processes such as $\B\to h\proton\antiproton $ (where $h$ is a light, stable meson) can proceed through intermediate resonances such as $\cquark\cquarkbar$ (charmonium) states and excited $\Lambdaresbar$ states and may also provide further insights into the \CP violation mechanisms. At the quark level the weak process, with both tree and penguin diagrams,  is essentially the same as those of the charmless \B decays discussed in the previous sections, and in particular the origin of the weak phase is the same. Diagrams involving charmonium states, however,  bring no associated weak phase. With $\proton\antiproton$ in the final state, there are interesting new features in the decay dynamics,  associated with the spin-1/2 particles, and potentially different hadronic interactions involving the production of the baryon pair. These features can be important to understand the role of FSI in the \CP asymmetry in non-leptonic B decays.  

The first evidence for \CP violation in this kind of decays was obtained by the LHCb collaboration in 2014 \cite{LHCb-PAPER-2014-034} in the decay mode $\Bp\to \Kp\proton\antiproton$, in a study which also included the Cabibbo-suppressed decay $\Bp\to \pip\proton\antiproton$. The sample comprised about 19 thousand decays of $\Bp\to \Kp\proton\antiproton$ and about 2 thousand decays of $\Bp\to \pip\proton\antiproton$.

\begin{figure}[htbp]
\centering
\includegraphics[width=.45\columnwidth,angle=0]{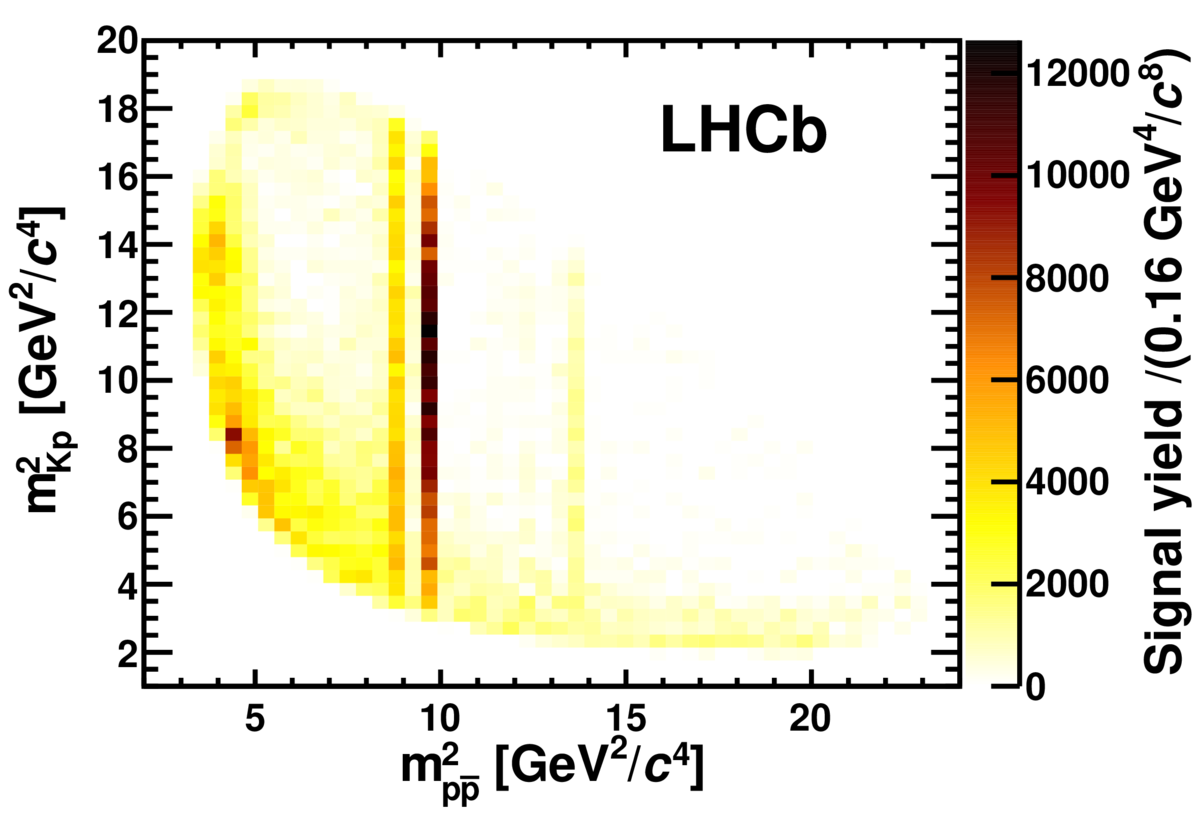}
\includegraphics[width=.45\columnwidth,angle=0]{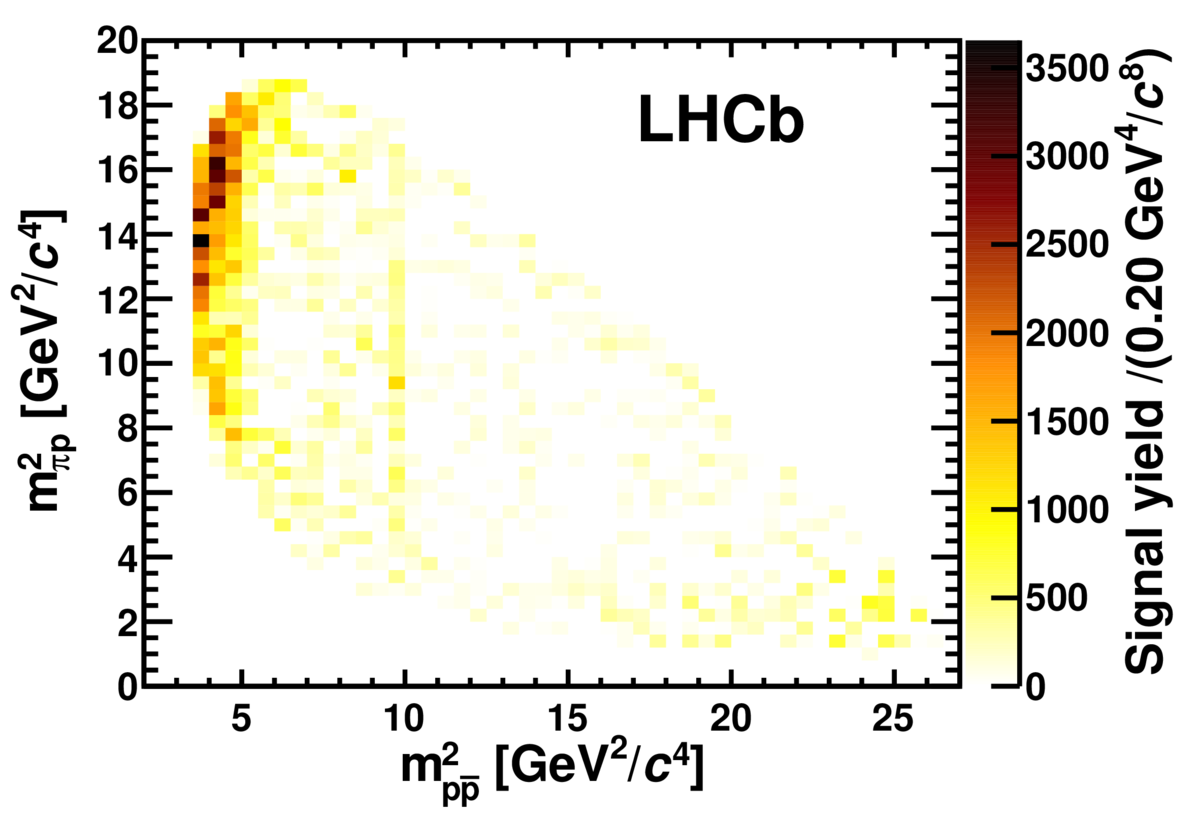}
\caption{ Dalitz plot distributions for the decays (left) $\Bp\to \Kp\proton\antiproton$ and (right) $\Bp\to \pip\proton\antiproton$  from LHCb run I data \cite{LHCb-PAPER-2014-034}. }
\label{Dalitz_ppk_pppi}
\end{figure}

The Dalitz plot distributions for these two channels are shown in Fig.~\ref{Dalitz_ppk_pppi}. Except for the charmonium resonances, which  play an important role in these decays, the event distribution accumulates at the edges of the phase space. Only one resonance was clearly identified in the $\Kp\antiproton$ mass spectrum, the $\Lbar(1520)$.  No resonance was identified at low $\proton\antiproton$ invariant mass.  There is no amplitude analysis for these decays yet.

\begin{figure}[!htb]
\centering
\includegraphics[width=.4\columnwidth,angle=0]{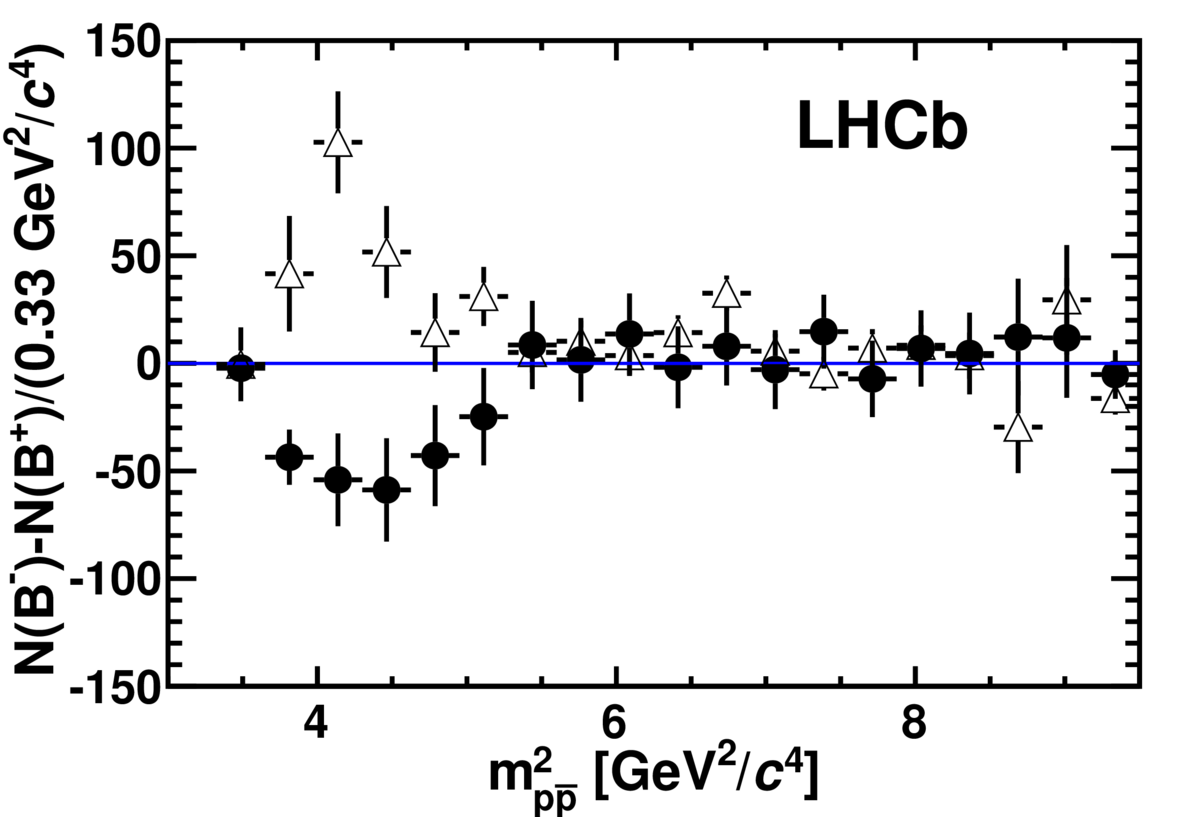}
\caption{ Distribution of $N(B^-) - N(B^+)$ in the squared $\proton\antiproton$ invariant mass for $ m^2_{Kp} < 10 \gevgev$ (black filled circles) and  $ m^2_{Kp} > 10 \gevgev$ (open triangles) \cite{LHCb-PAPER-2014-034}.}
\label{CP_ppk}
\end{figure}

The evidence for \CP violation in the $\Bp\to \Kp\proton\antiproton$ mode comes mainly from the low $\proton\antiproton$ mass region. Figure~\ref{CP_ppk} shows the asymmetry for two regions of  $ m^2_{\Kp\antiproton}$, above and below $10 \gev\gev$, dividing the phase space in two bands.  The \CP asymmetry as a function of the squared $\proton\antiproton$ invariant mass shows a clear sign inversion like that  observed at low $\pip\pim$ mass for $\Bp\to\pip\pip\pim$ and $\Bp\to\Kp\pip\pim$ decays, which has its origin from the strong phase difference between the S and P-waves.

In the region $m_{\proton\antiproton} < 2.85\gev$ and  $ m^2_{\Kp\antiproton} > 10\gev$, the \CP asymmetry  was  measured to be $0.096\pm 0.024 \pm 0.004$, representing an effect of about 4$\sigma$ significance. For the decay $\Bp\to \pip\proton\antiproton$ there were not enough data to perform a similar study. With LHCb data from run II such a study should be possible and in both decays amplitude analysis would be the ideal way to study the sources of \CP violation.

\subsubsection{Charmless $b$-baryon decays} 
\label{b-baryons}
Although \CP violation in baryons is an essential issue for baryogenesis, the study of these decays presents challenges from both the theoretical~\cite{Lu:2009cm,Hsiao:2014mua,Durieux:2016nqr,Zhao:2018zcb,Roy:2019cky} and experimental sides. 

The first difficulty for the experimental observation of \CP violation in beauty baryons comes from  production,  which is almost exclusive to hadronic machines. At the LHC, though, their production is favoured at low transverse momenta \cite{LHCb-PAPER-2014-004} which turns $b$-baryon selection efficiencies lower than those of \B-meson decays \cite{TimVava}.  
As in charmless \B decays, the underlying weak phase responsible for any potential \CP asymmetry comes from the $b\to u$ transition, and the branching ratios are of ${\cal O}(10^{-5})$ or less. 

In two-body decays, the LHCb collaboration, using Run I data, has searched for \CP violation in the modes $\Lb\to \proton\Km$ and $\Lb\to \proton\pim$, with signal yields of about respectively 9 and 6 thousand decays~\cite{LHCb-PAPER-2018-025}. No significant \CP asymmetries were found, 
\begin{eqnarray}
A_{\CP}( \Lb\to \proton\Km)  = & \!\!-0.020 \pm 0.013 \pm 0.019, \nonumber \\
A_{\CP}(\Lb\to \proton\pim)  = & \!\! -0.035 \pm 0.017 \pm 0.020, \nonumber
\end{eqnarray}
yet the measurements are about 4 times more precise then those performed previously by the CDF collaboration~\cite{CDF2014}. 

Similarly to \B-meson decays, \CP violation studies become richer in three- or four-body decays. In 2016, with data from run I, the LHCb collaboration reported a first evidence for \CP violation in the baryon sector by analysing the decay $\Lb\to \proton\pim\pip\pim$  \cite{LHCb-PAPER-2016-030}, with a sample of about 6 thousand events. The analysis made use of the triple-product observables discussed in Sect.~\ref{tripleproducts} such that a \CP asymmetry $a_\CP^{T-\rm odd}$ as in Eq.~\ref{a_CP_Todd} was measured as a function of phase space bins and the angle $\Phi$, which is the angle between the decay planes formed by the pairs $\proton\pim_{\rm fast}$ and $\pip\pim_{\rm slow}$ (slow and fast referring to the $\pip$ momenta). The distribution of $a_\CP^{T-\rm odd}$ in terms of $\Phi$ presented a p-value of $10^{-3}$ of compatibility with  \CP symmetry, translating to a 3.3$\sigma$ evidence for \CP violation in this decay. No evidence for \CP violation was found in the measurement of $a_\CP^{T-\rm odd}$ integrated through the phase space. 

An update of this measurement has just been released by the LHCb collaboration with Run I and part of Run II data, with a sample size comprising about 27 thousand $\Lb\to \proton\pim\pip\pim$ decays  \cite{LHCb-PAPER-2019-028}. The study was carried out by means of measuring the  $a_\CP^{T-\rm odd}$ observable -- following the same strategy as above -- as well as using the energy test method, sensitive to local \CP effects in the phase space distribution,  as outlined in Sect.~\ref{unbinned}. The evidence of \CP violation previously reported  \cite{LHCb-PAPER-2016-030} has not been confirmed, although altogether the results are marginally compatible with \CP conservation.

The search for \CP violation by means of the $a_\CP^{T-\rm odd}$ observable both integrated and in regions of the phase space has been extended to the $\Lb \to \proton \Km \pip\pim$, $\Lb \to \proton  \Km \Kp\Km$ and $\Xib \to \proton \Km \pip\Km$ modes in Ref.~\cite{LHCb-PAPER-2018-001} using Run I data from LHCb. The sample sizes were  about 20 thousand, 5 thousand and 700 decays, respectively. No evidence for \CP violation was found.

More recently, a comprehensive  study including four $\Lb$ and two $\Xib$ 4-body modes provided measurements of integrated \CP asymmetries \cite{LHCb-PAPER-2018-044}. Each \CP asymmetry was measured with respect to that of a $\Lc$ formed as an intermediate state and leading to the same final state, for example $\Delta A_\CP(\Lb\to \proton\pim\pip\pim)\equiv A_\CP(\Lb\to \proton\pim\pip\pim) - A_\CP(\Lb\to (\Lc\to\proton\pim\pip)\pim)$. In this way, corrections due to instrumental and $b$-baryon production asymmetries cancel out to first order. Further corrections, such as kinematic weighting, were also used to guarantee control over nuisance asymmetries. The resulting asymmetries are presented in Table~\ref{ACP_Lambdab}. No evidence for \CP violation was found.

\begin{table}[tb] 
\caption{Phase-space integrated \CP asymmetries for 4-body $\Lb$ and $\Xib$ charmless decays from LHCb collaboration  \cite{LHCb-PAPER-2018-044}.}
\begin{center} 
\begin{tabular}{lc}
\hline 
 Decay & $\Delta A_{\CP}(\%)$ \\ 
\hline 
$\Lb \to \proton \Km \pip\pim$  &  $+ 3.2 \pm 1.1 \pm 0.6$ \\ 
$\Lb \to \proton  \pim \pip\pim$ &   $+ 1.1 \pm 2.5 \pm 0.6$ \\ 
$\Lb \to \proton  \Km \Kp\pim$  & $-6.9  \pm 4.9  \pm 0.8$ \\ 
$\Lb \to \proton  \Km \Kp\Km$  & $+ 0.2 \pm 1.8 \pm 0.6$ \\ 
$\Xib \to \proton \Km \pip\Km$  &  $-6.8 \pm 8.0 \pm 0.8$ \\ 
$\Xib \to \proton \Km \pip\pim$  &  $17 \pm 11 \pm 1$ \\ 
\hline 
\end{tabular} \end{center} 
\label{ACP_Lambdab}
\end{table}

Overall, the current status is that while \CP violation is well established in the \B-meson sector, no compelling evidence exists for \CP violation in $b$-baryon decays to date. Yet most of the studies rely on LHCb data from run I. It is reasonable to expect that  signs of \CP violation from $b$-baryons may still arise with a comprehensive study using the full Run I and Run II data.

\section{Hadronic double-charm interaction in charmless \B decays}
\label{chapter5}

Charmless \B decays, as discussed in the previous section, can potentially produce \CP violation due to the tree level transition $ \bquark \to \uquark $, involving the weak phase $\gamma$, interfering with different weak transitions such as the penguin-type diagram depicted in Fig.~\ref{treepenguin}. Nevertheless, a charmless final state can also be reached through the formation of a double-charm.  By simply considering the CKM hierarchy, the inclusive process  $\bquark \to \cquark \cquarkbar  \quark$ is about two orders of magnitude more likely than  those with $\bquark\to \uquark  \quark\quarkbar'$ ($q, q'=\uquark,\dquark,\squark$) transitions.

The effective contribution of a double-charm amplitude to charmless \B decays can be twofold. The first one is by direct formation of a charmonium resonance decaying to a pair of pions or kaons, in a three- or four-body  decays. An example is  $ \Bp\to \chi_{c0}\pip$, with $\chi_{c0}\to\pip\pim$. The other is by production of a $\D\Dbar'$ pair rescattering to $h \bar h'$ ($h, h'$ being  light mesons), a long-distance process expected to be suppressed (otherwise $\B \to \D\Dbar$ and $ \B \to h \bar h'$ branching ratios would be similar) but nevertheless suitable to  have an impact giving the aforementioned higher inclusive process.

The possibility of this long-distance rescattering with the same quantum numbers, naturally providing a different strong phase, refers back to the discussion in Sect.~\ref{CPT}. Many authors have contributed to estimate the effect of double-charm component to charmless \B decays \cite{Suzuki2007,Soni2005,Smith2003,Buras1997}. 
Most of these models treat only the rescattering of the double-charm to  two- or quasi-two-body \B decays, while comparing to other models based on short-distance approximation or light meson  rescattering. But since the only observables are total rates (and asymmetries) there is no way to disentangle  the relative contribution of each process from the experimental results.

Here again three- and four-body  decays, due to their rich dynamics across the phase space, may provide signatures of possible contributions associated with double-charm intermediate states, as discussed in the following. 

\subsection{The charmonium $\Bp\to h^+ \chi_{c0}$ decay}

The main charmonium resonance contributing to a charmless \Bp three-meson final state is the scalar $\chi_{c0}$, with mass $(3414.7\pm 0.3)\mev$ and width $(10.5\pm 0.8)\mev$~\cite{PDG2019}. 
The  $\Bp\to h^+ \chi_{c0}$ ($h = \kaon, \pion$) decays weakly at tree level through a $\bquark \to \cquark \cquarkbar \quark$ ($\quark = \squark, \dquark$) transition,  with the $\chi_{c0}$ meson then decaying strongly to a pair of mesons. 
The branching fraction of the decay $\Bp \to \Kp \chi_{c0}$,  for example,  is ${\cal B}(\Bp \to \Kp \chi_{c0})= (1.49 \pm 0.15)\times 10^{-4} $  with ${\cal B}(\chi_{c0}\to\pip\pim) \sim 6 \times 10^{-3}$ \cite{PDG2019},.

\begin{figure}[!htb]
\centering
\includegraphics[width=0.4\columnwidth,angle=0]{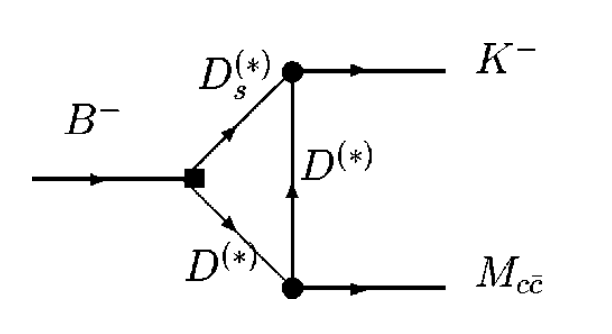}
\caption{Possible long-distance contribution to the decay $B^\pm \to K^\pm  \chi_{c0}$ \cite{Colangelo2003} .}
\label{B-chiK}
\end{figure}

In what concerns theoretical calculations, the critical factor here is that this decay does not include factorisable contributions; the $\chi_{c0}$, being scalar, cannot be produced by a  matrix element \mbox{$\langle \chi_{c0}|\cquarkbar\bar \gamma_\mu(1-\gamma_5)\cquark |0\rangle  = 0$} due to conservation of the axial-vector current \cite{Melic2004}. On the other hand, the non-factorisable contributions from the soft hadronic matrix elements, based on the light-cone approach, are too small to accommodate the experimental data \cite{Wang2003}. It has been shown, however,  that rescattering effects, involving charm meson intermediate states, can reproduce the observed branching ratio  through a mechanism  shown in Fig.~\ref{B-chiK} \cite{Colangelo2003}. 

\begin{figure}[!htb]
\centering
\includegraphics[width=0.45\columnwidth,angle=0]{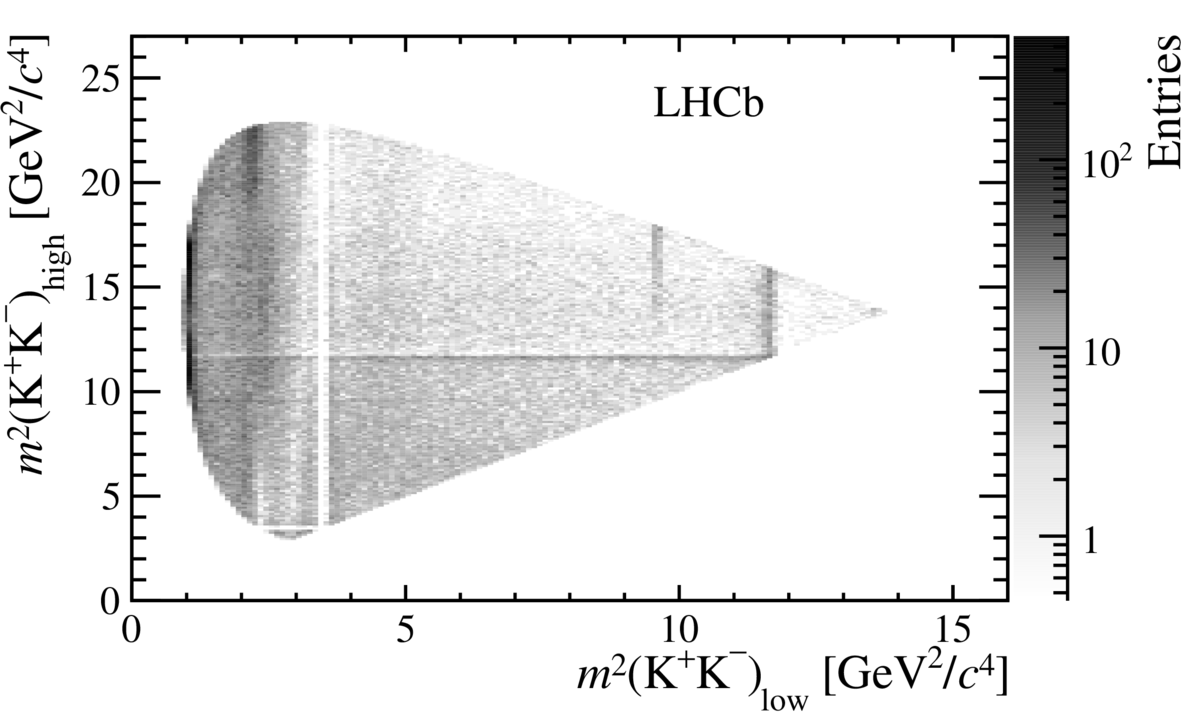}
\includegraphics[width=0.45\columnwidth,angle=0]{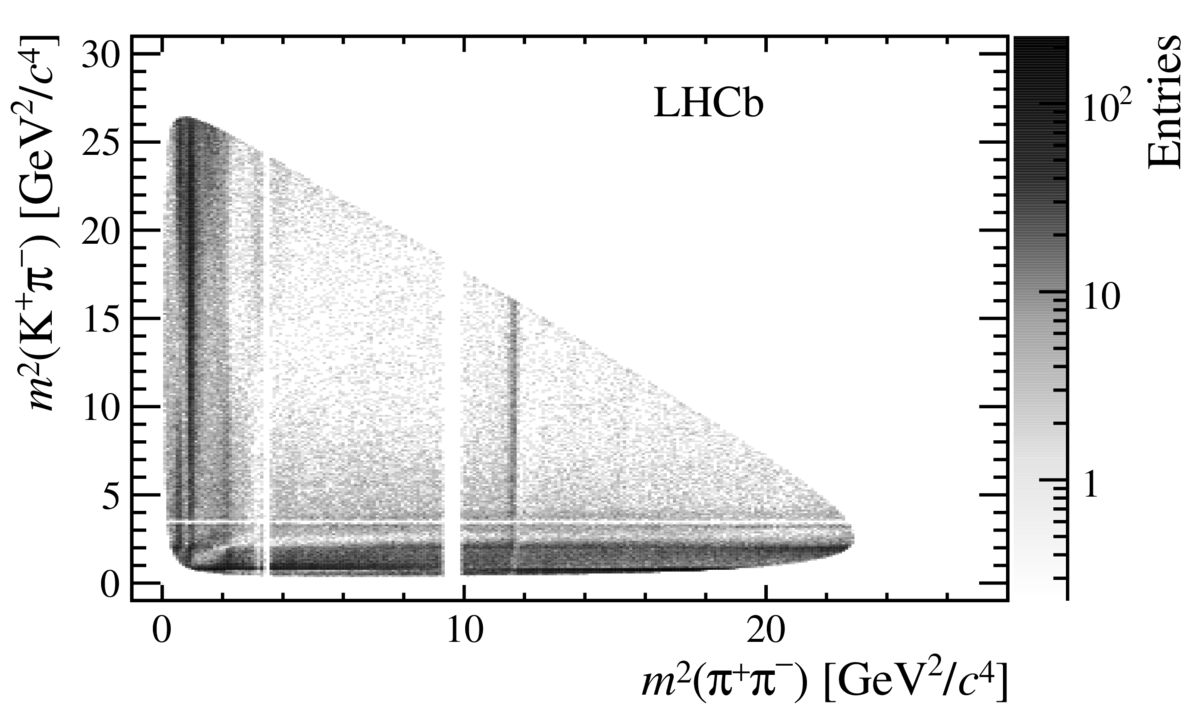}
\caption{Scalar $\chi_{c0}$ signature in the Dalitz plots of (left) $\Bp\to\Kp\Kp\Km$  and  (right) $\Bp\to\Kp\pip\pim$ decays  \cite{LHCb-PAPER-2014-044}.}
\label{Chi_c_LHCb}
\end{figure}

The contribution of the intermediate state $\Kp\chi_{c0}$ has been observed at a  fit fraction level of 1\% in amplitude analyses performed by the Babar collaboration in both final states $\Bp\to \Kp\pip\pim$~\cite{BaBar_bkpipi} and $\Bp\to \Kp \Kp\Km$~\cite{BaBar_B3K}. It has been seen to contribute to the same final states from LHCb run I data  \cite{LHCb-PAPER-2014-044} as can be seen in  Fig.~\ref{Chi_c_LHCb} as a clear narrow band at  around $11.6 \gevgev$ in $m^2_{\Kp\Km}$ and $m^2_{\pip\pim}$ in the Dalitz plot of the decays $\Bp\to \Kp \Kp\Km$ and $\Bp\to \Kp\pip\pim$, respectively.
It is also possible to see a band with lack of events just above the $\chi_{c0}$ mass indicating a pattern of interference with the non-resonant contribution underneath. 

The decay amplitude $\pip\chi_{c0}$ was  not identified in the Dalitz plot analysis of the decay $\Bp\to\pip\pip\pim$ made recently by the LHCb collaboration~\cite{LHCb-PAPER-2019-017} but it is likely to be observed from run  II data, at the level of a few hundred events or so. 

 A possible interference between the $\Bp\to \pip \chi_{c0} $ amplitude  and the dominant $\Bp\to\pip\rho(770)^0$ (which has a total \CP asymmetry consistent with zero, as discussed in Sect.~\ref{quasi2body}) could be used for a measurement of $\gamma$ \cite{Gobel1998}, neglecting contributions from other amplitudes in their interference region, otherwise other components may as well be included \cite{Gobel2000}. 
A similar argument  is proposed in Ref.~\cite{Gronau1995}, this time exploiting a measurement of  the \CP asymmetry for the $\Bp\to \pip\chi_{c0}$ through the interference with a non-resonant amplitude.
Since $\gamma\approx 70^\circ$, this interference term possibly changes sign by  charge conjugation, appearing as constructive interference for \Bp decay and destructive for \Bm decay (or {\it vice-versa}).  
They also have generalised this approach for other final states involving  vector light mesons and the charmonium $\eta_c$.

\subsection{Double-charm and $\Bp\to \Kp\Kp\Km$ final state }
\label{doublecharmKKK}
The amplitude analysis of  $\Bp\to\Kp\Kp\Km$ from both the BaBar and Belle collaborations  \cite{BaBar_B3K,Belle_3h_2004} showed a dominant  non-resonant contribution. In a recent paper \cite{Pat2017}  the authors  propose a non-resonant contribution to this decay which includes rescattering hadronic component involving double-charm-meson intermediate states, in a similar direction as proposed for the $\Bp\to \Kp\chi_{c0}$ in Ref.~\cite{Colangelo2003}. 

The LHCb experiment collected more than a hundred thousand $\Bp\to \Kp\Kp\Km$ decays from \mbox{run I} period  \cite{LHCb-PAPER-2014-044} and the Dalitz plot distribution can be seen in Fig.~\ref{Chi_c_LHCb}. 
The narrow resonant contributions are clearly seen,  $\Bp\to \Kp\phi(1020)$ and $\Bp\to \Kp\chi_{c0}$, together with a large non-resonant contribution, populating  the whole phase space. Although no amplitude analysis was yet performed, the distribution appears to be consistent with the results from BaBar and Belle amplitude analyses \cite{BaBar_B3K,Belle_3h_2004}, which had much smaller samples. This significant presence of the non-resonant amplitude makes this decay different from the other charmless three-body \B decays (with low occupation at higher masses), and potentially offers sensitivity, through interference, to study high-mass intermediate resonances, including   \CP signatures. 

\begin{figure}[htbp]
\centering
\includegraphics[width=0.45\columnwidth,angle=0]{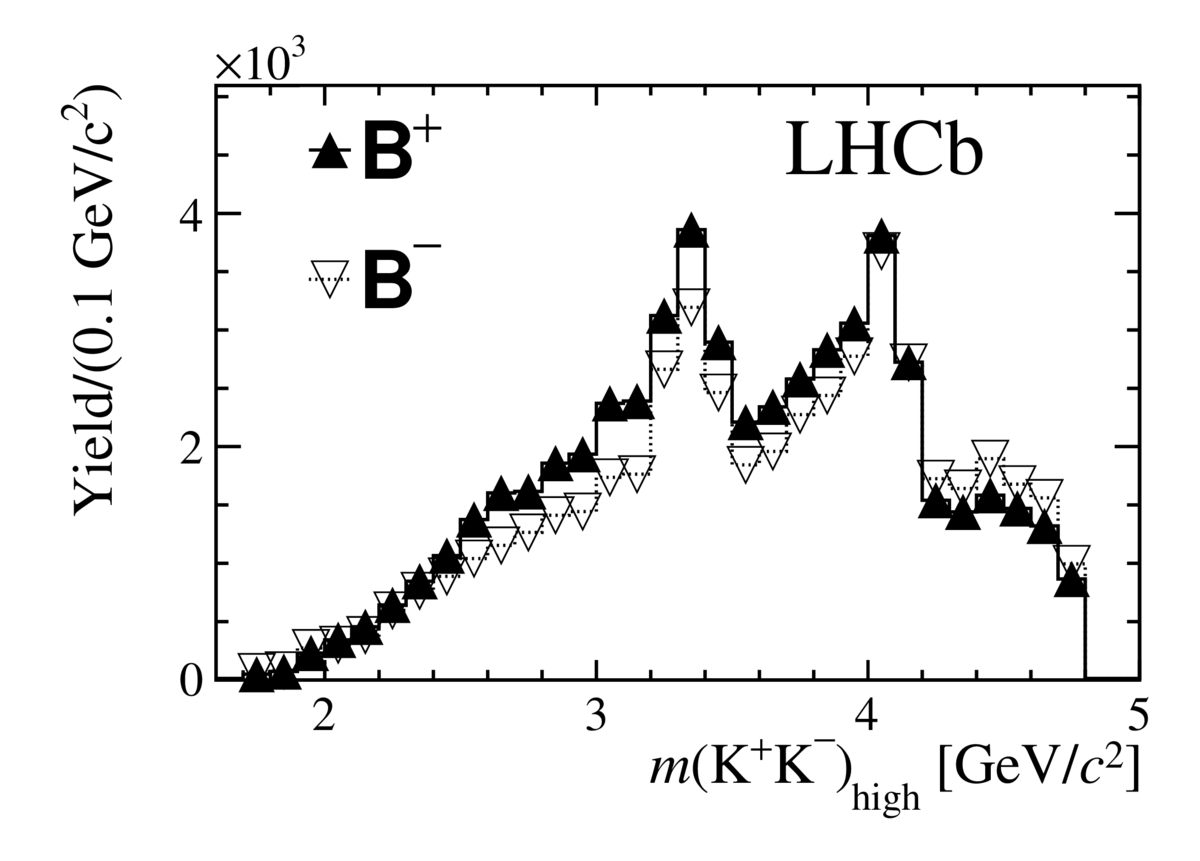}
\includegraphics[width=0.45\columnwidth,angle=0]{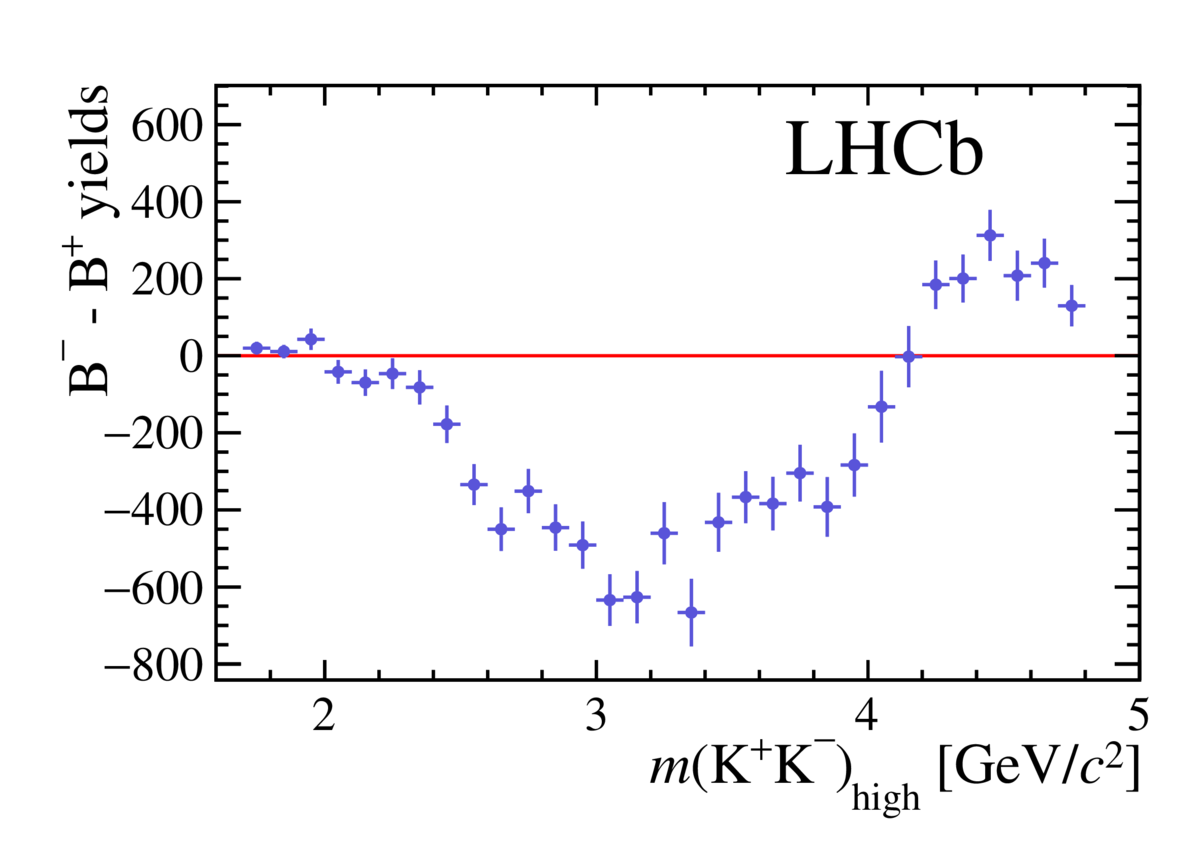}
\caption{ (left) Projection of $m(KK)_{\rm high}$ distribution for $B^\pm \to K^\pm K^+K^-$ decay  separated by charge. (right)  Subtracted distribution of the \Bm and \Bp $m(KK)_{\rm high}$ projections from the left plot. \cite{LHCb-PAPER-2014-044} }
\label{KKK_projection}
\end{figure}

Another interesting feature observed in the $\Bp\to \Kp\Kp\Km$ Dalitz plot distribution is the localised \CP asymmetry at low $\Kp\Km$ mass, already discussed in Sect.~\ref{kkpipiresc}. Figure~\ref{KKK_projection} (left) shows the events for $\Bp$ and $\Bm$ integrated in $m(\Kp\Km)_{\rm low}$  \cite{LHCb-PAPER-2014-044}. The two peaks correspond to the angular distribution associated with the vector resonance $\phi(1020)$. By subtracting both distributions, as shown in Fig.~\ref{KKK_projection} (right), the charge difference associated to \CP violation can be seen. It is possible to identify the negative \CP asymmetry located in the $\pi\pi-KK$ rescattering region, but also the fact that the \CP violation pattern changes sign,  crossing zero at about 4\gev, which is near the $\D\Dbar$ threshold. Moreover, the LHCb data distribution for $\Bp\to \Kp\pip\pim$ also shows a similar,  although opposite,  sign inversion around 4 \gev.\footnote{See the Supplemental material associated to \cite{LHCb-PAPER-2014-044} at https://cds.cern.ch/record/1751517/files.} The same behaviour appears in the pair of channels $\Bp\to\pip\pip\pim$ and $\Bp\to\pip\Kp\Km$.

\begin{figure}[htbp]
\centering
\includegraphics[width=.4\columnwidth,angle=0]{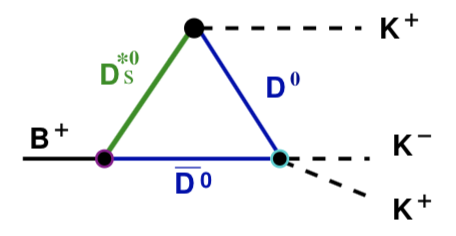}
\caption{Possible long-distance contribution to the $\Bp\to \Kp\Kp\Km$ decay \cite{Pat2017}. }
\label{LoopKKK}
\end{figure}

This  \CP-asymmetry sign inversion can be explained  by a hadronic charm loop process shown in Fig.~\ref{LoopKKK}, according to the model presented in Ref.~\cite{Pat2017}.
The corresponding amplitude  has a strong-phase sign change at the $\D\Dbar$ threshold: the phase changes from about $-\pi/2$ to $\pi/2$ close to the region where the \CP asymmetry changes sign in Fig.~\ref{KKK_projection}. Although this signature favours the interpretation of  double-charm rescattering, the model has more complexity and only a full amplitude analysis could test for the validity of this mechanism.

\subsection{Possible consequence for the $\Bp\to\pip\pip\pim$ decay}

As discussed previously, the short-distance contribution for the $\Bp \to \pip\pip\pim$ decay is dominated by the tree contribution with a weak phase $\gamma$. On the other hand, a  long-distance contribution can also be possible by a similar process to that in Fig.~\ref{LoopKKK} in the decay $\Bp\to\Kp\Kp\Km$. The interference between these short-distance tree component  and  double-charm rescattering, carrying no weak phase, may lead to a signature of \CP violation spread throughout the phase space. By inspecting the bottom-left plot in Fig.~\ref{Mirandizing4channels}, indeed there is a significant \CP asymmetry in the high $\pip\pim$ mass regions.

\begin{figure}[htbp]
\centering
\includegraphics[width=.5\columnwidth,angle=0]{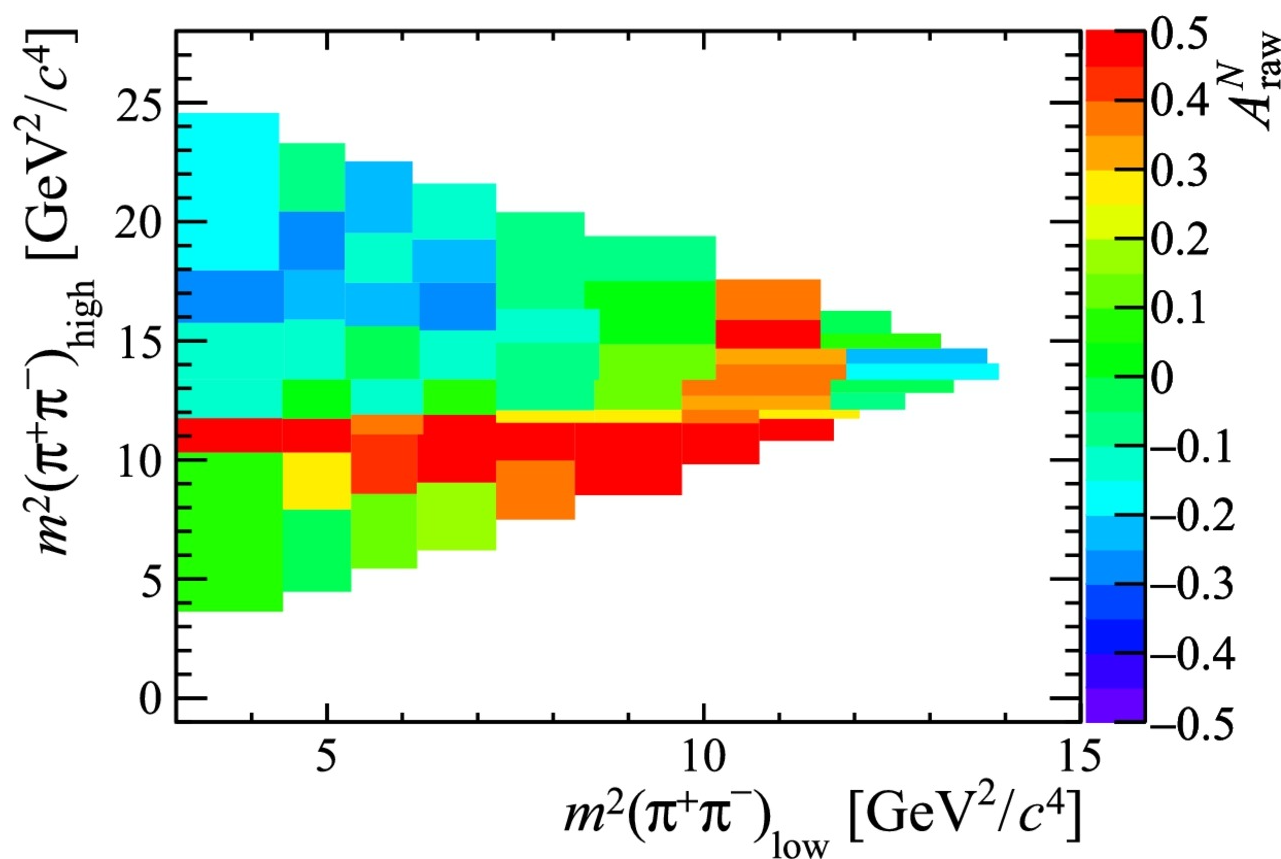}
\caption{\CP asymmetry pattern for the decay $\Bp\to\pip\pip\pim$, obtained through the Miranda method via simulation according to the model presented in Ref.~\cite{Bediaga:2020ztp}. }
\label{B3pi_DD}
\end{figure}

\begin{figure}[htbp]
\centering
\includegraphics[width=.5\columnwidth,angle=0]{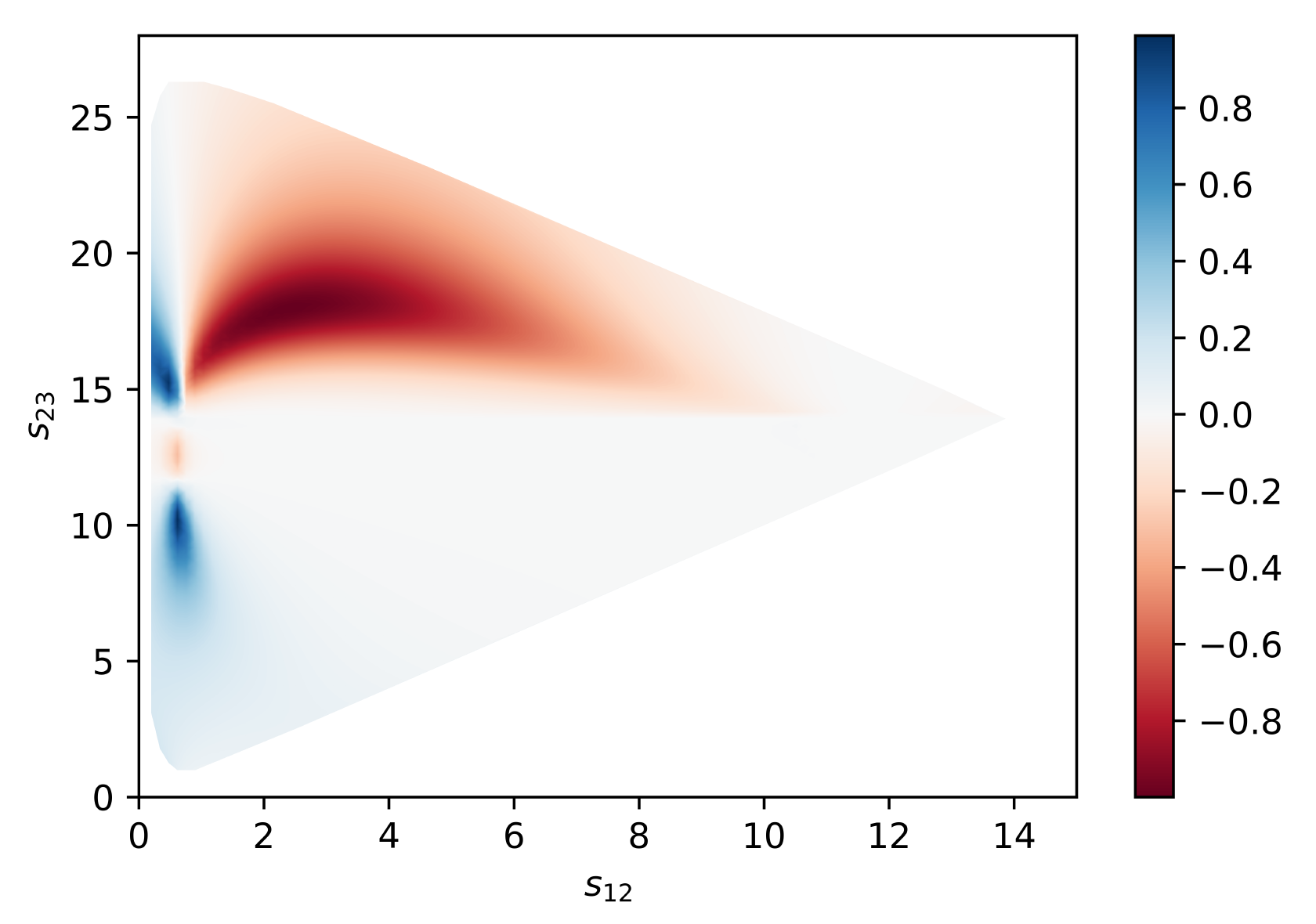}
\caption{\CP asymmetry pattern for the decay $\Bp\to\pip\pip\pim$ considering the interference between the $\rho(770)^0$ and $\chi_{c0}(3860)$ \cite{Mannel:2020abt}. The variables $s_{12}$ and $s_{23}$ here represent $m^2(\pip\pim)_{\rm low}$ and $m^2(\pip\pim)_{\rm high}$, respectively.}
\label{MannelB3pi}
\end{figure}

Recently a phenomenological study for the $\Bp\to\pip\pip\pim$ mode has explored the interference between the tree component, carrying the weak phase, and a modelled amplitude for the $\D\Db$ and $\pi\pi$ coupled channels, with   $\chi_{c0}$ as  complex pole in the $\D\Db$ channel \cite{Bediaga:2020ztp}. As discussed above in Sect.~\ref{doublecharmKKK}, the  latter amplitude has a strong-phase sign change when it passes from off-shell to on-shell, and this effect could be   responsible for the \CP asymmetry pattern  observed  in the bottom-left plot in Fig.~\ref{Mirandizing4channels} near the $\D\Db$ thresholds. A simulation for this model, using Laura++ \cite{Laura++} generator, is shown in Fig.~\ref{B3pi_DD}. 
Another recent study ~\cite{Mannel:2020abt}  also discusses the potential role  of open-charm threshold amplitudes to the observed \CP violation in this decay channel. In a similar way,  the model exploits  the interference between light resonances with both open-charm  $\D\Db$  amplitudes and high-mass charmonium  resonances.  The \CP asymmetry pattern obtained through the  interference between $\rho(770)^0$ and the propagator for $\chi_{c0}(3860)$ scalar resonance is illustrated in Fig.~\ref{MannelB3pi}.

A possible observation of the $B^\pm \to \pi^\pm \chi_{c0}$ mode with the LHCb run II data may show a \CP violation pattern in this charmonium state as predicted by Ref.~\cite{Gronau1995}, and, if so, one may infer the importance of a double charm re-scattering in this decay. In other words, if the \CP violation pattern for both  the $\chi_{c0}$ component and the non-resonant component (covering a wider mass region) are similar, this would point towards the interpretation of the non-resonant amplitude receiving an important contribution from the double-charm loop mechanism.

\subsection{Search for double-charm signature in $ \Bc\to\pip\Kp\Km$ decay} 

Recently LHCb, using Run I data,  presented evidence of the decay $\Bc \to \pip\chi_{c0}$,  with $\chi_{c0} \to \Kp\Km$ decay, with a branching ratio  of $ \left[\sigma(\Bc)/\sigma(\Bp)\right] \times {\cal B}(\Bp\to\pip\chi_{c0} )= (9.8 \pm 3.4) \times 10^{-6}$  \cite{LHCb-PAPER-2016-022}. The collaboration searched also for a possible non-charmonium component to $\Bc\to \pip\Kp\Km$ decay, but no significant contribution was found. From short-distance calculations  \cite{Liu2009}, this contribution indeed  should be helicity suppressed. This opens  the possibility to   observe $\Bc \to \pip\Kp\Km$  events coming from a long-distance double charm rescattering contribution similar to those shown in Figs.~\ref{B-chiK} and \ref{LoopKKK}.

\begin{figure}[!htb]
\centering
\includegraphics[width=0.5\columnwidth,angle=0]{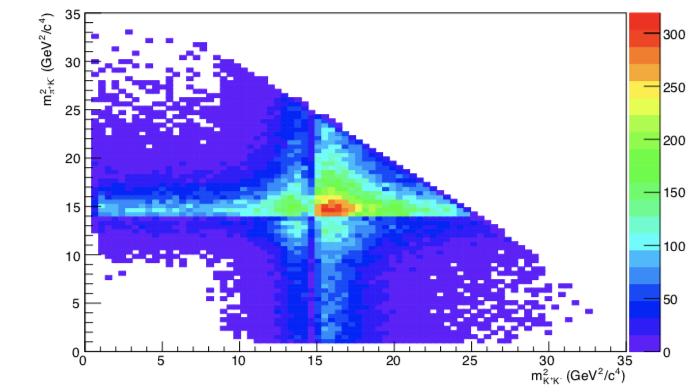}
\caption{Simulated Dalitz plot  for the decay $ \Bc\to  \pip\Kp\Km$ considering double-charm long-distance contributions \cite{Pat2018}, using Laura++ \cite{Laura++}.}
\label{B_cDalitz} 
\end{figure}

   A  simulation with  amplitudes corresponding to the long-distance contributions from the double-charm rescattering processes $\Dz\Dzb \to \Kp\Km$ and $\Dp\Dsm\to\pip\Km$, as from Ref.~\cite{Pat2018}, presents a  distinguished  signature for the Dalitz plot distribution, as shown in Fig.~\ref{B_cDalitz}. This is quite different from the distribution observed in charmless three-body \Bp decays where the events are concentrated at the edges of the Dalitz plot, as in Fig.~\ref{Chi_c_LHCb}.

\section{Direct \CP violation in \B decays to open charm}
\label{chapter6}

At tree level, the $\bquark$ quark dominantly decays to a $\cquark$ quark, with no weak phase.  This process makes decays to open charm those with higher branching ratios. Some of these final states, after the further weak decay of the $\cquark$ hadron,  can also be reachable through the quark coupling $ \bquark \to \uquark\cquarkbar\quark$ ($\quark = \squark, \dquark$) involving the CKM weak phase $\gamma$, which is also a tree-level process, with no penguin diagrams involved. The interference between them can  allow the experimental determination of  the angle $\gamma$. From the theoretical point of view, these decays are extremely clean \cite{BrodZupan_2013}.

The main mechanisms proposed to exploit this feature as a source of direct \CP violation are the Gronau--London--Wyler (GLW) \cite{Gronau1990,Gronau1991},  Atwood--Dunietz--Sony (ADS) \cite{Atwood1996,Atwood2000}, and Giri--Grossman--Soffer--Zuppan (GGSZ) \cite{GGSZ} methods, as described next, with their main associated experimental results. The value of $\gamma$ is obtained from the combination of these different methods.

 \subsection{The GLW method}
 
\begin{figure}[htbp]
\centering
\includegraphics[width=.8\columnwidth,angle=0]{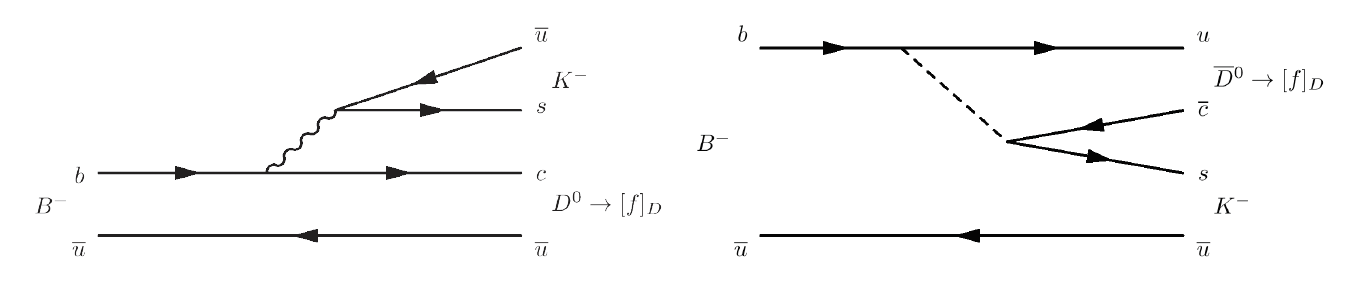}
\caption{Leading diagrams for the decays $\Bp\to \Dz \Kp $ and $\Bp \to \Dzb\Kp $, where $\left[f\right]_D$ is a final state accessible by both $\Dz$ and $\Dzb$ \cite{TimVava}.}.
\label{B-DK}
\end{figure}

The GLW method makes use of the interference between the processes $\Bp\to \Dz h^+$ and $\Bp\to \Dzb h^+$ ($h=\PK,\pi$). The first is dominant, proceeding through a $\bquark\to \cquark$ transition, while the second is suppressed given by the process $ \bquark \to \uquark \cquarkbar \squark$. 
  These decays can interfere  if  $\Dz$ and $\Dzb$ (referred to as $\D$ -- a general superposition of both) decay to \CP eigenstates, such as  $\D\to \Kp\Km$ (\CP-even) or $\D\to\KS\piz$ (\CP-odd),  as shown in Fig.~\ref{B-DK} \cite{TimVava}. 
  The resulting \CP asymmetry, as in Eq.~\ref{ACP_sindelta_sinphi}, depends on the ratio $r_\B$ of the sub-leading $\Bp\to \Dzb\Kp$ to the leading $\Bp\to \Dz\Kp$ amplitude as well as on their relative strong phase $\delta_\B$,
\begin{equation}
A_\CP = \frac{2\eta^\CP r_\B \sin (\delta_\B) \sin (\gamma)}{1+r_\B^2 +2\eta_\CP r_\B\cos(\delta_\B)\cos(\gamma)}~,
\label{ACP_GLW}
\end{equation}
where $\eta^\CP=\{+1,-1\}$ depending on whether the $\Dz$ decays to a \CP-even or \CP-odd final state, respectively. The experimental analyses can obtain $r_\B$, $\delta_B$ and $\gamma$ using measured \CP asymmetries from different $\Dz$ final states. 

\begin{figure}[htbp]
\centering
\includegraphics[align = c, width=0.7\columnwidth,angle=0]{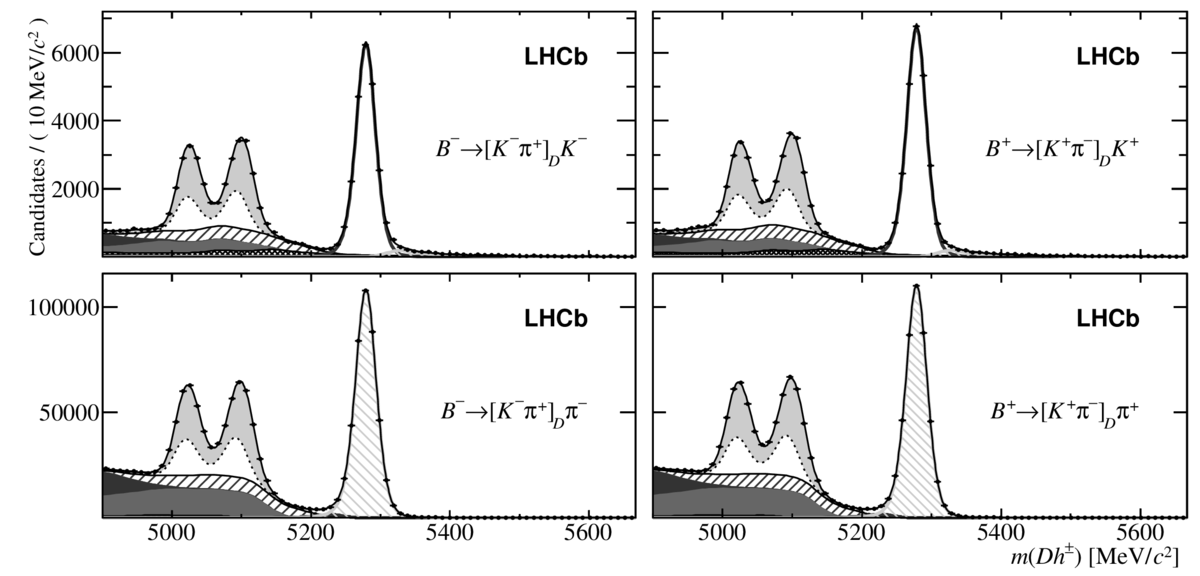}
\includegraphics[align = c, width=.25\columnwidth,angle=0]{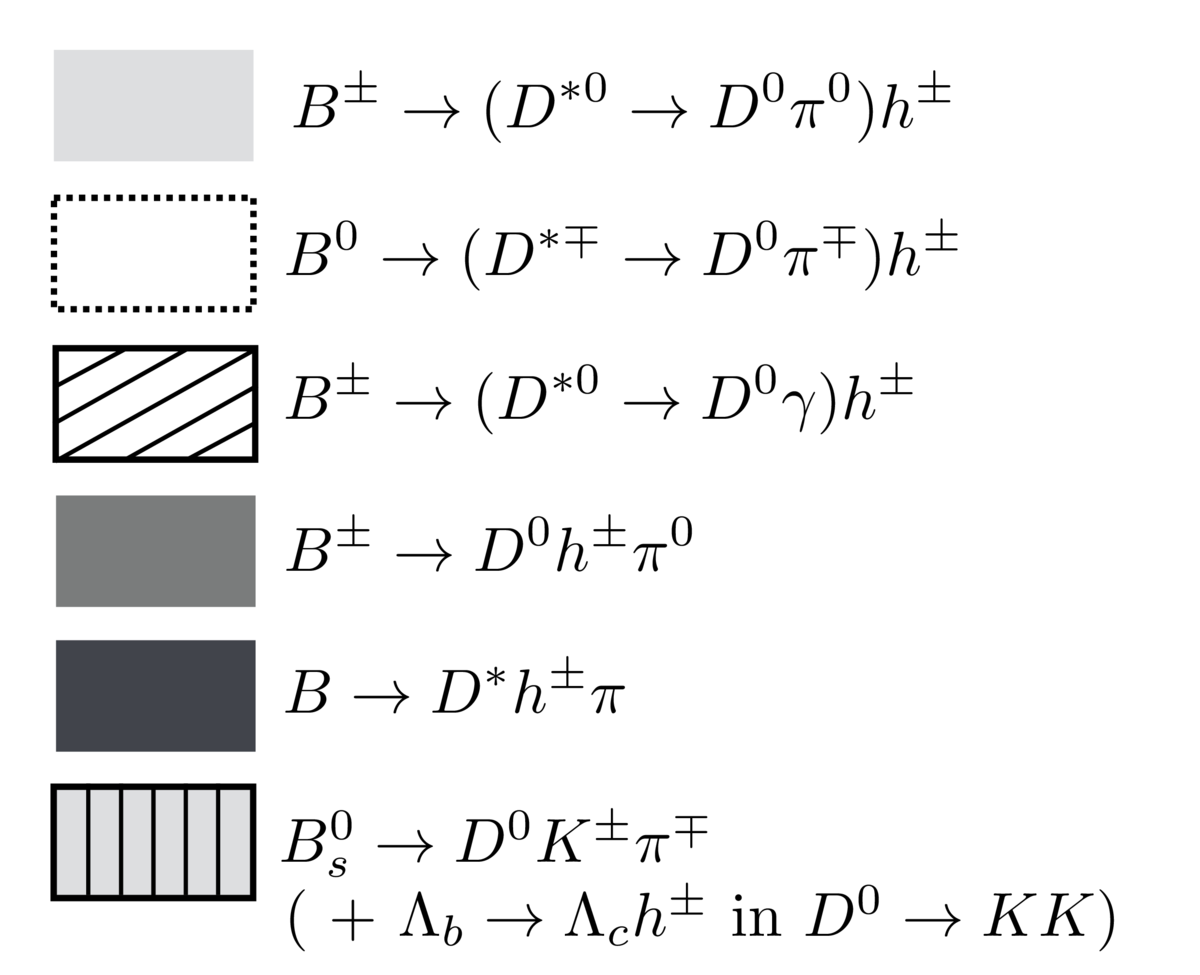}
\caption{Invariant mass distributions of selected  (top) $\B^\pm \to [\Kp\Km]_\D\PK^\pm$ and (bottom) $\B^\pm \to [\pip\pim]_D\PK^\pm$ decays, separated by
charge \cite{LHCb-PAPER-2017-021}. The different fit components are displayed in the legend on the right. }
\label{B-DK-hh}
\end{figure}

The Babar, Belle, CDF and LHCb collaborations have reported $A_\CP$ and $\gamma$ measurements based on GLW modes (see for example \cite{Aaltonen:2009hz,delAmoSanchez:2010ji,Aihara:2012aw,LHCb-PAPER-2014-041}).  The latest and most precise measurement of $A_\CP$ in \mbox{$\Bp\to\Dz\Kp$} comes from LHCb using full run I and part of run II data~\cite{LHCb-PAPER-2017-021}. Figure~\ref{B-DK-hh} shows the relevant mass distributions for $\Bp\to\D\Kp$ where $\D\to\Kp\Km$ and $\D\to\pip\pim$. A visible difference in signal yields for \Bp and \Bm is seen in both final states. The combined  \CP asymmetry of the two final states was measured to be 
\begin{equation}
A_{\CP} = - (12.4\pm  1.2\pm 0.2)\%
\end{equation}
with \CP violation clearly established in this mode. 

  \subsection{The ADS method}
  
The ADS method \cite{Atwood1996} is similar to GLW but instead uses the interference between Cabibbo-favoured (CF) $\Dz$ decays and Doubly--Cabibbo-suppressed (DCS) $\Dzb$ decays (and vice versa), for example $\Dz\to\Km\pip$ and $\Dzb\to\Km\pip$. In this case, two new parameters get in place: $r_\D$ and $\delta_\D$, respectively  the ratio of the DCS to the CF amplitude and their relative phase ($\bar D$ to $\D$ to the same final state). Equation~\ref{ACP_GLW} gets modified to
\begin{equation}
A_\CP = \frac{ 2r_\D r_\B \sin (\delta_\D+\delta_\B) \sin (\gamma)}{r_\D^2+r_\B^2 +2r_\D r_\B \cos(\delta_\D+\delta_\B)\cos(\gamma)}~.
\label{ACP_ADS}
\end{equation}

\begin{figure}[htb]
\centering
\includegraphics[width=.8\columnwidth,angle=0]{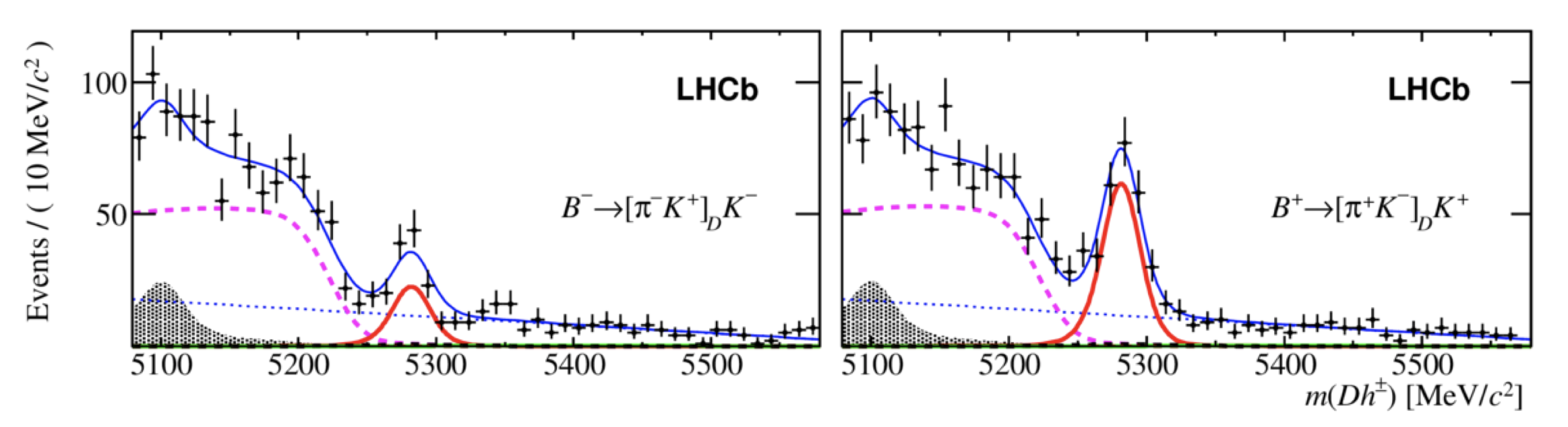}
\caption{Invariant mass distributions of selected  $\Bp\to[\Km\pip]_{\D} \Kp $ separated by
charge. The signal contribution to the fits3 is displayed through the red curve. For details, see \cite{LHCb-PAPER-2016-003}.}
\label{B-DK-Kpi}
\end{figure}

Many ADS modes have been studied, such as in Refs.~\cite{PhysRevD.84.091504,PhysRevLett.106.231803, LHCb-PAPER-2016-003}. 
The measured \CP asymmetry obtained for the $\D\to \PK^\mp\pi^\pm$ channels is quite large \cite{LHCb-PAPER-2016-003}, 

\begin{equation}
A_{\CP} = - 0.403 \pm  0.056~,
\end{equation}
which can also be seen from Fig.~\ref{B-DK-Kpi}.
  
The ADS method can also be applied for three- or four-body modes \cite{Atwood2000}, for example $\D\to\Km\pip\piz$ and $\D\to\Km\pip\pim\pip$, with the introduction of an extra parameter, the coherence factor $\kappa$, multiplying the $2r_\D r_\B$ terms in Eq.~\ref{ACP_ADS} varying between 0 and 1 depending on the resonant structure of the decay. It has been also used by several analyses such as in Refs.~\cite{PhysRevD.84.012002}  (using $\D\to\Km\pip\piz$) and \cite{LHCb-PAPER-2016-003} (using $\D\to\Km\pip\pim\pip$).

\subsection{The GGSZ method}

\begin{figure}[htbp]
\centering
\includegraphics[width=0.4\textwidth]{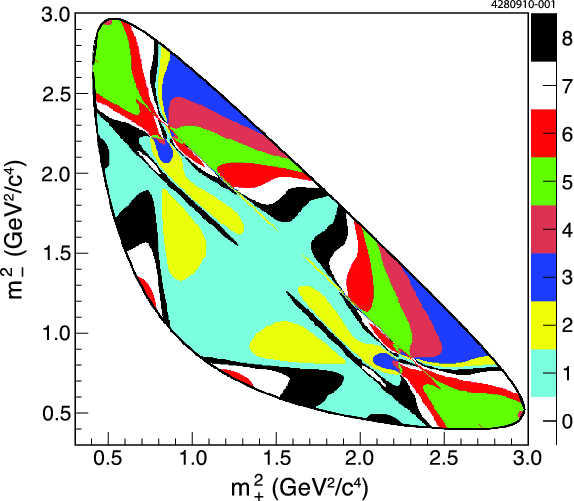}
\caption{Map of the strong phase difference across the Dalitz plot between \Dz and \Dzb decays in the final state $\KS\pip\pim$, as obtained by CLEO. The Dalitz variables are $m^2_+ = m^2(\KS\pip)$ and $m^2_- = m^2(\KS\pim)$~\cite{CLEO_DKSpipi_phase_2010}.}
\label{cleo_Dkspipiphasemap}
\end{figure}

The GGSZ method \cite{GGSZ} exploits the strong phase variation across the phase space of a three- or four-body self-conjugate \D decay in the process $\Bp\to \D h^+$. The benchmark mode is $\D\to \KS\pip\pim$ but there are  others such as, for example,  $\D\to\KS\Kp\Km$ and $\D\to\KS\pip\pim\piz$. It compares the distribution of events in the phase space for \Bp and \Bm  and is the most sensitive method for obtaining $\gamma$. It depends on the prior -- external -- knowledge of the strong phase which could have been measured through an amplitude analysis (model dependent) or  direct measurement (model independent) via the quantum correlation of $\Dz\Dzb$ pairs from $\psi(3770)$, as obtained by the CLEO collaboration \cite{CLEO_DKSpipi_phase_2010} for $\D\to \KS\pip\pim$ for instance.\footnote{Just recently, the BESIII collaboration released their results on the strong phase across the Dalitz plot for $\D\to \KS\pip\pim$ decays \cite{BESIII_Kspipi2020_1,BESIII_Kspipi2020_2}, with the same binning scheme as CLEO, and which is expect to improve the corresponding uncertainty on $\gamma$ by approximately a factor of 3.} 
In Fig.~\ref{cleo_Dkspipiphasemap} the model-independent strong phase difference for \Dz and \Dzb decays is shown across the Dalitz plot, where the eight (coloured) regions represent different phase intervals. Although the model-dependent approach may provide better statistical uncertainties, it relies on amplitude models (full Dalitz plot analysis) which suffer from irreducible systematics. The model-independent approach, on the other hand, is only statistically limited.  

The observables in the GGSZ method are the variables $x_\pm$ and $y_\pm$ which related to the physical observables as
\begin{eqnarray}
x_\pm =&\!\!  r_B\cos(\delta_\B\pm\gamma) \nonumber \\
y_\pm =&\!  r_B\sin(\delta_\B\pm\gamma) ~.
\end{eqnarray}

The GGSZ method has been applied for many modes, both using model-dependent (see for instance \cite{PhysRevLett.105.121801,PhysRevD.81.112002,LHCb-PAPER-2014-017})  and model-independent approaches~\cite{PhysRevD.85.112014,LHCb-PAPER-2014-041,10.1093/ptep/ptw030,LHCb-PAPER-2016-006,LHCb-PAPER-2019-044}. The most precise measurement of $\gamma$ from a single analysis comes from the LHCb collaboration, using the full run I and part of run II data,  with a model-independent approach for the decay modes $\D\to\KS\pip\pim$ and $\D\to\KS\Kp\Km$  \cite{LHCb-PAPER-2018-017}, to get  $\gamma=(87^{+11}_{-12})^\circ$.
Belle collaboration has very recently performed the first GGSZ analysis using the four-body mode $\D\to\KS\pip\pim\piz$ \cite{Resmi:2019ajp}.

\subsection{The combination result of $\gamma$}

\begin{figure}[htb]
\centering
\includegraphics[width =0.5\textwidth]{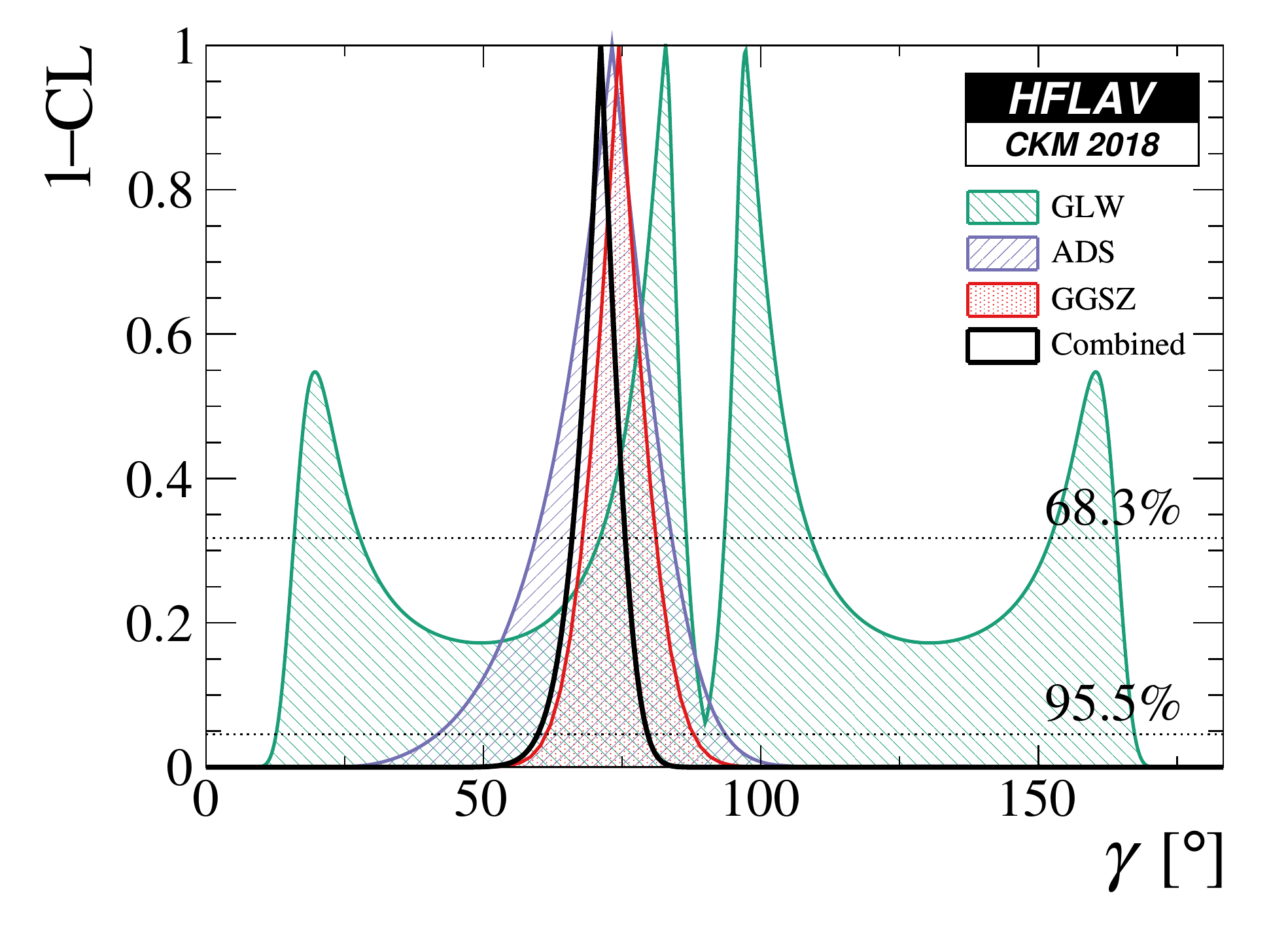}
\caption{World average of $\gamma$ split by method ~\cite{HFLAV19}.}
\label{gamma_hflav}
\end{figure}

From combining the values of $\gamma$ obtained by the different methods and different final states, the world average from HFLAV \cite{HFLAV19} is
\begin{equation}
\gamma = (71.1^{+4.6}_{-5.3})^\circ~, 
\end{equation}
which is also shown in Fig.~\ref{gamma_hflav} separated for GLW, ADS and GGSZ methods. The best sensitivity is clearly seen to come from the GGSZ method. Overall the LHCb measurements largely dominate the result. 

When compared to indirect determinations coming from fits with CKM constraints, such as from CKMFitter~\cite{CKMfitter2015} and UTFit~\cite{UTfit-UT} collaborations, the central value of $\gamma$ above is  higher and marginally compatible. The smaller indirect values are driven by the determination of the mass split in $\B-\Bb$ system through mixing measurements~\cite{BlankeBuras_2018,KingLenz_2019}. If NP effects appear at tree level, the value of $\gamma$ from direct determination can differ substantially from that of indirect measurements~\cite{BrodLenz_2014}. The expected precision of about $1^\circ$ from the LHCb Upgrade and Belle II in the upcoming years will be able to tackle this apparent tension.

\section{Direct \CP violation in Charm}
\label{chapter7}

As discussed in Sect.~\ref{charm_cap2},  \CP violation in the charm sector is highly suppressed from the CKM ansatz. Nevertheless, during the last two decades, one could witness the charm available samples to grow enormously: in the early 2000s there were just a few thousand reconstructed $\Dp\to \Kp\Km$ in FOCUS \cite{FOCUS_CPV_2000} and CLEO \cite{CLEO_CPV_2002} experiments,  while the current available sample from LHCb in run II comprises tens of millions decays in the same channel \cite{LHCb-PAPER-2019-006,LHCb-PAPER-2019-032}. 

By the end of 2011, there appeared some evidence for \CP violation through the \CP asymmetry difference in the decays of \Dz\to\Kp\Km and \Dz\to\pip\pim, first announced by the LHCb experiment \cite{LHCb-PAPER-2011-023} then by CDF~\cite{CDF_2_DCPV_D_KK_pipi_2012} and Belle~\cite{Belle_DACP_2012}. Although the effect soon after faded  with new measurements from LHCb~\cite{LHCb-PAPER-2013-003, LHCb-CONF-2013-003} it created a lot of excitement in the theoretical community in order to try to understand whether the effect --  almost at the percent level -- could still be explained by the SM, via penguin enhancements, $SU(3)_F$ breaking, long-distance effects \cite{Brod_ConsistentPenguins_2012,Brod_SizePenguins_2012,Grossman_Robinson_JHEP_2013,Atwood_Soni_2013,Cheng_Chiang_2012_2,Cheng_Chiang_2012_1,Cheng_2012,Pirtskhalava:2011va,Bhattacharya:2012ah,Li_2012,Hiller_2013,Feldmann:2012js,Franco:2012ck}
 or would require the contribution of new physics dynamics ~\cite{DaRold:2012sz,Hiller:2012wf,Altmannshofer:2012ur,Isidori:2011qw,Rozanov:2012zz,Grossman:2012eb,Wang:2011uu,Hochberg:2011ru,Chang:2012gna,Giudice:2012qq,Delaunay:2012cz,Chen:2012am,Chen:2012usa}. This huge amount of effort, the ``$\Delta A_\CP$ saga'', has stimulated further the quest for \CP violation in charm. 
 
Now direct \CP violation has finally been established from the same observable, by the LHCb experiment in 2019~\cite{LHCb-PAPER-2019-006}. This is a milestone in Particle Physics and it is natural to expect that the upcoming years, with further analyses in different decay modes, and increasing data samples from LHCb and Belle II,  will bring more surprises in the charm sector. 
In the following we present the state-of-the-art on study and search for direct \CP violation.

\subsection{Two-body modes}
\label{Dto2body}

The decays of \D mesons to two particles offer the simplest topology for \CP violation studies in charm. The majority of the two-particle final states  are composed of pseudo-scalar (stable) mesons; a few ``quasi'' two-body modes can also be addressed. Although a few decay modes, such as $\Dp\to \phi\pip$ and $\Dp\to\Kp\Km$, have reached the impressive sensitivity of a few to several $10^{-4}$, most of the measurements are still testing for asymmetries at the level of percent. There is yet plenty of room for searches of NP effects  with run II LHCb data as well as with the upcoming data from  Belle II and the upgraded LHCb.

\subsubsection*{$\Dz\to h^-h^+$}

As mentioned previously, a big breakthrough has come recently with the announcement, from the LHCb collaboration, of the first observation of \CP violation in the charm sector 
\cite{LHCb-PAPER-2019-006}. The measurement was made through the Cabibbo-suppressed decay modes \decay{\Dz}{\Km\Kp} and \decay{\Dz}{\pim\pip}. Since the final states are \CP symmetrical -- and thus reachable from both \Dz and \Dzb  mesons -- these decay modes permit, in principle, direct \CP violation as well as through mixing and interference between mixing and decay. 

The \Dz\to\Km\Kp and \Dz\to\pim\pip decays are among the most important channels for the study of \CP violation due to their relatively large branching ratios and, from the experimental side, the simplicity of two charged particles in the final state. Within the past decade or so, the search for direct \CP violation with \decay{\Dz}{\Km\Kp} and \decay{\Dz}{\pim\pip} decays had contributions from Babar~\cite{BaBar_DCPV_D2h_2008}, Belle~\cite{Belle_DCPV_D2h_2008}, CDF~\cite{CDF_1_DCPV_D_KK_pipi_2012, CDF_2_DCPV_D_KK_pipi_2012} and LHCb~\cite{LHCb-PAPER-2011-023, LHCb-PAPER-2013-003, LHCb-PAPER-2014-013, LHCb-PAPER-2015-055}.  

As for any neutral mode, the key experimental ingredient is the determination of the \D flavour (\Dz or \Dzb) at production -- the {\it tagging} strategy. Usually the decay  $\Dstarp\to \Dz\pi^+_s$ is used ($ \pi^+_s$ notation used since this emerging pion is slow) and this is referred to as $\pi$-tagging. LHCb also uses \D decays coming from semi-leptonic \B decays, $B\to\Dz\mu^-X$ (where $X$ is whatever particles not reconstructed in the decay) where the  Z.                                                                                                                                                                                                      muon charge gives the flavour of the \D ~-- referred to as $\mu$-tagging.

When studying these modes, an interesting observable is the difference of the time-integrated \CP asymmetries 
\begin{equation}
\Delta A_\CP(\kaon\kaon-\pion\pion) = A_\CP(\decay{\Dz}{\Km\Kp}) - A_\CP(\decay{\Dz}{\pim\pip})~,
\label{dacp}
\end{equation}
which is a robust measurement for two reasons: production asymmetries are expected to cancel out since they do not depend on the final state; and detection asymmetries, due only to the flavour-tagging particle (since the final states are \CP symmetric) are also expected to cancel out. Besides it is quite possible the difference in the \CP asymmetries to be enhanced comparatively to that of the individual asymmetries. For instance, under the limit of exact  U-spin symmetries the decays \decay{\Dz}{\Km\Kp} and \decay{\Dz}{\pim\pip} would have \CP asymmetries equal in size and opposite in sign \cite{Grossman_CPV_CS_2007,Hiller_2013}.  The two final states are also connected through final-state rescattering, as discussed in Sect.~\ref{chapter2_CP_CPT}, and the level of \CP violation in one may be correlated to the other.

The time-integrated \CP asymmetry is expected to be mostly sensitive to direct \CP violation \cite{Gersabeck_Interplay_2012,CDF_1_DCPV_D_KK_pipi_2012}, 
\begin{equation}
A_\CP(f) = a_\CP^{\rm dir}(f) - \frac{\langle t(f)\rangle}{\tau(\Dz)}A_\Gamma~.
\end{equation}
The direct \CP-violating term $a_\CP^{\rm dir}(f)$ depends on the final state $f$; $\langle t(f)\rangle$ is the measured (reconstructed) mean decay time, and $A_\Gamma$ is the indirect \CP violation parameter, currently compatible with zero at the per-mille level \cite{LHCb-PAPER-2016-063} and not expected to be final-state dependent. Thus, one can write 
\begin{equation}
\Delta A_\CP(KK-\pi\pi) = \Delta a_\CP^{\rm dir}(KK-\pi\pi) - \frac{\Delta\langle t\rangle}{\tau(D^0)}A_\Gamma~,
\end{equation}
where $\Delta\langle t\rangle$ is the difference of the reconstructed decay times of \decay{\Dz}{\Kp\Km} and \decay{\Dz}{\pim\pip} modes. 

\begin{figure}
\begin{center}
\includegraphics[width=0.3\textwidth]{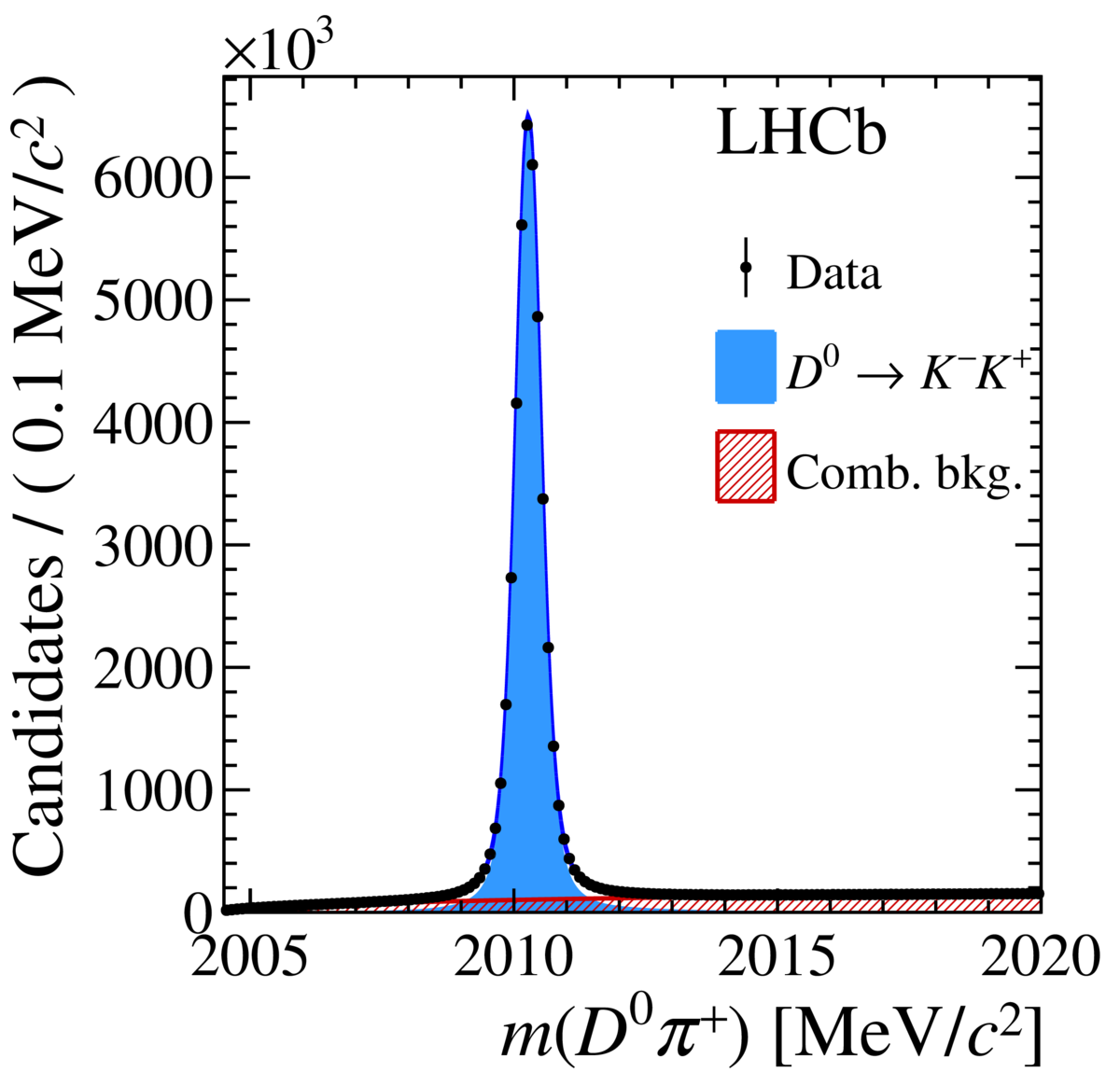} \includegraphics[width=0.3\textwidth]{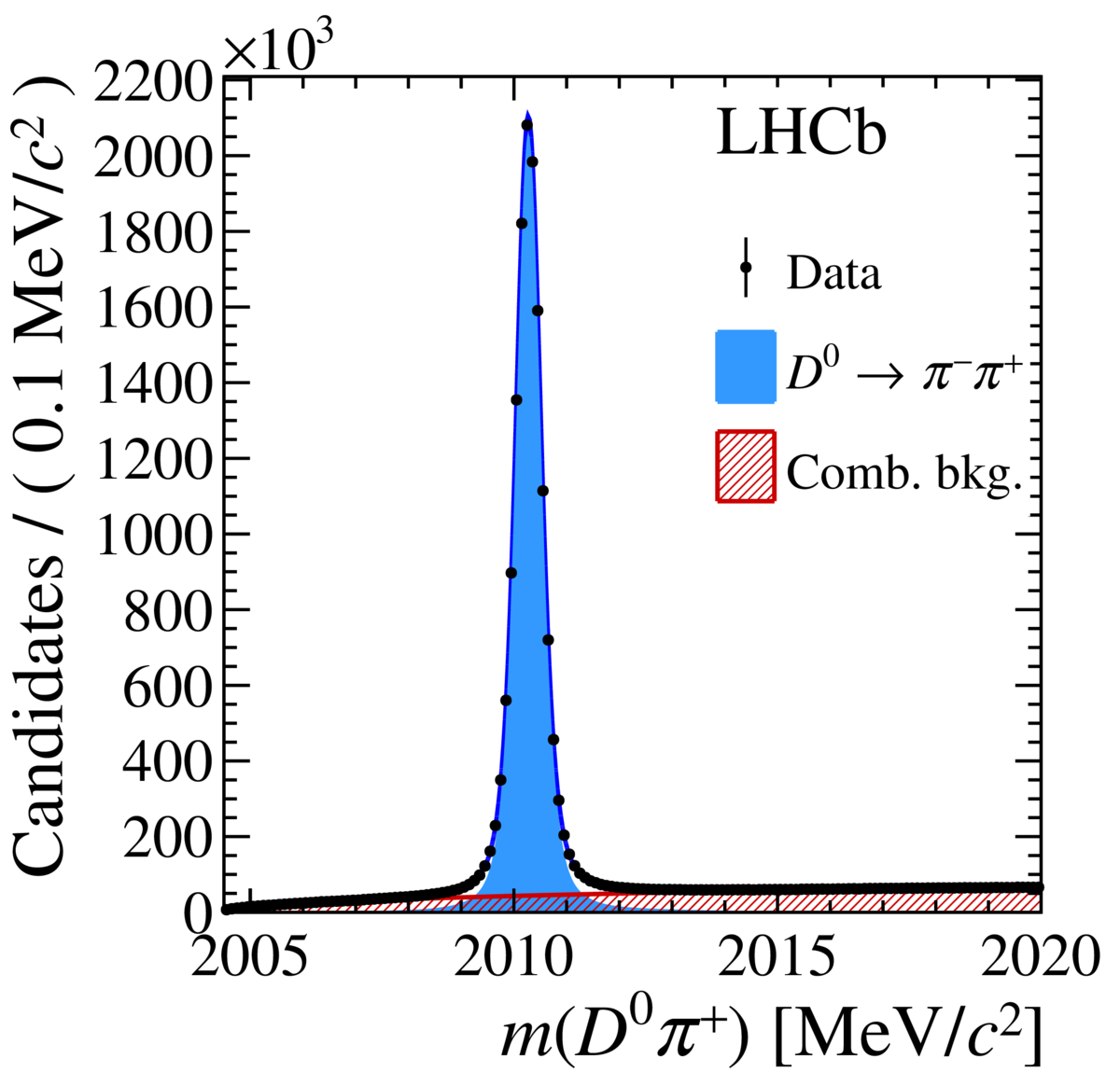}

\includegraphics[width=0.3\textwidth]{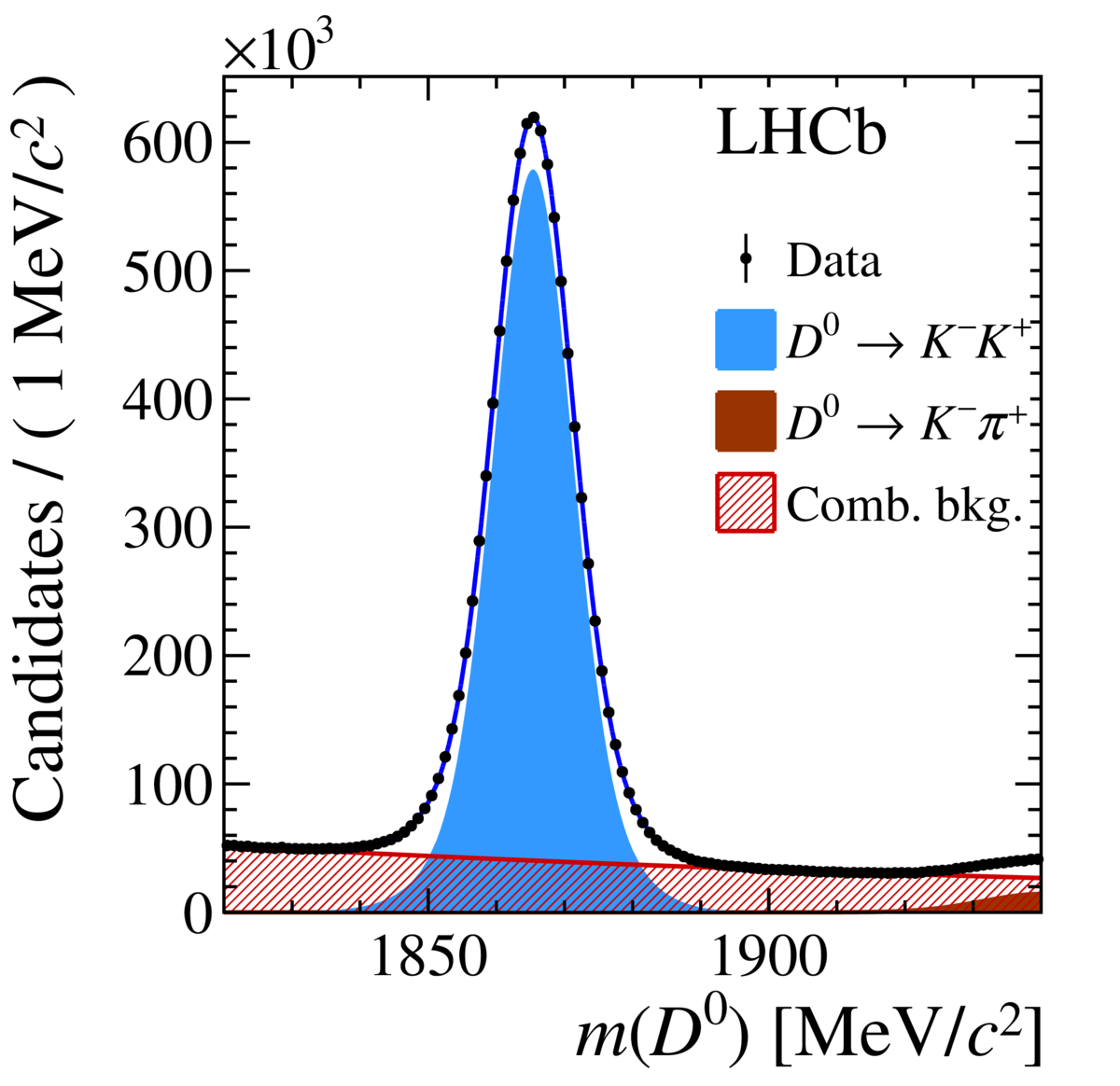} \includegraphics[width=0.3\textwidth]{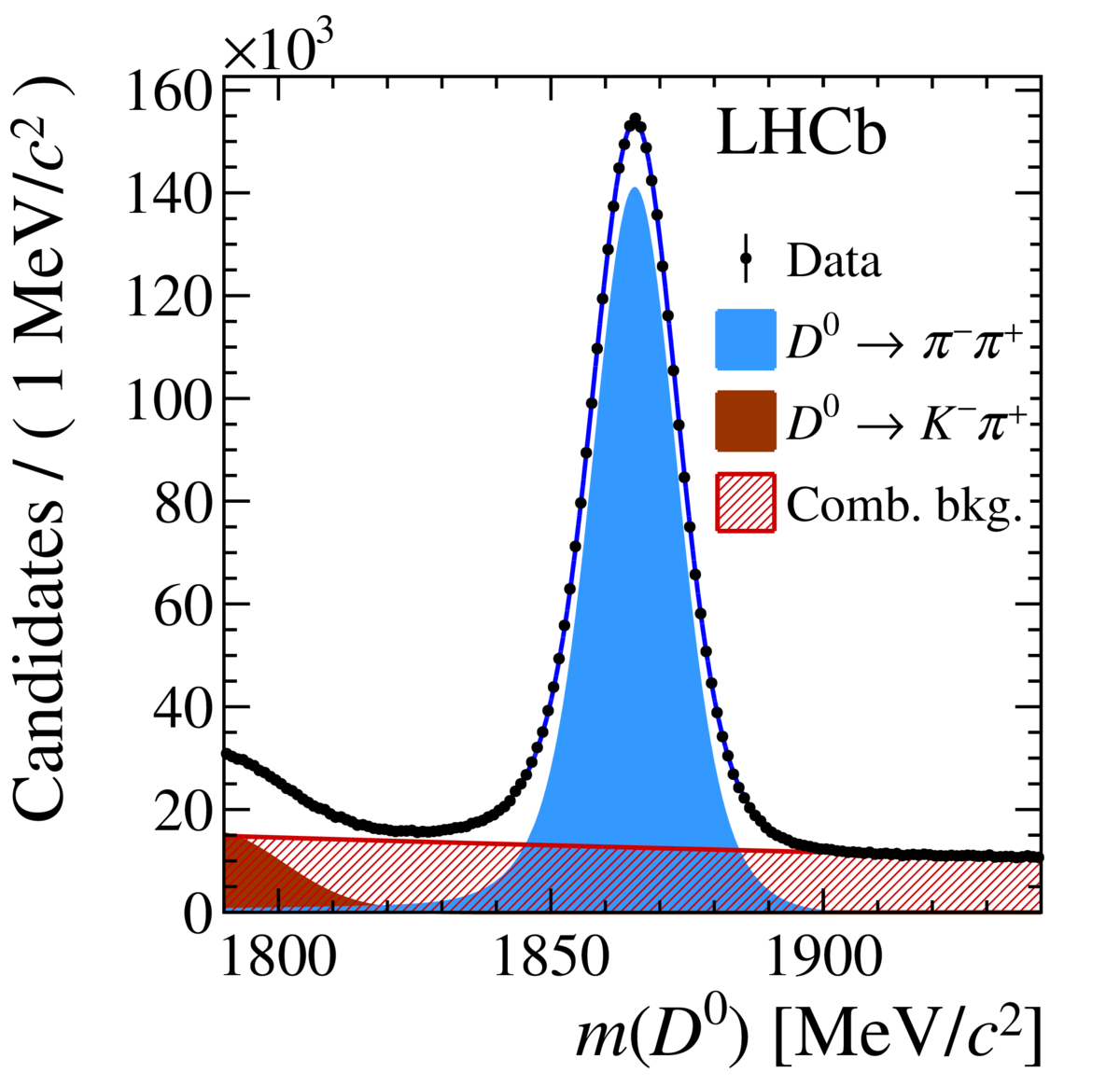}
\end{center}
\caption{Invariant mass distributions of (left) $\Dz\to\Kp\Km$ and (right)  $\Dz\to\pip\pim$ candidates from (top) $\pi$-tagged and (bottom) $\mu$-tagged samples. For the $\pi$-tagged samples the invariant mass corresponds to that of the $\Dstarp\to \Dz\pip$ candidate, while for the $\mu$-tagged sample it corresponds to that of the $\Dz$ candidates~\cite{LHCb-PAPER-2019-006}.}
\label{LHCb_DKK_Dpipi}
\end{figure}

Using run II data, the LHCb collaboration analysed large samples of 44 (9) million decays of $\Dz\to\Kp\Km$ and 14 (3) million decays of $\Dz\to\pip\pim$ for the $\pi$-tagged ($\mu$-tagged) samples, as shown in Fig.~\ref{LHCb_DKK_Dpipi}. 

\begin{figure}
\centerline{\includegraphics[width=0.5\textwidth]{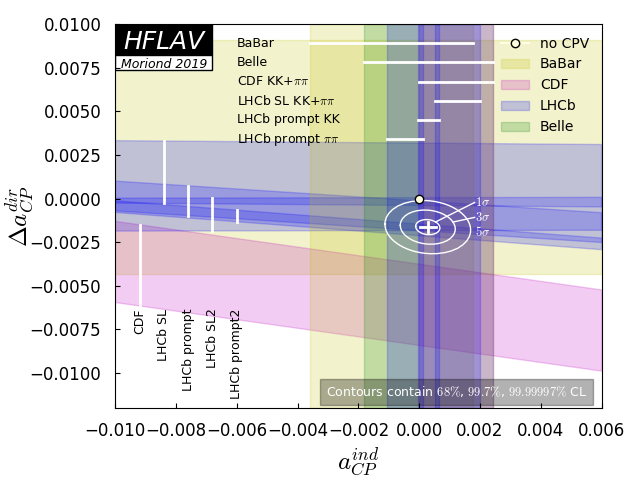}}
\caption{The combination plot for $\Delta a_\CP^{\rm dir}(KK-\pi\pi)$ against $a_\CP^{\rm ind}(KK-\pi\pi)$ from the HFLAV group~\cite{HFLAV19}.}
\label{DeltaaCP_hflav}
\end{figure}

The measured values obtained are  $\Delta A_\CP(KK-\pi\pi)^{\rm \pi-tagged} =(-18.2\pm 3.2\pm 0.9)\times 10^{-4}$ and $\Delta A_\CP(KK-\pi\pi)^{\rm \mu-tagged} =(-9\pm 8\pm 5)\times 10^{-4}$, for $\pi$-tagged and $\mu$-tagged samples, respectively. By performing a full combination of these results with previous LHCb measurements~\cite{LHCb-PAPER-2011-023, LHCb-PAPER-2013-003, LHCb-PAPER-2014-013, LHCb-PAPER-2015-055},  the final result is
\begin{equation}
\Delta A_\CP(KK-\pi\pi) =(-15.4\pm 2.9)\times 10^{-4}, 
\end{equation}
which is a $5.3\sigma$ effect and constitutes the first observation of \CP violation in the charm sector. 
After correcting by the small effect of indirect \CP violation, using the experimental value of $\Delta\langle t\rangle$ and the measurement of $A_\Gamma$ from LHCb~\cite{LHCb-PAPER-2016-063}, the value for direct \CP violation obtained is $\Delta a_\CP^{\rm dir}(KK-\pi\pi) = (-15.7\pm 2.9)\times 10^{-4}$. By performing a world average combination the HFLAV group~\cite{HFLAV19} gets
\begin{equation}
\Delta a_\CP^{\rm dir}(KK-\pi\pi) = (-1.64\pm 0.28)\times 10^{-3},
\end{equation}
which can also be observed from the combination plot of direct $\Delta a_\CP^{\rm dir}(KK-\pi\pi)$ and indirect $a_\CP^{\rm ind}(KK-\pi\pi)$ \CP asymmetry results  in Fig.~\ref{DeltaaCP_hflav}.

Besides the \CP asymmetry differences, it is clearly important to access the individual \CP asymmetries.
The LHCb collaboration measured $A_\CP(\Km\Kp) $ \cite{LHCb-PAPER-2016-035} with $\pi$-tagged samples using run I data, with 3.0\invfb  -- but results are yet to come from run II.  

The  value for $\Delta a_\CP^{\rm dir}(KK-\pi\pi)$  puts \CP violation in the charm sector at the level of $10^{-3}$, which is higher than naive expectations at ${\cal O}(10^{-4})$ ~\cite{Grossman_CPV_CS_2007,Khodjamirian_Petrov_2017} and some authors consider contributions from NP dynamics \cite{Nir_Implications_CharmCPV_2019,Chala_2019,Calibbi_2019}. Nevertheless it is
consistent with models that include non-perturbative effects, with QCD enhancements and $SU(3)_F$ symmetry breaking ~\cite{Buccella_2019,Cheng_2019,Li_ImplicationsCPV_2019,Grossman_2019}.  An interesting effect discussed in ~\cite{Soni_CPenhancement_2019} considers  \CP violation enhancement due to a nearby resonance, $f_0(1710)$ playing the role of the {\it communicator} between the channels and, being close to the \Dz mass, providing the necessary strong phase.

 One should consider more generally the potential effects from $KK\leftrightarrow\pi\pi$ rescattering~\cite{Bianco_Bigi_2020}. This was discussed in Sect.~\ref{kkpipiresc} in the context of charmless three-body \B decays and, for charm decays, rescattering can be expected to be even more important. In this sense, three-body  Cabibbo-suppressed decays, especially those with  large samples such as  $\Dp\to \Km\Kp\pip$ and $\Dp\to\pim\pip\pip$, are excellent candidates for observation of \CP violation with the already available data from LHCb, as it is discussed below in Sect.~\ref{Dto3body}.  

 In any case, this outstanding result from $\Dz\to\Kp\Km$ and $\Dz\to\pip\pim$ decays calls for a  comprehensive study of potential charm processes where direct \CP violation might appear at a similar level. 

\subsubsection*{$\D^+_{(s)}\to h^0 h^+$}

Recently, the LHCb collaboration has searched for CP violation in the Cabibbo-suppressed decays $\Dsp\to\KS\pip$, $\Dp\to \KS\Kp$ and $\Dp\to \phi(1020)\pip$ with $3.8\,\rm fb^{-1}$ from run II $pp$ collision data \cite{LHCb-PAPER-2019-002}. The  data samples for these decays are huge, comprising respectively 600 thousand, 5.1 million and 53.3 million candidate decays. The \KS meson is reconstructed through its \pip\pim final state, which is \CP symmetric. The $\Dp\to \phi\pip$ constitutes a quasi-two-body mode; the measurement is made selecting \Kp\Km pairs in the vicinity of the \Pphi mass with no attempt to separate from other contributions of the decay $\Dp\to\Km\Kp\pip$. To allow the measurement of the \CP asymmetry, LHCb uses a set of Cabibbo-favoured control modes, $D^+\to K_S^0\pi^+$, $D_s^+\to K_S^0K^+$ and $D_s^+\to \phi(1020)\pi^+$, in order to provide a cancellation of production and detection asymmetries.  As such, the \CP asymmetries to a good approximation are given by
\begin{eqnarray}
A_{\CP}(\Dsp\to \KS\pip)&  \approx&  A_{\rm raw}(\Dsp\to \KS\pip) - A_{\rm raw}(\Dsp\to \phi\pip)\nonumber  \\
A_{\CP}(\Dp\to \KS\Kp) &  \approx&   A_{\rm raw}(\Dp\to \KS\Kp) - A_{\rm raw}(\Dp\to \KS \pip)\nonumber \\  
 & - & A_{\rm raw}(\Dsp\to \KS\Kp)  + A_{\rm raw}(\Dsp\to \phi\pip)  \\
A_{\CP}(\Dp\to \phi\pi^+) & \approx&    A_{\rm raw}(\Dp\to \phi\pip) - A_{\rm raw}(\Dp\to \KS\pip)~, \nonumber
\end{eqnarray}
where $A_{\rm raw}$ are the raw asymmetries from  $\D_{(s)}^+$ and $\D_{(s)}^-$ yields. 
A weighting procedure to match the distribution of kinematic quantities is performed to guarantee accurate cancellations. In the final states with a \KS state, both its detection and \CP asymmetries are subtracted from the measured values \cite{LHCb-PAPER-2014-013}. The results obtained for 2015--2017 data \cite{LHCb-PAPER-2019-002} are combined to previous LHCb results from run I, leading to the measurements
\begin{eqnarray}
A_{\CP}(\Dsp\to \KS\pip) & =& (1.6\pm 1.7\pm 0.5)\times 10^{-3} \nonumber\\
A_{\CP}(\Dp\to \KS\Kp)  & =& (-0.04\pm 0.61\pm 0.45)\times 10^{-3} \nonumber\\
A_{\CP}(\Dp\to \phi\pi^+)& =& (0.03\pm 0.40\pm 0.29)\times 10^{-3}~,
\end{eqnarray}
which  are all consistent with \CP conservation. For the $D^+\to \phi\pi^+$ mode, the result represents  the most precise \CP asymmetry measurement obtained from a single measurement, with a combined statistical and systematic uncertainty of only $5\times 10^{-4}$. The decays of \Dp and \Dsp to \KS\Kp have  also been studied by the BESIII collaboration in 2019~\cite{BESIII_DsKSLKp_2019} with a much smaller sample, so LHCb dominates these measurements. The novelty is the inclusion of \KL\Kp final states -- with results consistent with \CP conservation at $\sim 3\%$ level. 

Other interesting modes for charged \D's involve the production of \etapr, such as \Dp\to\etapr\pip, which is Cabibbo-suppressed, and \Ds\to\etapr\pip, which is Cabibbo-favoured. A few years ago, Belle~\cite{Belle_Detaprpi_2011} and CLEO \cite{CLEO_Dsetaprpi_2013} have measured \CP asymmetries for \Dp\to\etapr\pip and \Ds\to\etapr\pip above the 1\% and 2\% level, respectively. Despite the difficulty of reconstructing  \etapr in a hadron machine, the  LHCb collaboration has performed a study for these decays with 3\invfb from run I data \cite{LHCb-PAPER-2016-041}. Control modes \Dp\to\KS\pip and \Dsp\to\Pphi\pip are used to deal with detection and production asymmetries. The results
\begin{eqnarray}
A_\CP(\Dp\to\etapr\pip)  = (-0.61\pm 0.72\pm 0.53\pm 0.12)\% \nonumber \\
A_\CP(\Dsp\to\etapr\pip)  = (-0.82\pm 0.36\pm 0.22\pm 0.27)\%
\end{eqnarray}
are the most precise to date. The last uncertainty is due to external measurements of the asymmetry of the control modes.

The Cabibbo-suppressed mode $\Dp\to\pip\piz$ has the same dominant tree diagram as \Dz\to\pim\pip, with the only change in the spectator quark. 
But since the final state has isospin $I=2$, it cannot proceed through a gluonic-penguin amplitude. Thus, in this case there is no weak phase contributing from the SM and the expected \CP asymmetry is null;  any sign of \CP violation would be indication of New Physics mechanisms.
 In 2018, Belle searched for  \CP violation in this mode with its full  data sample, with an integrated luminosity of 921\invfb sample \cite{Belle_Dpipi0_2018}. The $\Dp\to\pip\piz$ decay was reconstructed both from $\Dstarp\to\Dp\piz$ (tagged) and directly from the interaction point (not tagged). The raw asymmetry $A_{\rm raw}^{\pip\piz}$ was obtained from signal yields, and corrected by the \pip detection asymmetry and forward--backward asymmetry.\footnote{Forward--backward asymmetry arises due to the interference of amplitudes mediated by virtual photon, \Z and higher orders in \en\ep collisions.} 
The result,
\begin{equation}
A_\CP(\Dp\to\pip\piz) = (+2.3\pm 1.2\pm 0.2)\%,  \nonumber
\end{equation}
is consistent with the SM and with \CP conservation at the 1\% level, which still places no strong constraint. The LHCb collaboration may be able to provide results from run II with precision at the level of Belle. Belle II, with the full $50{\rm ab}^{-1}$, will be able to reach the 0.2\% level~\cite{BelleII_PhysicsBook_2018}.

\subsubsection*{$\Dz\to h^0h^0$}

The $\Dz\to\KS\KS$ channel has a small branching fraction, $(1.41\pm 0.05)\times 10^{-4}$~\cite{PDG2019}, about 30 times smaller than the $\Kp\Km$ mode. This is due to the fact that it can only occur through  $W$-exchange diagrams with $\cquark\uquarkbar\to \squark\squarkbar$ and  $\cquark\uquarkbar\to \dquark\dquarkbar$ which would  cancel in the limit of exact  $SU(3)_F$  symmetry, or via a penguin-type annihilation. The contributions, with different  weak phases, are expected to be of similar size, so the interference term is comparatively larger than for other charm modes. According to Ref.~\cite{Nierste_DKSKS_2015}, the \CP asymmetry could be as large as 1\% within the SM. 

Within the last 5 years, the search for \CP violation in this mode  was performed by LHCb~\cite{LHCb-PAPER-2015-030, LHCb-PAPER-2018-012} and by Belle~\cite{Belle_D0KSKS_2017}  -- the latter leading with best precision so far, $A_\CP(\Dz\to\KS\KS) = (-0.02\pm 1.5)\%$. The combined uncertainty, though, is still not sufficient to test the SM prediction. Belle II expects to lower this uncertainty \cite{BelleII_PhysicsBook_2018} to 0.66\% and 0.23\% with integrated luminosities of $5\,\rm ab^{-1}$ and  $50\, \rm ab^{-1}$, respectively.

More recently the authors of Ref.~\cite{Nierste_DKSKS_2015} extended the work to study the quasi two-body decay \mbox{$\Dz\to\KS\Kstar$} \cite{Nierste_DKKst_2017} to estimate a \CP asymmetry that can be as high as  $0.3\%$. In any case, the adequate treatment for this mode should be addressed through a full Dalitz plot investigation to the \Dz\to\KS\Km\pip final state. This is discussed in Sect.~\ref{Dto3body}. 

The $\Dz\to\piz\piz$ is another interesting channel and it gains interest also with its connection  to $\Dz\to\Kp\Km$ and $\Dz\to\pip\pim$ final states (see for instance  \cite{MullerNierste_2015}).
Belle studied the decay \Dz\to\piz\piz with 966\invfb, with a sample of 34 thousand signal events, and also studied the Cabibbo-favoured \Dz\to\KS\piz, with a sample of 467 thousand signal events~\cite{Belle_Dpi0pi0_2014}. The results are $A_\CP(\Dz\to\piz\piz) = (-0.03\pm 0.64\pm 0.10)\%$ and $A_\CP(\Dz\to\KS\piz) = (-0.10\pm 0.16\pm 0.07)\%$, where the last result is already corrected for \Kz asymmetries. The \piz\piz final state represents a challenge for LHCb but in the upcoming years Belle II is expected to bring improvement, with an expected sensitivity of $0.3\%$ with the first $5 \,\rm ab^{-1}$.

\subsection{Three-body modes}
\label{Dto3body}

As discussed in Sect.~\ref{multibody}, for multi-body decay modes \CP violation  effects can be studied through the phase space of the decay, and can be potentially larger than integrated ones. Even if, when compared to \B decays, the available phase space is relatively small, there are many resonances with masses below the \D mass, and usually the decay dynamics are very rich and with rather interesting interference patterns. \CP violation effects may be enhanced in local regions and these local asymmetries can be higher than the phase-space integrated results \cite{bigi1,Bigi_CP-2-3-4Body_2011,Bigi_Charm_2015}.

For \D decaying to stable pseudo-scalar mesons in a three-body final state,  the dynamics are directly read from the distribution of events in the Dalitz plot. In particular, interference effects are visually apparent. If sizeable localised \CP violation effects exist, they would be clearly visible, as for the case of three-body  charmless \B decays  discussed in Sect.~\ref{Bhhh}. For \D decays, with very small asymmetries expected within the SM, the search for \CP violation within the phase space requires very large samples and a very careful control of nuisance asymmetries, due to production and detection effects. 

Although a few Dalitz analyses searched for \CP violation more than 10 years ago with limited number of events, it was in 2008 when the Babar collaboration presented a search for \CP violation with about 82 thousand  $\Dz\to \pim\pip\piz$ and 11 thousand $\Dz\to\Km\Kp\piz$ decay candidates, with high purity~\cite{BaBar_CPVD3body_2008}. The study  was performed through both  amplitude analysis and model-independent search, by calculating the residuals of the difference in population in bins of the Dalitz plot for \Dz and \Dzb (similar to the Miranda technique presented in Sect.~\ref{chapter3}). No sign for \CP violation was found. Through the model-dependent approach, for the $\Dz\to \pim\pip\piz$ (with larger samples) the uncertainties for \CP violation in phases and magnitudes were about 2$^\circ$ and higher, and about $2\%$ and higher, respectively.

A more recent model-dependent study where \CP violation  was carried out by the LHCb collaboration in the channels $\Dz\to \KS\Km\pip$ and $\Dz\to\KS\Kp\pim$ \cite{LHCb-PAPER-2016-026}, using  3\invfb  (full run I data) with signal yields of about 113 thousand and 76 thousand, respectively.  No evidence for \CP violation was found, probing at or higher than 2$^\circ$ and $2\%$  for phase and magnitude differences, respectively.

\begin{figure}
\centerline{\includegraphics[width = 0.4\textwidth]{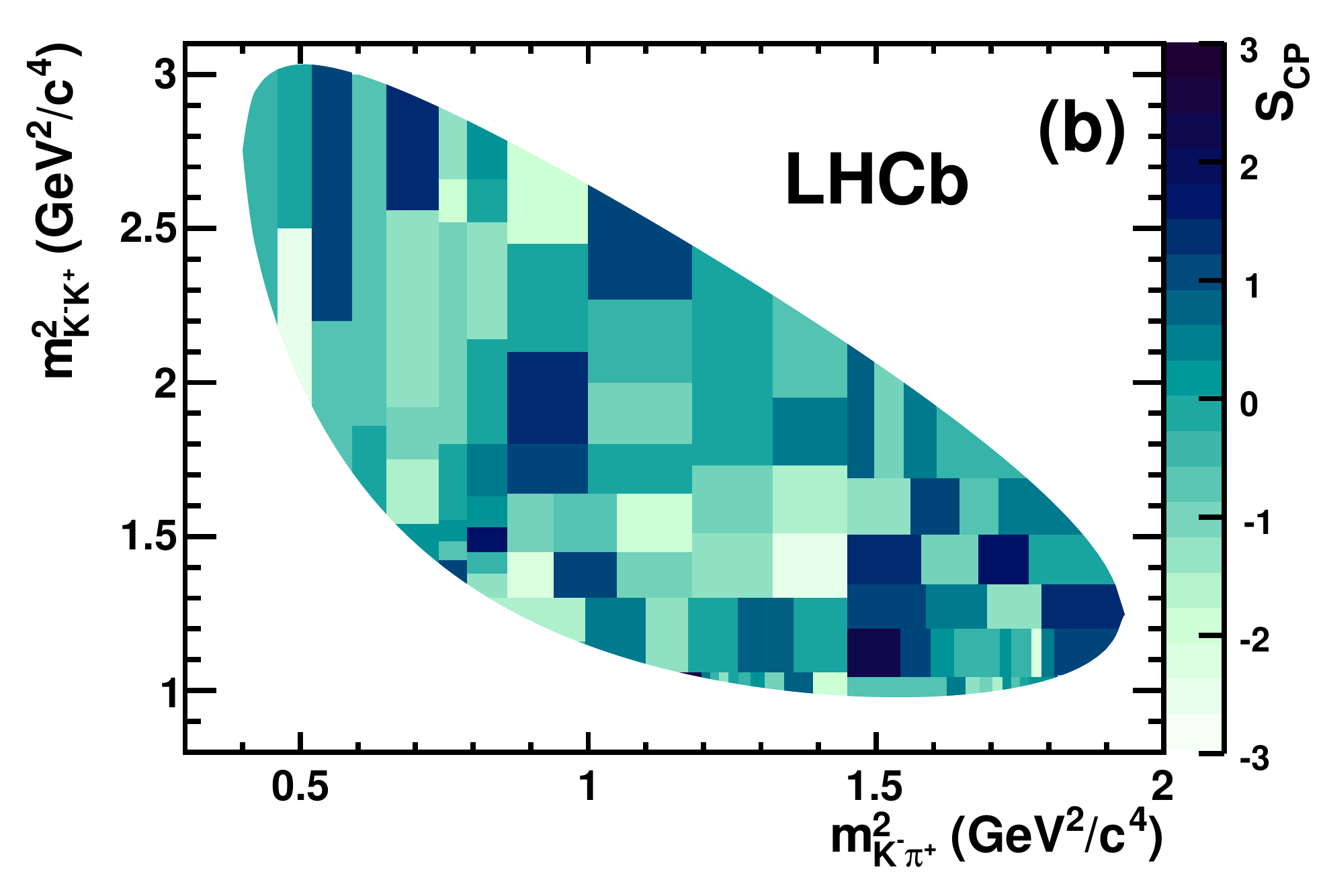} \includegraphics[width = 0.4\textwidth]{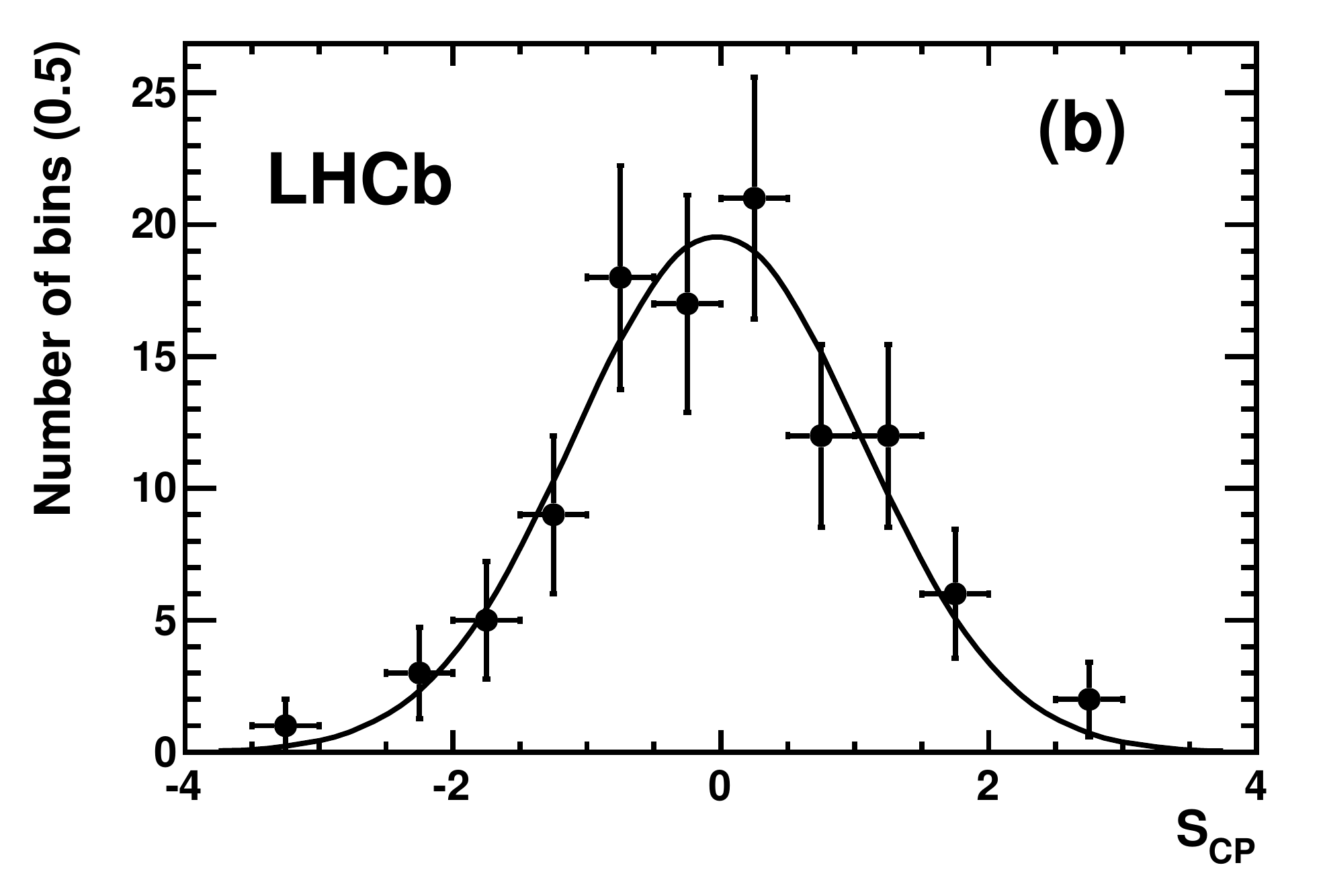}}
\caption{(Left) \CP asymmetry significances, $S_\CP$, across the Dalitz plot for candidate decays of $\Dp\to \Km\Kp\pip$ in 2010 data from the LHCb experiment. (Right) Distribution of the $S_\CP$ values. See \cite{LHCb-PAPER-2011-017} for details.}
\label{LHCb_KKpi_2011}
\end{figure}

Using the first set of data collected from the LHCb experiment in 2010, a search for \CP violation was performed for $\Dp\to \Km\Kp\pip$ decays  \cite{LHCb-PAPER-2011-017}. A total of about 370 thousand signal candidates were analysed. The approach used the Miranda technique ~\cite{bigi1} described in Sect.~\ref{miranda} where the Dalitz plot was divided in bins, both uniformly or adaptively. The Cabibbo-favoured decays $\Dsp\to \Kpm\Kp\pip$ and $\Dp\to\Km\pip\pip$, for which \CP violation is not expected,  were used as {\it control channels} to check for nuisance asymmetries, with no observable effects. The $\Dp\to \Km\Kp\pip$ Dalitz plot was then studied for a few combinations of binning schemes. In Fig.~\ref{LHCb_KKpi_2011} the significances are shown across the Dalitz plot for 106 adaptive bins; the corresponding $\chi^2\equiv \sum_i [S_\CP(i)]^2$ and number of degrees of freedom  resulted  in a p-value of about 10\%, compatible with the hypothesis of \CP conservation. Other variations in binning schemes lead to the same conclusion. 

\begin{figure}
\centerline{\includegraphics[width= 0.4\textwidth]{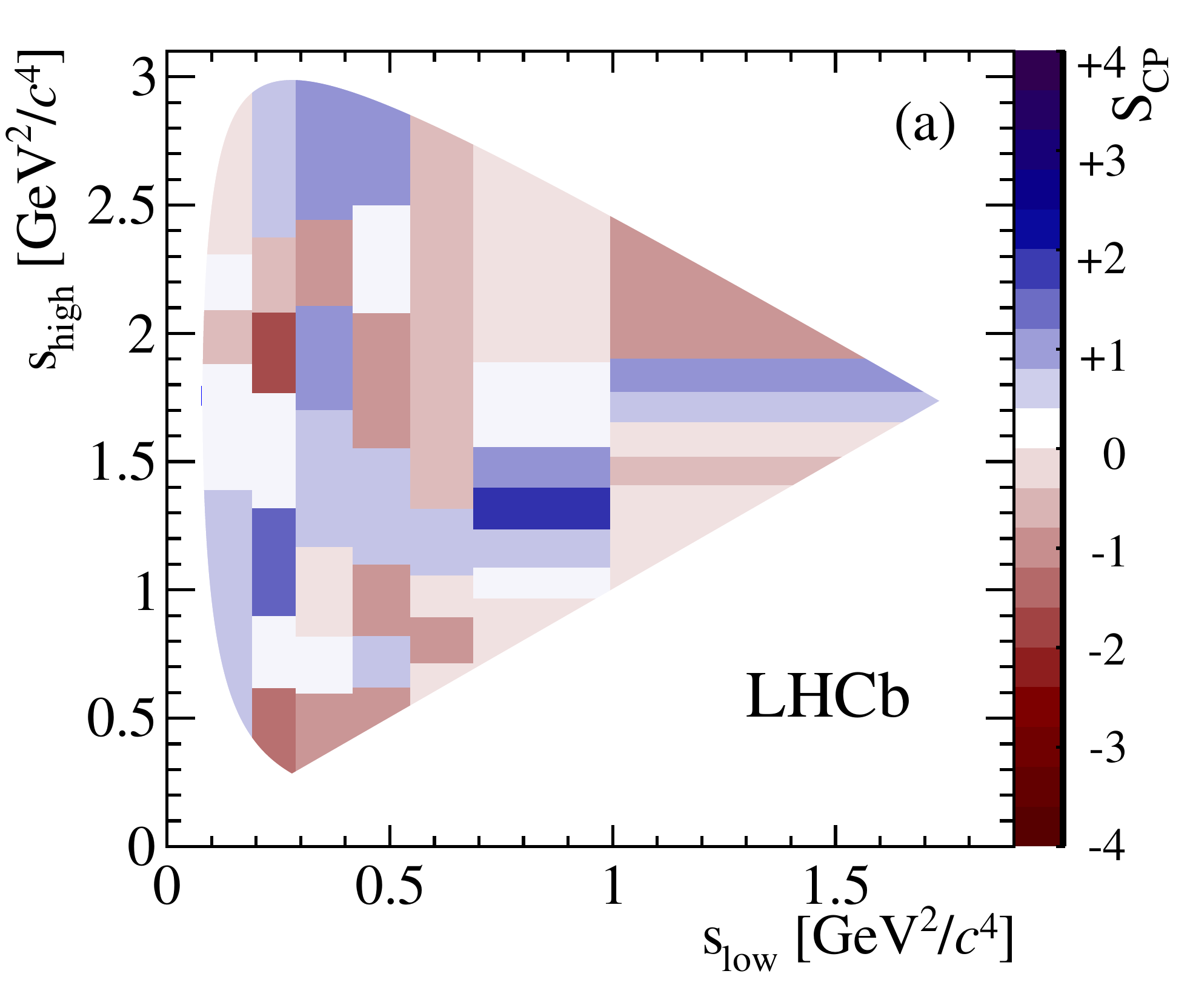}\includegraphics[width= 0.4\textwidth]{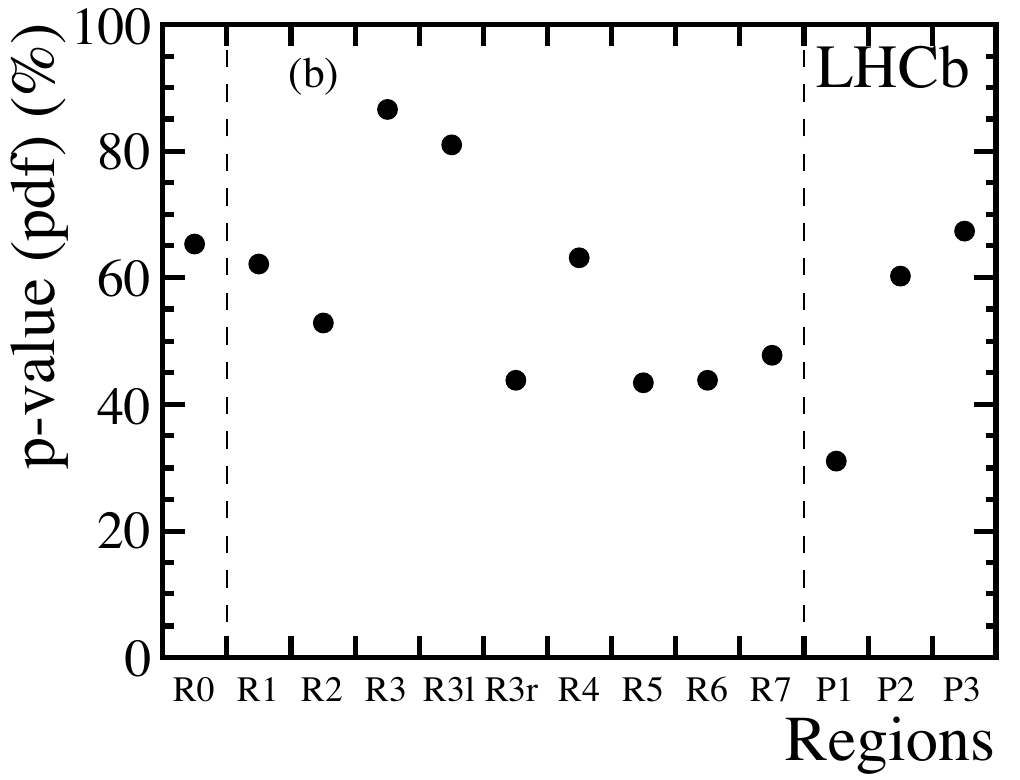}}
\caption{Results for the \CP violation search in $\Dp\to\pim\pip\pip$ decays by LHCb collaboration with 2011 data. (Left) $S_\CP$ distribution with 106 adaptive bins. (Right) Distribution of p-values of the test statistic $T$ for different regions in the Dalitz plot. See \cite{LHCb-PAPER-2013-057} for details.}
\label{LHCb_D3pi}
\end{figure}

The same technique was applied to $\Dp\to\pim\pip\pip$, with a sample of about 3 million candidates with 82\% purity~\cite{LHCb-PAPER-2013-057} collected by LHCb during 2011. Besides the binned Miranda approach, the kNN  unbinned technique, described in Sect.~\ref{unbinned},  was also used. Sensitivity studies based on pseudo-experiments showed that with the given data sample size  both methods would be sensitive to asymmetries of about 2\% and phase differences of about $1^\circ-2^\circ$ for specific resonant contributions ($\Dp\to \rho^0(770)$ for example). The modes $\Dsp\to \pim\pip\pip$ and $\Dp\to\Km\pip\pip$ were used as control modes, with no sign of local asymmetries. When analysing the signal region, no evidence for \CP violation in  $\Dp\to\pim\pip\pip$ was found. In Fig.~\ref{LHCb_D3pi}  the $S_\CP(i)$ distribution across the Dalitz plot is shown for 106 bins, as well as  the resulting p-values for the kNN method.

\begin{figure}
\centerline{\includegraphics[width = 0.5\textwidth]{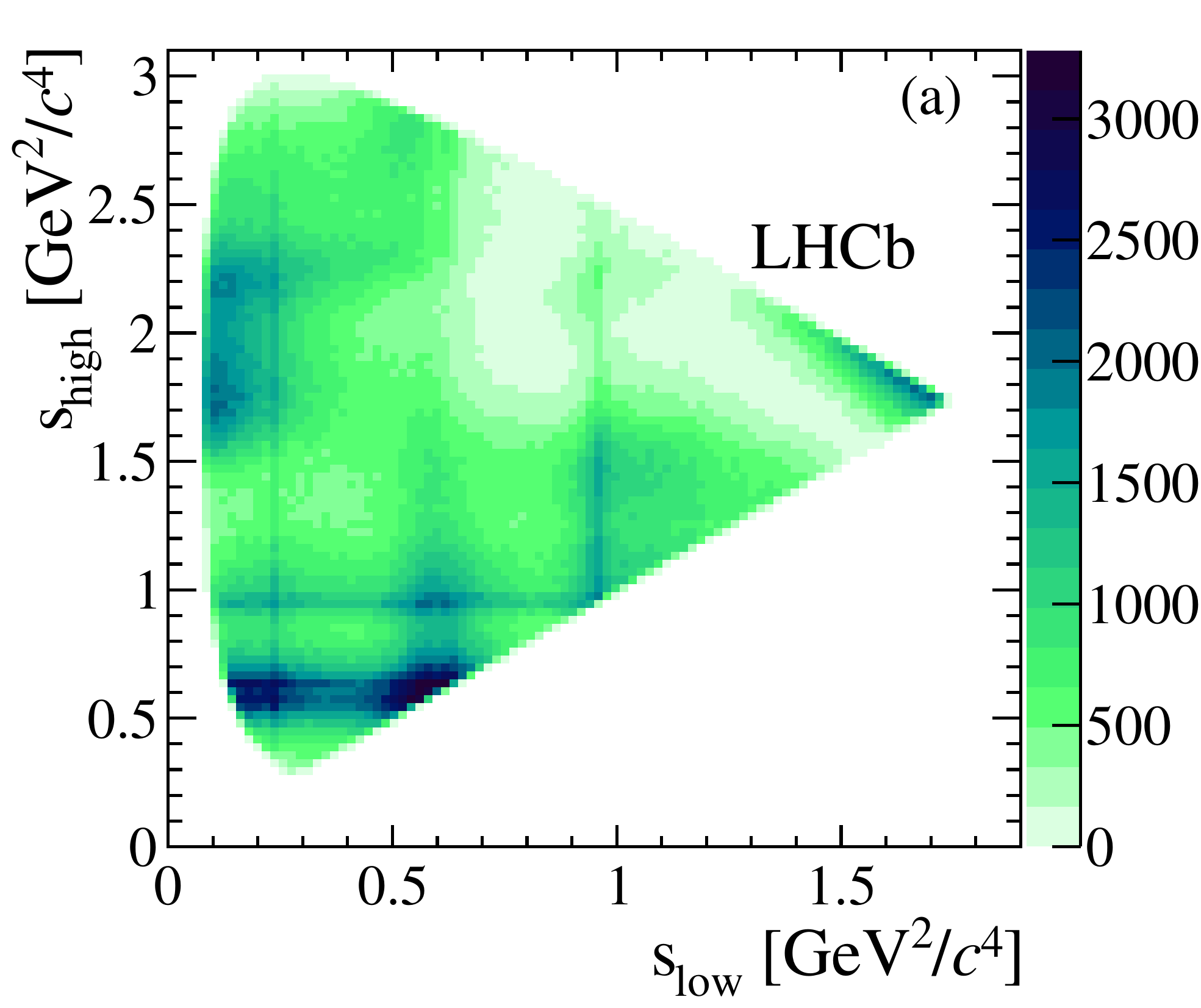}}
\caption{Dalitz plot distribution of $\Dp\to\pim\pip\pip$ candidate decays, from 2011 LHCb data. The Dalitz variables $s_{\rm low}$ and $s_{\rm high}$ correspond to $m^2(\pim\pip)_{\rm low}$ and $m^2(\pim\pip)_{\rm high}$, respectively ~\cite{LHCb-PAPER-2013-057}.}
\label{DalitzD3pi_LHCb}
\end{figure}

There are no results yet with LHCb run II data for \CP violation searches across the Dalitz plot for $\Dp\to\pim\pip\pip$ and $\Dp\to \Km\Kp\pip$ decays. The expected sample sizes are ${\cal O}(10^8)$ events~\cite{LHCb-PII-Physics} which would allow \CP asymmetries to be tested at the $10^{-3}$ level. 
Given the recent observation of \CP violation in $\Dz\to\Kp\Km$ and $\Dz\to\pip\pim$ it is likely  that direct  \CP violation would be soon observed in these decays too. Moreover, the study across the phase space, being potentially more sensitive to local asymmetries, may provide information regarding the eventual sources of \CP asymmetries,  e.g. resonance enhancements, S- and P- wave interference, and $\PK\PK-\pi\pi$ rescattering \cite{Bigi3,Bigi_Charm_2015,Bianco_Bigi_2020,Li_ImplicationsCPV_2019}. In Fig.~\ref{DalitzD3pi_LHCb} the (diagonal-folded) Dalitz plot distribution for $\Dp\to\pim\pip\pip$ is shown evidencing the richness of the dynamics of the decay --- in particular one can see the large interference between the low-mass S-wave, dominated by the $\sigma(500)\pip$ amplitude, and the $\rho(770)^0\pip$ amplitude which makes the higher mass lobe of the $\rho(770)^0$ to almost disappear. 

Another important and promising decay is $\Dz\to \pim\pip\piz$. With  data collected in 2012, the LHCb collaboration searched for  \CP violation in this decay. The sample comprised 660 thousand candidate decays and used the $\Dstar\to \Dz\pip$ decay to tag the flavour of \Dz at production.  The search was carried out in the Dalitz plot via the energy test technique described in Sect.~\ref{unbinned}. The metric function $\psi_{ij}$, defined in Eq.~\ref{T_energytest} was chosen as a Gaussian, $\psi_{ij} = e^{-d^2_{ij}/2\delta^2}$ where $d_{ij}$ is the ``distance'' (measured in \gevgevcccc) between the $i$ and $j$ events in the 3D-space spanned by the two-body  mass combinations of the three final-state particles.  From pseudo-experiments with the same size as the data, the energy test showed sensitivity of about 2--4\% in magnitudes and 1--3$^\circ$ in phases, depending on which intermediate amplitude \CP violation were introduced. When compared to the Miranda technique, the energy-test method showed similar or better precision. When applied to the $\Dz\to \pim\pip\piz$ data sample, the $T$ value obtained is consistent with \CP conservation, with a p-value of $(2.6\pm 0.5)\%$. A limitation of the method is the computation price: with samples of ${\cal O}(10^{7-8})$ events  (current sample sizes for some Cabibbo-suppressed channels in LHCb run II) the permutations needed to access the p-value can be very  time consuming.

\subsection{Four-body modes}
\label{Dto4body}

The main four-body  hadronic modes for the search for \CP violation are the Cabibbo-suppressed decays $\Dz\to\Kp\Km\pip\pip$ and $\Dz\to\pip\pim\pip\pip$.
The dynamics of four-body decay modes involving only mesons in the initial and final state are fully described by  a five-dimensional phase space. As in the case for three-body  modes, the search for \CP violation can be pursued in model-dependent and model-independent ways, with the further complication of dealing with 5 independent variables. Clearly, given the limited sizes of decay samples, the sensitivity to charge asymmetries may vary substantially depending on the phase space region. Since there is no hint in which region  \CP violation may be potentially enhanced, the study is challenging. 
For instance, for the Miranda technique bins should in principle be defined in 5 dimensions, preferably with adaptive binning to control the minimum number of events in each bin, with a careful study of the binning scheme in order not  to lose sensitivity. This also applies for the  energy-test technique where the parameter $\delta$ defines an effective radius in phase space. Model-dependent searches rely on a full amplitude analysis, usually using the Isobar model to disentangle the different intermediate contributions by fitting their magnitudes and phases separately for particle and antiparticle. 

As discussed in Sect.~\ref{tripleproducts} in a four-body  decay other \CP-violating observables can be accessed, such as the $a_\CP^{T-\rm odd}$ asymmetry defined in terms of a triple-product of momenta of the final-state particles, defined in Eqs.~\ref{C_T} to \ref{a_CP_Todd}. 
The first searches for \CP violation using this T-odd observable were carried out in the  2000s by the FOCUS~\cite{FOCUS_KKpipiTodd_2005} and BaBar~\cite{Babar_D0KKpipi_Todd_2010} collaborations with the final state $\Dz\to\Kp\Km\pip\pim$. In 2014, the LHCb collaboration   also performed a search for T-odd correlations with about 170 thousand $\Dz\to\Kp\Km\pip\pim$ signal decays from run I data with three complementary approaches~\cite{LHCb-PAPER-2014-046}: by measuring the phase-space integrated value of  $A_T$, $\bar A_T$ and $a_\CP^{T-\rm odd}$, as well as in different regions of the phase-space and in different decay-time bins. The quantity $C_T$ was defined with particles 1, 2 and 3 being respectively \Kp, \pip and \pim. The value obtained for $a_\CP^{T-\rm odd}$ was $(0.18\pm 0.29\pm0.04)\%$, consistent with \CP conservation, as well as the values obtained in bins of phase space and decay time. Interestingly enough, the integrated values of $A_T$ and ${\bar A}_T$ are $\approx (-7\pm0.5)\%$, with also relatively large variations across the phase space, pointing to the significant final-state interactions due to the rich resonant contributions. Belle has performed a similar study, for the first time for the decay $\Dz\to \KS\pip\pim\piz$, with a total of about 1.7 million candidates~\cite{Belle_D0Kspipipi0_2017}. This final state receives contributions from  both Cabibbo-favoured and doubly-Cabibbo-suppressed transitions. The result for phase-space integrated T-odd \CP asymmetry, as well in different regions of the phase space, has shown consistency with \CP conservation with a precision of 0.14\%. 

In a very recent analysis of the $\Dz\to\Kp\Km\pip\pip$ mode, Belle~\cite{Belle_D0KKpipiCPV_2019} has extended the concept of the asymmetry  $A_X$ for different kinematic variables $X$, so that 
\begin{equation}
a_X^\CP \equiv \frac{1}{2}\left( A_X - \eta_X^\CP {\bar A}_{\bar X}\right)~, 
\end{equation}
with $\bar A_{\bar X}$ defined the same way as $A_X$ (changing sign convention with respect to that  in Eq.~\ref{ATbar}) and $\eta_X^\CP$ is the \CP eigenvalue specific to $X$. The kinematic variables used were $\sin2\Phi$, $\cos\theta_1\cos\theta_2\sin\Phi$, $\sin\Phi$ (these with $\eta_X^\CP=-1$) , and $\cos\Phi$ and $\cos\theta_1\cos\theta_2\cos\Phi$ (with $\eta_X^\CP=+1$), where $\theta_1$ ($\theta_2$) is the angle of the \Kp (\pip) momentum and the direction opposite of the \Dz in the \Kp\Km (\pip\pim) rest frame, and $\Phi$ is the angle between the decay planes of \Kp\Km and \pip\pim pairs in the \Dz rest frame. This set of variables is expected to be sensitive to \CP violation in the interference between the S- and P-wave productions of the $\Kp\Km$ and $\pip\pim$ pairs. With a sample of about 108 thousand signal decays, all $a_X^\CP$ asymmetries were found consistent with zero, at the level of 0.4\% (including both statistical and systematic uncertainties).

Searches for \CP violation using model-dependent and model-independent analysis over the phase space were also performed for both $\Dz\to\Kp\Km\pip\pip$ and $\Dz\to\pip\pim\pip\pip$ modes. Full amplitude analyses for these decays were published in 2017 making use of CLEO data~\cite{dArgent_D0hhhh_2017} with limited samples sizes of a few thousand events, resulting in integrated asymmetries probed only at the level of 1.5--2\%, and in different sub-channels at 5--10\% or higher. 

\begin{figure}
\centerline{\includegraphics[align=c, width = 0.4\textwidth, height = 0.4\textwidth]{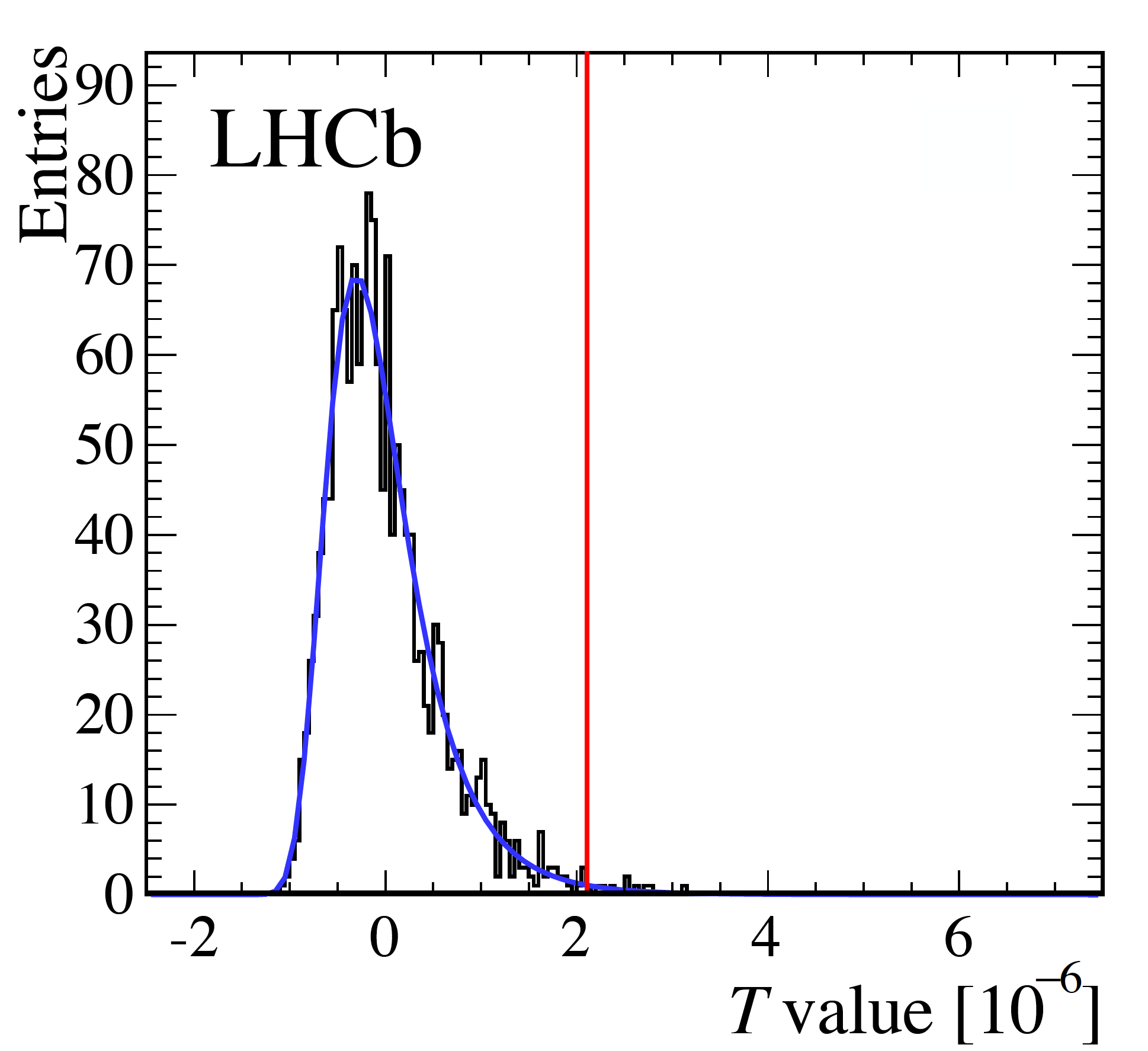} \includegraphics[align=c, width = 0.4\textwidth, height = 0.42\textwidth]{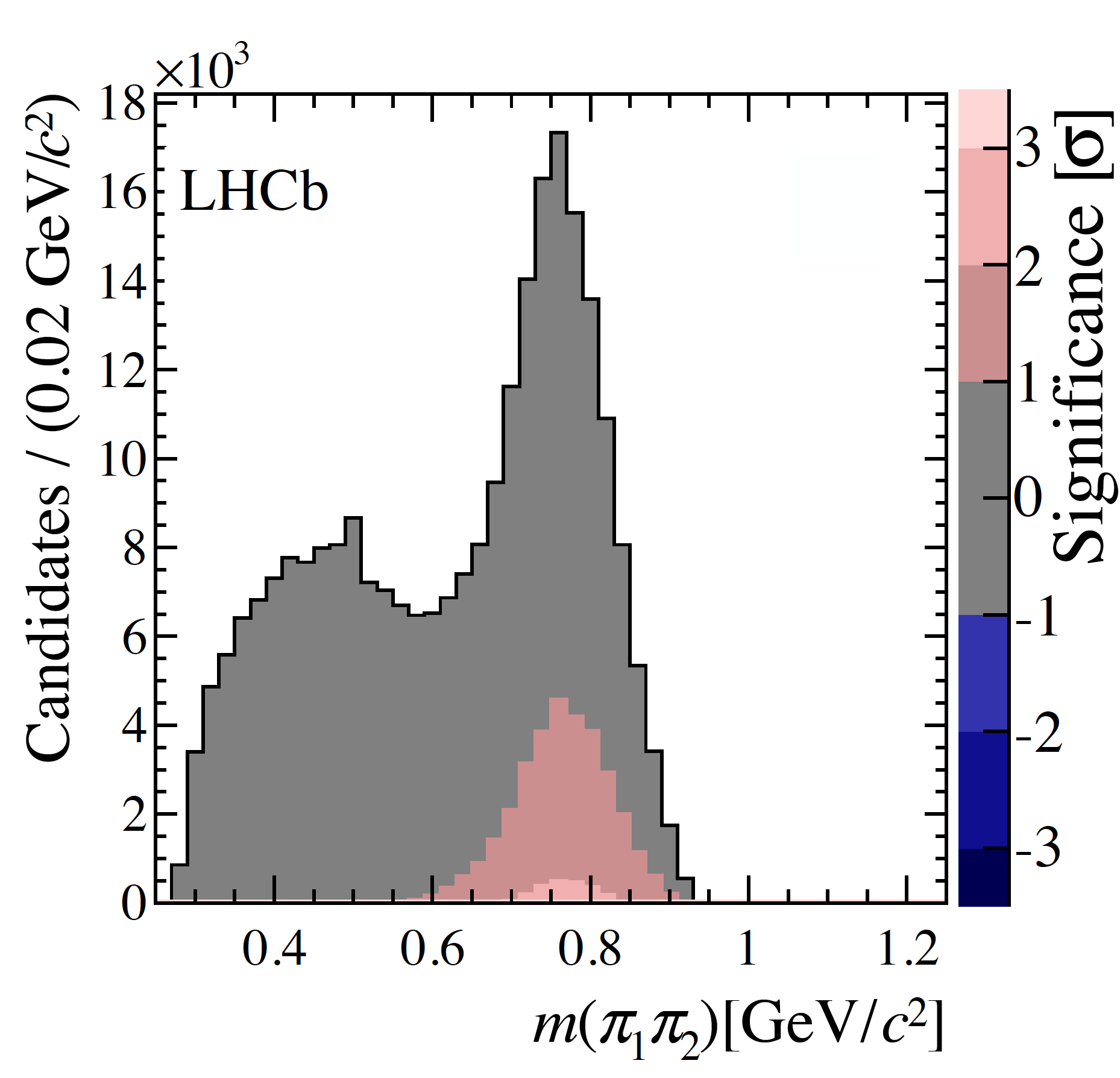}}
\caption{Results of the energy test for $\Dz\to \pim\pip\pim\pip$ decays by the LHCb collaboration \cite{LHCb-PAPER-2016-044}.  (left) Permutations of T-value for the $P$-odd \CP violation test. (right) The $\pi^-\pi^+$ invariant-mass projection showing the significance of the $T$-value contribution in different regions.}
\label{LHCb-D4pi-energytest}
\end{figure}

The LHCb collaboration studied both decays with much larger samples. In a first work with 1\invfb of data from run I ~\cite{LHCb-PAPER-2013-041},  $5.7\times 10^4$ and $3.3\times 10^5$  signal decays for $\Dz\to\Kp\Km\pip\pip$  and $\Dz\to \pim\pip\pim\pip$ modes, respectively, both tagged through $\Dstarp\to\Dz\pip$,  were analysed and \CP violation was searched for  via the Miranda technique over the phase space  defined by five invariants chosen as a set of  2- and three-body  mass combinations. The p-values under the no \CP violation hypothesis were found to be 9.1\% and 41\%, indicating consistency with \CP symmetry. 
With the full run I data sample, comprising about 1 million signal decays, the  $\Dz\to \pim\pip\pim\pip$ mode was then studied by LHCb via the energy test technique --- and also dividing the total sample into four sub-samples  depending on  the sign of a triple-product $C_T$  and on the  \Dz flavour~\cite{LHCb-PAPER-2016-044}. As such, the analysis could be sensitive to both $P$-even and $P$-odd \CP asymmetries. From simulation, sensitivity for \CP violation effects is found to be typically 4--5\% in amplitudes or 3--4$^\circ$ in phases for the main intermediate states. The results have  p-values of $(4.3\pm 0.6)\%$ and $(0.6\pm 0.2)\%$ for $P$-even and $P$-odd \CP violation tests, respectively, obtained from the distribution of permutation T-values.  Interestingly,  the $P$-odd \CP-violation test was found to be only marginally consistent with \CP conservation --- this can be seen in Fig.~\ref{LHCb-D4pi-energytest}, where the right plot indicates the $\pi^-\pi^+$ invariant-mass region where the discrepancies are found. 

The decay $\Dz\to\Kp\Km\pip\pip$ was also studied very recently by the LHCb collaboration through a full amplitude analysis from the full run I data sample (3\invfb) using semileptonic b-hadron decays into $\Dz\mun X$ final states \cite{LHCb-PAPER-2018-041}. The sample has almost 200 thousand candidates with 83\% purity. In total, 26 different intermediate states were considered for the phase-space fit and, for each, \CP magnitude asymmetries, relative phase differences, and \CP fraction asymmetries were obtained. The sensitivity for \CP violation varied depending on the sub-channel, from about 1\% to 15\%, with results compatible with \CP conservation.

There are yet no results using run II data from LHCb. The sensitivity expected for $a_\CP^{T-\rm odd}$, for example, could be as low as $2.4\times 10^{-4}$~\cite{LHCb-PII-Physics}.

\subsection{Charm Baryons}

    In general, the charm baryon sector is so far much less studied than the charm meson sector. Within the last 5 years or so, though, this scenario began to change, with many new results  on production and branching ratios \cite{BESIII_LambdaCS_2016, BESIII_LambdaAbsolute_2016, BESIII_LambdaAbsoluteSL_2017, BESIII_Lambdac_nKSpi_2016, BESIII_eeLcLcbarProd_2018, Acharya:2017lwf, Acharya:2017kfy,LHCb-PAPER-2012-041,LHCb-PAPER-2017-026, LHCb-PAPER-2018-040, LHCb-PAPER-2019-035}, lifetimes~\cite{LHCb-PAPER-2018-019,LHCb-PAPER-2018-028,LHCb-PAPER-2019-008}, decay asymmetries and polarisation \cite{BESIII_LambdaDecayAsymmetries_2019}, and new states\cite{LHCb-PAPER-2017-002, LHCb-PAPER-2017-018}. In what concerns \CP violation, there is a long way to go \cite{Bigi_CharmBaryons_2012} but the direct \CP violation observed in the meson sector is now adding fuel and pushing towards the study of charm baryons \cite{Wang_2019,Shi_Bigi_2019,Grossman_USpin_baryons_2019}. 
    
    Within the CKM ansatz, the underlying mechanism for the appearance of a weak phase in Cabibbo-suppressed charmed baryon decays is the same as for charmed mesons, $c\to du\bar u$ and $c\to su\bar u$ transitions. 
The main actors for the search of \CP violation are, in principle,  the two-body  $\Lc\to \Lz\Kp$ and three-body  $\Lc\to \proton\Kp\Km$ and  $\Lc\to \proton\pip\pim$.  From both the theoretical and experimental sides, the study involving three-body  baryon decays is challenging due to polarisation, five-dimensional phase space and short lifetime.  

\begin{figure}
\centering
\includegraphics[width = 0.35\textwidth]{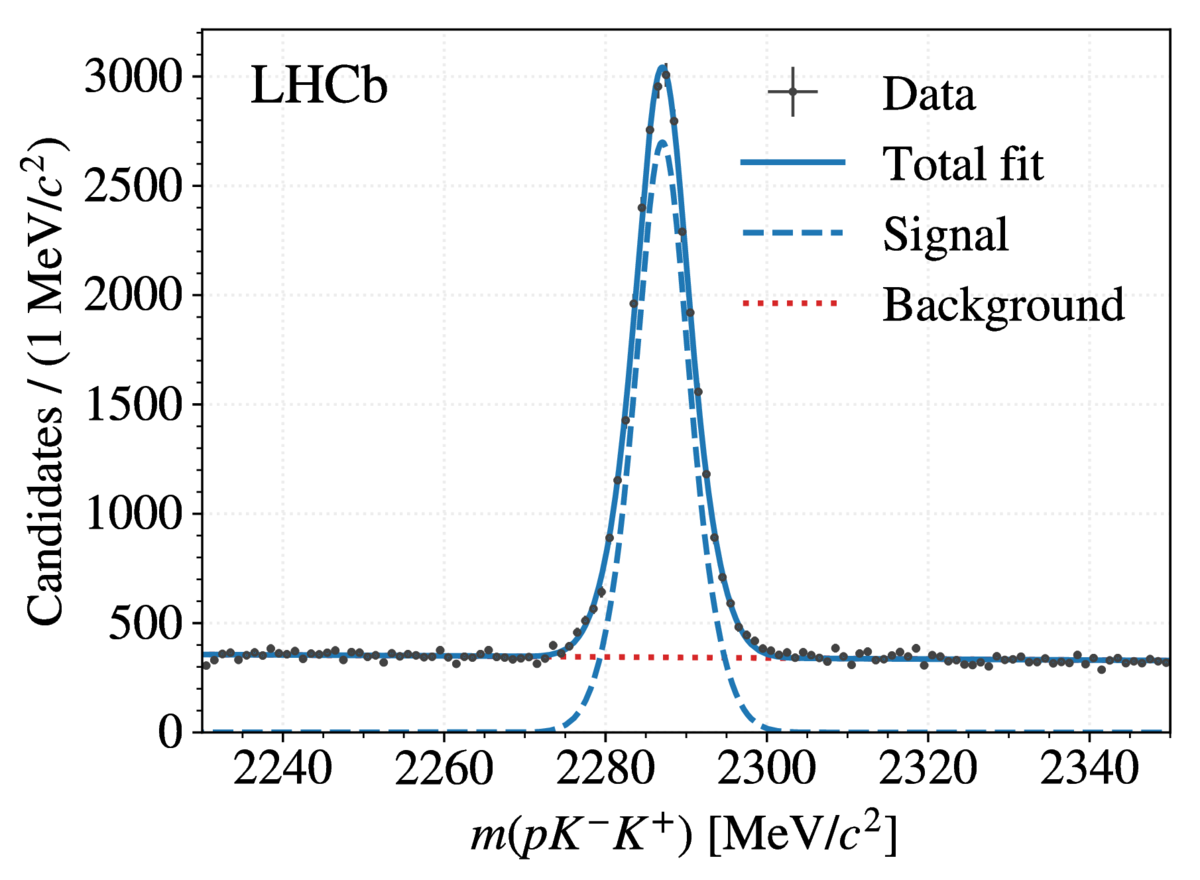}
\includegraphics[width = 0.35\textwidth]{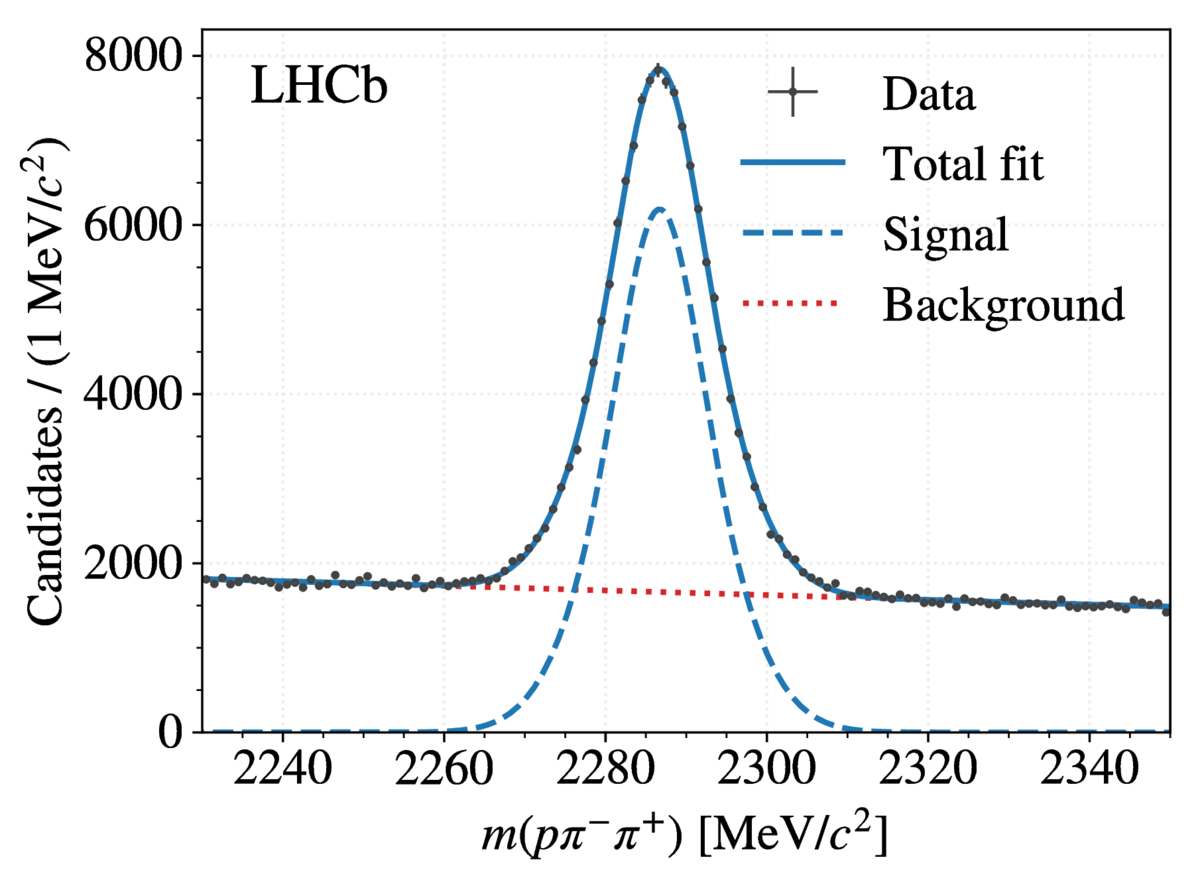} \\
\includegraphics[width = 0.35\textwidth]{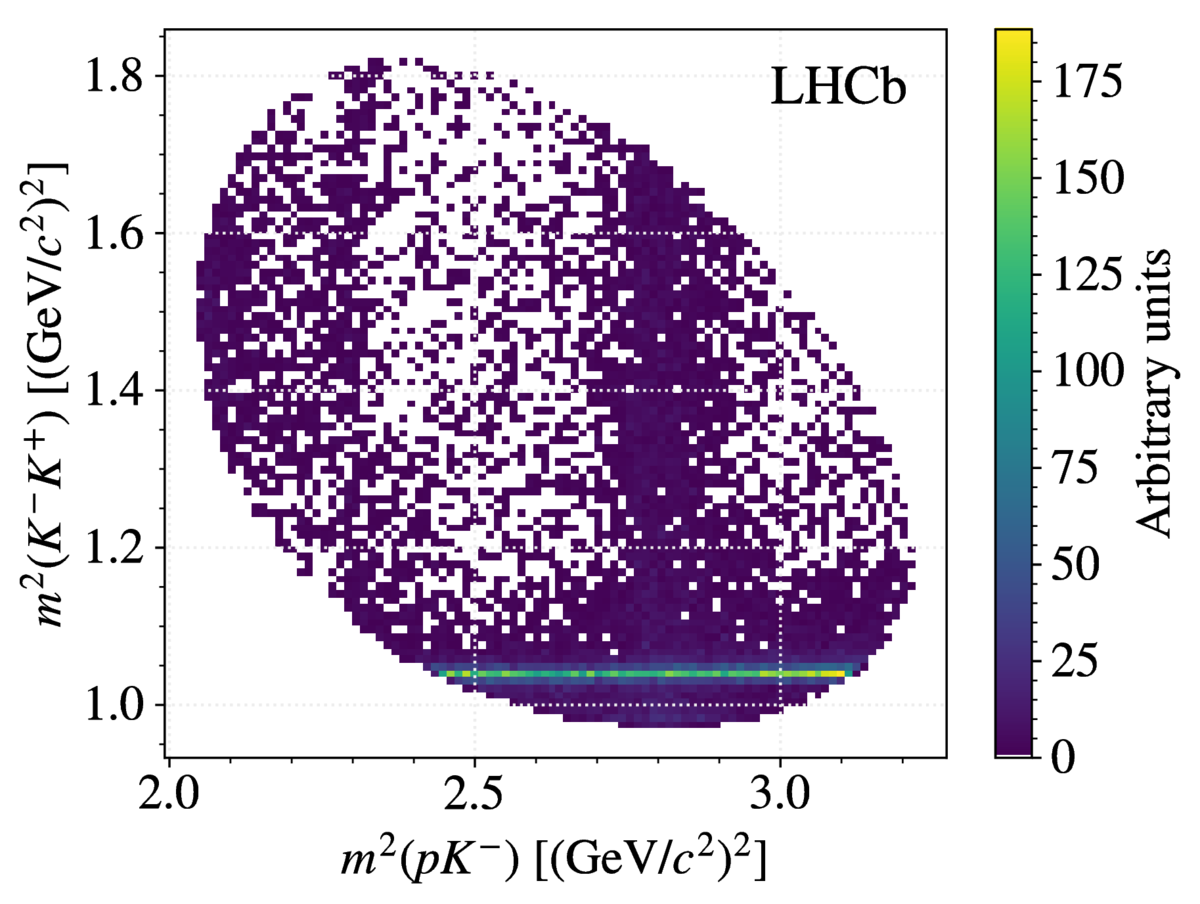}
\includegraphics[width = 0.35\textwidth]{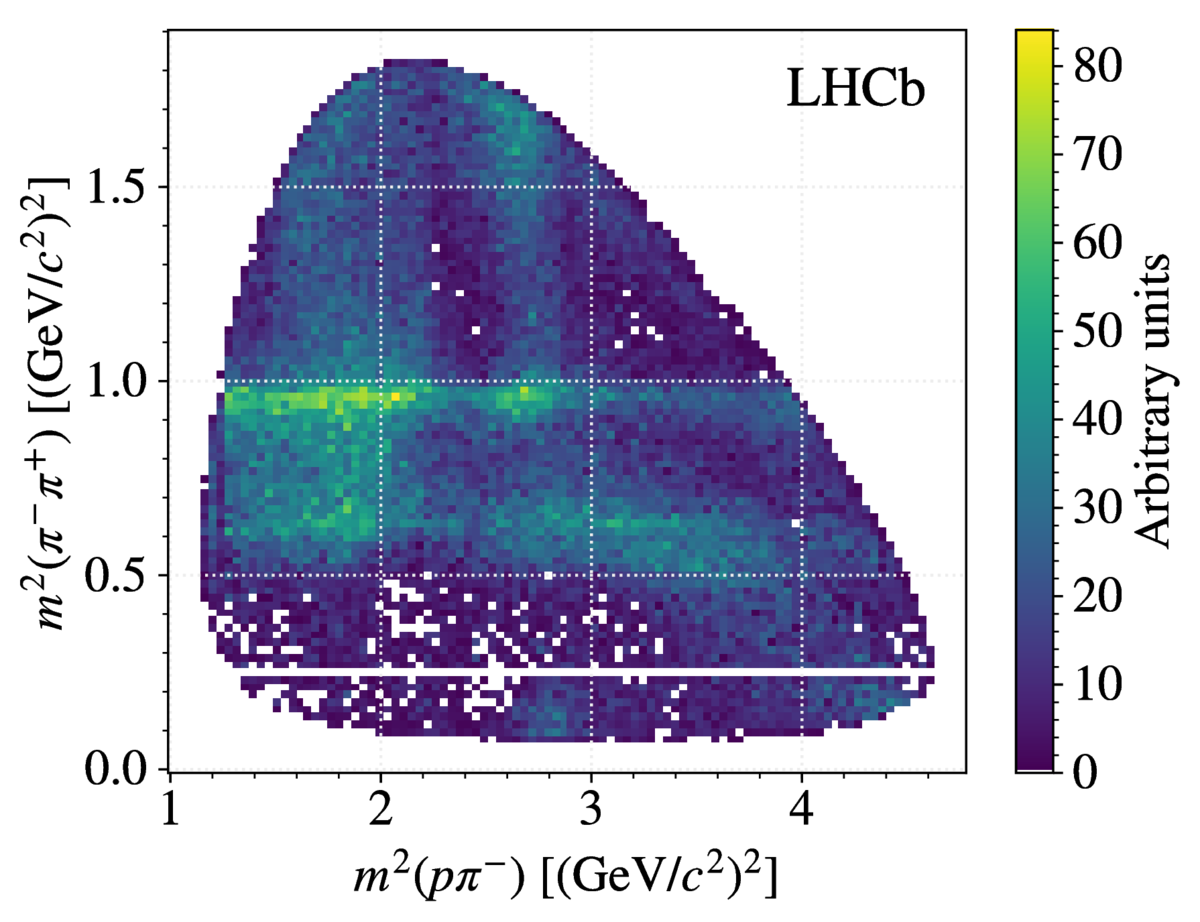} 
\caption{Invariant mass distributions for (top left) $\Lc\to \proton\Kp\Km$ and (top right) $\Lc\to \proton\pip\pim$ candidates. The corresponding Dalitz plots (background-subtracted and efficiency-corrected) are shown in the bottom plots~\cite{LHCb-PAPER-2017-044}.}
\label{LcDeltaACP}
\end{figure}

The LHCb collaboration performed the first search for \CP violation in the decays $\Lc\to \proton\Kp\Km$ and  $\Lc\to \proton\pip\pim$, using run I data~\cite{LHCb-PAPER-2017-044}. The decays were selected through the decay chain $\Lb\to \Lc\mun X$ (with $X$ being any additional unreconstructed particles) to diminish the level of background. The sample consisted of about 25 thousand $\Lc\to \proton\Kp\Km$ decays and 160 thousand $\Lc\to \proton\pip\pim$ decays. The invariant mass distributions and the Dalitz plots for both final states are shown in Fig.~\ref{LcDeltaACP}.
To cancel the $\Lb$ production asymmetry as well as muon and proton detection asymmetries, the differences of raw asymmetries are measured, once the kinematics of the $\Lc\to \proton\Kp\Km$ decays are matched to those of $\Lc\to \proton\pip\pim$. This procedure may alter the numerical value of the true physics asymmetry, thus the measured asymmetry for $\Lc\to \proton\Kp\Km$ is called a  {\it weighted} asymmetry, and the measurement is defined as
\begin{equation}
\Delta A_\CP^{\rm wgt} \equiv A_\CP(\proton\Kp\Km) - A_\CP^{\rm wgt} (\proton\pip\pim) \approx A_{\rm raw}(\proton\Kp\Km) - A_{\rm raw}^{\rm wgt}(\proton\pip\pim) ~.
\end{equation}
The result is
\begin{equation}
\Delta A_\CP^{\rm wgt}  = (0.30\pm 0.91\pm 0.61)\%~,
\end{equation}
showing consistency with \CP symmetry at the level of 1\%. It represents the first search for \CP violation in a three-body  baryon mode. New results are expected from LHCb using run II data not only on $\Lc$ but also, potentially,  $\Xic$ decay modes. Yet, to match the current precision in mesons the required amount of data will be possible only by the end of run V of LHCb, with 300\invfb ~\cite{LHCb-PII-Physics}.

\section{Conclusion}
\label{chapter8}

       More than half a century after the discovery of \CP violation,  many theoretical and experimental achievements led to a better understanding of this phenomenon. The CKM mechanism has proved to be  very successful in describing the current experimental data, in particular by the consistency among many different measurements leading to a  coherent  picture of the Unitarity Triangle. The measurement of $\gamma$ with a current precision of about $5^\circ$ is a remarkable result and $\gamma$ is, indeed, the only angle that have been measured exclusively from a direct \CP violation measurement, involving only tree-level amplitudes in \B decays to \D plus light mesons. 
              
        However, direct \CP violation in non-leptonic decays is in general a complex subject due to the interplay between the weak process and the underlying strong interactions. Although, on  the one hand, the strong phases are as much important as the weak phases for the observability of \CP violation, from the other hand it is very difficult to know a priori the size of the strong phases and, as a consequence, the expected level of \CP violation  effects in exclusive modes. Indeed, there is a long-term discussion about the relative importance of the short- versus long-distance contributions as sources of the strong phases. On this subject, much progress was obtained  within the  last years, mainly due to the recent experimental results. 

Some results in the \B-meson sector are particularly interesting. \CP violation is well established in  the two body modes  $ \BdorBs \to\Km\pip$ and $\Bz\to\pip\pim$. In quasi two-body modes,  first observations of \CP asymmetry were made for $\Bp\to\sigma(500)\pip$, $ \Bp\to  f_2(1270) \pip$ and $ \Bz\to \Kstarp \pim$ decays while no significant asymmetry was found in the $ \Bp\to \rho(770)\pip$  mode.
Big surprises came from three-body charmless \Bp decays. Rich patterns of large \CP asymmetries across the Dalitz plot were observed, with regions of positive and negative $A_\CP$. Since the source of weak phase from the SM in these decays comes only from the angle $\gamma$, the structures  need to arise from the variation of the strong phase. Moreover, these results point towards  the importance of long-distance hadronic phases. A particular result is the large \CP asymmetry of about 66\% associated to the rescattering region $\PK\PK\leftrightarrow \pi\pi$ in the decay $\Bp\to\Km\Kp\pip$. Another unforeseen result was the pattern of large \CP asymmetries at high $\pi\pi$ masses in the decay $\Bp\to\pip\pip\pim$, in a region not dominated by resonances. 

In the charm sector, a new era is starting with the observation of direct \CP violation  in the decays $\Dz\to\Kp\Km$ and $\Dz\to\pip\pim$, with a  \CP-asymmetry difference at the level of $10^{-3}$. The result, higher than expected from naive short-distance expectations, can be accommodated by models predicting strong, hadronic enhancements, although approaches based on NP models are also suggested. 
This outstanding result pushes further forward the study of other modes where effects of \CP violation may be at a similar level. This is the case of three- and four-body Cabibbo-suppressed modes such $\Dp\to\Km\Kp\pip$,  $\Dp\to\pim\pip\pip$ and $\Dz\to\Kp\Km\pip\pip$. From the beautiful lesson learnt from charmless \Bp decays, localised \CP asymmetries in the phase space can be potentially larger than integrated ones. The role of $\PK\PK\leftrightarrow \pi\pi$ rescattering can be potentially more important for charm-hadron decays too, and deserves attention.

An interesting and relevant discussion is the impact of \CPT symmetry -- assuming it to be exact --  on constraining \CP violation in exclusive modes. Final-state interactions, fundamentally hadronic,  are the carriers of this constraint: any observed \CP violation in one mode should be compensated by other(s) mode(s) with the same quantum numbers and, in particular, flavour content. Although the concept is clear, its practical consequences are not obvious when the multiplicity of decay channels is large, which is the case of \B decays. Yet, the experimental data, in general with a concentration near the edges of the Dalitz plot, may seem to indicate that \CPT can be more useful than anticipated \cite{Bigi_Charm_2015}. While quantitative statements are hard to make, the idea here is to  use it to enlighten  \CP violation studies  by trying to correlate different measurements or to search for new paths.  

A not-so-welcome surprise is the absence, so far, of a compelling evidence for \CP violation in baryons. The LHCb latest results in $\Lb\to\proton\pim\pip\pim$ decay with a larger sample have neither confirmed nor denied the hint for \CP violation in an earlier study. It is curious that \CP violation, being a necessary condition for baryogenesis, has only shown up clearly in the meson sector. 

The last years were naturally dominated by results from the LHCb collaboration, with  BaBar, Belle, and  BES III also contributing. New results, and hopefully  some surprises, are awaited from  run II LHCb  data, since  many decay modes are still being analysed. This is the case in particular of multi-body decays discussed in this review, including baryons. Belle II is now entering the game, with the ability to provide good physics impact to final states with neutrals. One, from many examples, is the possibility to soon bring new elements to the $\Delta A_\CP(\B\to \PK\pi)$ puzzle from the analyses of decay modes $ \Bz \to \Kz\piz$, $ \Bz \to \Kz\eta $, and  $ \Bz \to \Kz\eta'$. Similarly, three-body \B decay modes with neutrals in the final states may help to better understand the sources of direct \CP violation, such as rescattering contributions and the S- and P-wave interference.

In a near future, with further accumulation of Belle II data and the start of the upgraded LHCb from run III (and beyond), flavour physics and in particular \CP violation will be entering into an era of high precision.  Besides testing the validity of the SM and representing a tool for NP searches, precise measurements of direct \CP violation, especially coming from multi-body decays, could be a very interesting instrument for the understanding of hadronic interactions. In any scenario, \CP violation is expected to maintain a leading role in Particle Physics.

\section*{Aknowledgements}
The authors want to thank S. Amato, J. Miranda, I. Nasteva, E. Polycarpo and A. Reis for the useful  discussions while preparing this review. 
This work was partially supported from the Brazilian funding agencies {\it Coordenação de Aperfeiçoamento de Pessoal de Nível Superior} (CAPES) [Finance Code 001], {\it Conselho Nacional de Desenvolvimento Científico e Tecnológico} (CNPq), and {\it Fundação Carlos Chagas Filho de Amparo e Pesquisa do Estado do Rio de Janeiro} (FAPERJ).

\end{document}